\documentclass[a4paper,12pt]{article}

\usepackage[english]{babel}
\usepackage[utf8x]{inputenc}
\usepackage[T1]{fontenc}

\usepackage[a4paper,top=2.5cm,bottom=2cm,left=2.5cm,right=2.5cm,marginparwidth=1.75cm]{geometry}

\usepackage{etoolbox}
\makeatletter
\patchcmd{\maketitle}{\@makefntext}{\fakecommand}{}{}
\patchcmd{\maketitle}{\rlap}{\hbox}{}{}
\patchcmd{\@maketitle}{\@author}{\hspace*{5pt}\@author}{}{}
\makeatother

\usepackage{amsmath}
\usepackage{graphicx}
\usepackage[colorinlistoftodos]{todonotes}
\usepackage[colorlinks=true, allcolors=blue]{hyperref}
\usepackage{setspace}
\usepackage{apacite}
\usepackage{dsfont}
\usepackage{booktabs}
\usepackage[titletoc]{appendix}
\usepackage{tabularx} 
\usepackage{colortbl}
\usepackage{longtable}
\usepackage{threeparttablex} 
\usepackage{threeparttable}
\usepackage{chngcntr}		
\usepackage{subcaption}
\usepackage[section]{placeins}
\usepackage{microtype}     
\usepackage{caption}        
\usepackage{rotating}
\usepackage{pdflscape}    
\usepackage{afterpage}
\usepackage{geometry}
\usepackage{color} 
\usepackage{float} 
\usepackage{lmodern}
\usepackage{hyperref}

\usepackage[hang,flushmargin]{footmisc}

\DeclareMathOperator*{\argmin}{\arg\min}
\newcommand{\bigCI}{\mathrel{\text{\scalebox{1.07}{$\perp\mkern-10mu\perp$}}}} 

\newcommand\blfootnote[1]{%
  \begingroup
  \renewcommand\thefootnote{}\footnote{#1}%
  \endgroup
}

\title{Machine Learning Estimation of Heterogeneous Causal Effects: Empirical Monte Carlo Evidence\blfootnote{Financial support from the Swiss National Science Foundation (SNSF) is gratefully acknowledged. The study is part of the project "Causal Analysis with Big Data" (grant number SNSF 407540\_166999) of the Swiss National Research Programme "Big Data" (NRP 75). A previous version of the paper was presented at the the Economics Departments of University of Wisconsin at Madison, the Universities of California at Berkeley, Davis, Irvine, Los Angeles, and San Diego, at Amazon, Seattle, \textit{Econometrics in the Castle: Machine Learning in Economics and Econometrics} in Munich, and the \textit{Statistical Week} in Linz. We thank participants, in particular Bryan Graham, Andres Santos, Alejandro Schuler, Stefan Wager and Michael Zimmert for helpful comments and suggestions. The usual disclaimer applies.}
} 

\author{Michael C. Knaus\thanks{University of St. Gallen. Michael C. Knaus is also affiliated with IZA, Bonn, \href{mailto:michael.knaus@unisg.ch}{michael.knaus@unisg.ch}. } \and Michael Lechner\thanks{University of St. Gallen. Michael Lechner is also affiliated with CEPR, London, CESifo, Munich, IAB, Nuremberg, and IZA, Bonn, \href{mailto:michael.lechner@unisg.ch}{michael.lechner@unisg.ch}.} \and
Anthony Strittmatter\thanks{University of St. Gallen, \href{mailto:anthony.strittmatter@unisg.ch}{anthony.strittmatter@unisg.ch}.}}

\date{First version: October 31, 2018 \\ \smallskip This version: \today }

\begin{document}
\maketitle

\onehalfspacing

\begin{abstract}
We investigate the finite sample performance of causal machine learning estimators for heterogeneous causal effects at different aggregation levels. We employ an Empirical Monte Carlo Study that relies on arguably realistic data generation processes (DGPs) based on actual data. We consider 24 different DGPs, eleven different causal machine learning estimators, and three aggregation levels of the estimated effects. In the main DGPs, we allow for selection into treatment based on a rich set of observable covariates. We provide evidence that the estimators can be categorized into three groups. The first group performs consistently well across all DGPs and aggregation levels. These estimators have multiple steps to account for the selection into the treatment and the outcome process. The second group shows competitive performance only for particular DGPs. The third group is clearly outperformed by the other estimators.


\bigskip



\end{abstract}

\newpage
\doublespacing
\section{Introduction}

Economists and many other professionals are interested in causal effects of policies or interventions. This has triggered substantial advances in microeconometrics and statistics in understanding the identification and estimation of different average causal effects in the recent decades \cite<see, e.g.,>[and references therein]{Imbens2009,Athey2017}. However, in most applications it is also interesting to look beyond the average effects in order to understand how the causal effects vary with observable characteristics. For example, finding those individuals who benefit most from active labor market policies, promotion campaigns or medical treatments is important for the efficient allocation of public and private resources.

In recent years, methods for the systematic estimation of heterogeneous causal effects have been developed in different research disciplines. Those methods adapt standard machine learning methods to flexibly estimate heterogeneity along a potentially large number of covariates. The suggested estimators use regression trees \cite{Su2009,Athey2016}, Random Forests \cite{Wager2017,Athey2017a}, the least absolute shrinkage and selection operator (Lasso) \cite{Qian2011,Tian2014,Chen2017}, support vector machines \cite{Imai2013}, boosting \cite{Powers2018}, neural nets \cite{johansson2016learning,Shalit2016EstimatingAlgorithms,Schwab2018PerfectNetworks} or Bayesian machine learning \cite{Hill2011,Taddy2016}.\footnote{\citeA{Hastie2009} introduce the underlying machine learning algorithms. \citeA{Athey2018TheEconomics} and \citeA{belloni2014high} provide an overview how those methods might be used in the estimation of average causal effects and other parameters of interest.} Recently, the first applied studies using these methods appeared in economics \cite<e.g.,>{Bertrand2017,Davis2017,Knaus2017,Andini2018,Ascarza2018,Strittmatter2018WhatEvaluation}.

In contrast to the rather mature literature about the estimation of average causal effects, the literature on the estimation of effect heterogeneity is still lacking guidance for practitioners about which methods are well suited for their intended applications. Theoretical asymptotic approximations are currently either not available, incomplete, or they are based on non-overlapping assumptions preventing comparisons of estimators. Furthermore, the available information about the finite sample performance is of limited use to practitioners. Most comparisons are based on data generating processes (DGPs) that are very unrealistic for real applications. One exception is \citeA{Wendling2018} who base their simulation study on data from medical records. However, they focus in their study on the special case of binary outcomes and data structures that are unusual in economics.

In this study, we categorize major approaches from different fields. We distinguish between generic approaches that can be combined with a variety of different off-the-shelf machine learning estimators and estimator specific approaches that modify an existing method. The generic approaches are combined with the machine learning estimators Random Forest and Lasso. This leads to eleven different causal machine learning estimators under investigation. As opposed to standard simulation methods that rely on a synthetic DGP, we investigate the finite sample performance of these estimators in an Empirical Monte Carlo (EMCS) approach \cite<e.g.,>{Huber2013,lechner2013sensitivity}. An EMCS informs the DGPs as much as possible by real data and reduces synthetic components in the DGP to a minimum. We consider six different specifications of the heterogeneous causal effects, two different sample sizes, and DGPs with and without selection into treatment.\footnote{We focus on point estimation and leave the investigation of inference methods for further research.}
 
Our contribution to the aforementioned literature is three-fold: First, we provide a comprehensive comparison of different estimators and DGPs. Second, we consider the finite sample properties of causal machine learning estimators for effect heterogeneity under DGPs that are arguably realistic at least in some fields of economics. Third, this is the first simulation study that considers also different aggregation levels of the heterogeneous effects. In particular, we consider an intermediate aggregation level between the most individualized causal effects and the average population effect. Such intermediate aggregation levels are important as feasible action rules for practitioners. 

Our findings suggest that no causal machine learning estimator is superior for all DGPs and aggregation levels. However, four estimators show a relatively good performance in all settings: Random Forests combined with a doubly robust outcome modification \cite<based on>{Chernozhukov2018}, Causal Forest with local centering \cite{Athey2017a}, Lasso combined with a covariate modification and efficiency augmentation \cite{Tian2014}, and Lasso with R-learning \cite{Nie2017}. All those methods use multiple estimation steps to account for the selection into treatment and the outcome process. Several other estimators may be suitable in specific empirical settings but their performance is unstable across different DGPs. Lasso estimators tend to be more unstable than Random Forests, which frequently prevents them from achieving a normal distribution. 

In the next section, we introduce the notation and the estimation targets. In Section \ref{sec:cml}, we describe and categorize causal machine learning approaches to estimate heterogeneous causal effects. In Section \ref{sec3}, we explain the implementation of the estimators. In Section \ref{sec4}, we discuss the EMCS approach. In Section \ref{sec5}, we provide our simulation results. The final Section concludes and hints at some topics for future research. Appendices \ref{sec:App_Dat}-\ref{sec:final_app} contain supplementary statistics and results. Appendix \ref{sec:App_Dat} describes the data that are used for the EMCS. Appendices \ref{sec:DGP} and \ref{sec:app-impl} provide details about the DGPs and the implementation of estimators, respectively. Finally, Appendix \ref{sec:final_app} shows and discusses the full simulation results of all DGPs. We provide code that implements the estimators under investigation in the R package \href{https://github.com/MCKnaus/CATEs}{\texttt{CATEs}} on GitHub.

\section{Notation and estimation targets}

We describe the parameters of interest using Rubin’s \citeyear{Rubin1974} potential outcome framework. The dummy variable $D_i$ indicates a binary treatment, e.g. participation in a training program. Let $Y_i^1$ denote the outcome (e.g., employment) if individual $i$ ($i=1, ..., N$) receives the treatment ($D_i=1$). Correspondingly, $Y_i^0$ denotes the outcome if individual $i$ does not receive the treatment ($D_i=0$). Each individual can either receive the treatment or not. This means that only one of the two potential outcomes ($Y^d_i$) is observable: 
\begin{equation}
	\label{OR}
	Y_i = D_i Y_i^1 + (1-D_i) Y_i^0.
\end{equation}

Thus, the \textit{individual treatment effect} (ITE) $\xi_i = Y_i^1 - Y_i^0$ of $D_i$ on $Y_i$ is never observed. However, the identification of expectations of $\xi_i$ may be possible under plausible assumptions. For example, the identification of the average treatment effect (ATE), $\tau = E[\xi_i]$, or the average treatment effect on the treated (ATET), $\theta = E[\xi_i \mid D_i=1]$, are standard in microeconometrics \cite<see, e.g.,>{Imbens2009}.

The focus of this study is on conditional average treatment effects (CATEs). CATEs take the expectations of $\xi_i$ conditional on exogenous pre-treatment covariates.\footnote{Covariates are also called features or predictors in parts of the machine learning literature.} We call the finest conditioning level that uses all available covariates $X_i$ \textit{individualized average treatment effect} (IATE),

\begin{equation}
	\label{IATE}
	\tau(x) = E[\xi_i \mid X_i=x] = \mu^1(x) - \mu^0(x),
\end{equation}

where $\mu^d(x) = E[Y_i^d \mid  X_i=x]$ denotes the conditional expectation of the unobserved potential outcomes. IATEs provide an approximation of ITEs for the set of covariates that are at the disposal of the researcher in a specific application. However, researchers may additionally be interested in intermediate aggregation levels that are coarser than IATEs but finer than ATEs. Especially groups based on a smaller set of pre-defined characteristics, $G_i$, may be of interest if the estimated IATEs need to be summarized for the research community, communicated to practitioners, or acted upon.\footnote{For example, if interest is in gender differences, $G_i \in \lbrace female, male \rbrace$.} We call the effects defined on this aggregation level \textit{group average treatment effects} (GATE),

\begin{equation}
	\label{GATE}
	\tau(g) = E[\xi_i \mid G_i = g] = \int\tau(x)f_{X_i \mid G_i=g}(x)dx.
\end{equation}

The identification of any aggregation level of individual treatment effects in observational studies is complicated by non-random treatment assignment. However, identification of the IATE and any coarser aggregation level is still possible if the observable covariates $X_i$ contain all confounders.\footnote{$X_i$ represents the union of confounders and heterogeneity variables for notational convenience. In principle, they may be completely, partly or non-overlapping \cite<see, e.g.,>{Knaus2017}.} These are covariates that jointly affect the treatment probability and the potential outcomes. Although there are alternative ways to identify the various effects, here we focus on the case where all the confounders are contained in the data available to the researcher. This means that we operate throughout the paper under the following assumptions.

\textbf{Assumption 1:} (Conditional independence): $Y_i^1,Y_i^0 \bigCI D_i \mid X_i=x$, for all $x$ in the support of $X_i$.

\textbf{Assumption 2:} (Common support): $0 < P[D_i = 1 \mid X_i =x] = p(x) < 1$, for all $x$ in the support of $X_i$.

\textbf{Assumption 3:} (Exogeneity of covariates): $X_i^1 = X_i^0$.\footnote{The potential confounders $X_i^d$ are defined equivalently to potential outcomes.}

\textbf{Assumption 4:} (Stable Unit Treatment Value Assumption, SUTVA): $Y_i=Y_i^1 D_i + Y_i^0 (1-D_i)$.

Assumption 1 states that the potential outcomes are independent of the treatment conditional on the confounding covariates. According to Assumption 2, the conditional treatment probability (often called propensity score) is bounded away from zero and one. Assumption 3 requires that the covariates are not affected by the treatment. Assumption 4 excludes spillover effects between treated and non-treated. Under Assumptions 1-4, 
\begin{align}
    E[Y_i^d \mid X_i=x, D_i = 1 - d] &  = E[Y_i \mid X_i = x , D_i = d] = \mu(d,x) \\
	\Rightarrow ~ \tau(x) & = \mu(1,x) - \mu(0,x),
	\label{Ident}
\end{align}

and thus IATEs, GATEs and ATE are identified from observable data. We denote the conditional expectations of the outcomes in one treatment arm by $\mu(d,x) = E[Y_i \mid X_i = x , D_i = d]$, the conditional expectation of the outcome as $\mu(x) = E[Y_i\mid X_i =x]$, and the conditional treatment probability by $p(x) = P[D_i = 1 \mid X_i =x]$.

\section{Causal machine learning of effect heterogeneity} \label{sec:cml}

Equation \ref{Ident} shows that the fundamental task is to estimate the difference of two conditional expectations. However, we never observe the differences at the individual level and have to estimate them in two different subpopulations. Thus, the estimation of IATEs is a non-standard machine learning problem. In this section, we present different approaches to target the estimation of IATEs. We distinguish between generic approaches and one estimator specific approach. Generic approaches split the causal estimation problem into several standard prediction problems and may be combined with a large variety of supervised machine learning estimators. On the other hand, Causal Forest \citeA{Athey2017a} is a modification of a specific machine learning estimator to move the target from the estimation of outcomes to the estimation of IATEs.

\subsection{Generic approaches}\label{sec:ga}

A straightforward generic approach follows directly from Equation \ref{Ident}. \textit{Conditional mean regressions} takes the difference of conditional expectations that are estimated in the two samples of treated and non-treated separately using off-the-shelf machine learning methods to estimate the conditional outcome means $\hat{\mu}(d,x)$:\footnote{This approach is also referred to as T-Learner \cite{Kunzel2017,Nie2017} or Q-Learning \cite{Qian2011}.}

\begin{equation}
	\label{CMR}
	\hat{\tau}_{CMR}(x) = \hat{\mu}(1,x) - \hat{\mu}(0,x).
\end{equation}

Conditional mean regressions are straightforward to implement. Any supervised machine learning methods for conditional mean estimation may be used. However, their usual target is to minimize the mean squared error (MSE) in two separate prediction problems and they are not tailored to estimate IATEs.\footnote{For an intuition why this is not optimal: Biases that for the same value of $x$ go in the same direction are less harmful than if they go in opposite directions. However, this cannot be accounted for by separate MSEs that are not directly linked (for a new Causal Forest estimator that takes up this theme directly, see \citeA{Lechner2018PenalizedEffects}).} This suggests that they may be outperformed by more specialized methods for this causal problem. Three generic multi-step approaches that target IATE estimation are presented in the following and a framework to summarize them is provided.

\subsubsection{Modified outcome methods}
\citeA{Abadie2005SemiparametricEstimators} introduces the idea of modifying the outcome to estimate conditional average treatment effects on the treated in studies based on difference-in differences. For IATEs in the experimental and observational setting, the idea is formulated by \citeA{Signorovitch2007} and \citeA{Zhang2012}, respectively. The latter discuss two modifications of the outcome, which we summarize as modified outcome methods (MOM). The first is based on inverse probability weighting (IPW) \cite<e.g.,>{Horvitz1952,Hirano2003} where the modified outcome is

\begin{equation}
	\label{MOMIPW}
	Y_{i,IPW}^* = Y_i \dfrac{D_i - p(X_i)}{p(X_i)(1-p(X_i))}.
\end{equation}

The second is based on the doubly robust (DR) estimator of \citeA{Robins1995SemiparametricData},

\begin{equation}
	\label{MOMDR}
	Y_{i,DR}^* = \mu(1,X_i) - \mu(0,X_i) + \dfrac{D_i (Y_i - \mu(1,X_i))}{p(X_i)} - \dfrac{(1 - D_i) (Y_i - \mu(0,X_i))}{(1-p(X_i))}.
\end{equation}

The crucial insight here is that $\tau(x) = E[Y_{i,IPW}^* \mid X_i = x] = E[Y_{i,DR}^* \mid X_i = x]$. This means that a regression with one of these modified outcomes and covariates $X_i$ can be used to obtain estimates of IATEs, $\hat{\tau}_{IPW}(x)$ or $\hat{\tau}_{DR}(x)$. In practice, the researcher has no access to the true parameters $p(x)$ and $\mu(d,x)$, the so-called nuisance parameters. The conditional expectations need to be approximated in a first step and plugged into Equations \ref{MOMIPW} and \ref{MOMDR}. Any suitable prediction method can be used to estimate the nuisance parameters as well as the IATEs.

The asymptotic properties of $E[Y_{i,DR}^*]$ as estimator for ATEs are well understood \cite{Belloni2014InferenceControls,Farrell2015,Belloni2017,Chernozhukov2017,Chernozhukov2018}. Furthermore, \citeA{Abrevaya2015EstimatingEffects} and \citeA{Lee2017} analyze the properties of estimating $\tau(z) = E[Y_{i,IPW}^* \mid Z_i = z]$ and $\tau(z) = E[Y_{i,DR}^* \mid Z_i = z]$ for a low-dimensional subset of covariates ($Z_i$), respectively. However, both do not consider machine learning to estimate the nuisance parameters. We are currently not aware of theoretical results for the case where nuisance parameters and IATEs are estimated with machine learning. 

Simulation evidence of \citeA{Powers2018} suggests that estimators based on $Y_{i,IPW}^*$ may exhibit high variance due to potentially extreme values of the propensity score in the denominator. Estimators based on $Y_{i,DR}^*$ might be more stable, because of the double-robustness property, but this is unexplored until now.

\subsubsection{Modified Covariate Method}

\citeA{Tian2014} introduce the modified covariate method (MCM) for experiments and \citeA{Chen2017} extend it to observational studies. They show that we can estimate IATEs by solving the objective function

\begin{equation}
	\label{MCMnp}
	\min_{\tau} \left[ \dfrac{1}{N} \sum_{i=1}^N T_i \dfrac{D_i - p(X_i)}{p(X_i)(1-p(X_i))} \left( Y_i -  \frac{T_i}{2} \tau(X_i) \right)^2  \right],
\end{equation}

where $T_i = 2D_i-1 \in \lbrace -1,1 \rbrace$. The name MCM results from the practice to replace the non-parametric function of the IATE with a linear working model, $\tau(x) = x \beta$. This enables us to rewrite the minimization problem \ref{MCMnp} as 

\begin{equation}
	\label{MCMli}
	\hat{\beta}_{MCM} = \argmin_{\beta} \left[ \dfrac{1}{N} \sum_{i=1}^N T_i \dfrac{D_i - p(X_i)}{p(X_i)(1-p(X_i))} \left( Y_i -  X_i^{MCM} \beta \right)^2  \right],
\end{equation}

where $X_i^{MCM}=T_i/2X_i$ are the modified covariates. The estimated IATEs are then obtained by $\hat{\tau}_{MCM}(x) = x\hat{\beta}_{MCM}$. The nuisance parameter $p(x)$ needs to be estimated in a first step using any suitable method. 

In principle, rewriting \ref{MCMnp} as 

\begin{equation}
	\label{MCMgen}
	\min_{\tau} \left[ \dfrac{1}{N} \sum_{i=1}^N T_i \dfrac{D_i - p(X_i)}{4p(X_i)(1-p(X_i))} \left( 2 T_i Y_i - \tau(X_i) \right)^2  \right],
\end{equation}

allows to apply any machine learning estimator that is able to solve weighted minimization problems. However, we are not aware of any study that notices and pursues this possibility. 

MCM does not require to specify any model of the so-called main effects $\mu(x)$ or $\mu(d,x)$. However, \citeA{Tian2014} describe that an estimate of $\mu(x)$ might be useful to increase efficiency. The efficiency augmented version replaces the outcome $Y_i$ in Equations \ref{MCMnp} to \ref{MCMgen} by the residuals $Y_i - \mu(X_i)$. Thus, MCM with \textit{efficiency augmentation}  (EA) requires additionally to estimate the nuisance parameter $\mu(x)$ in the first step. \citeA{Tian2014} show that MCM with a linear working model provides the best linear predictor of the potentially non-linear $\tau(x)$. However, we are not aware of any further theoretical analyses of the statistical properties of this approach.

\subsubsection{R-learning}

\citeA{Nie2017} propose R-learning that is based on the partially linear model of \citeA{Robinson1988Root-n-consistentRegression}. It is equivalent to MCM with EA for 50:50 randomization but solves otherwise a different minimization problem to estimate IATEs:\footnote{\citeA{Murphy2003} and \citeA{Robins2004} develop a similar method for optimal dynamic treatment regimes called A-Learning. It is adapted for the binary treatment case by \citeA{Chen2017}. We focus on R-Learning because it explicitly applies machine learning methods in all estimation steps.}

\begin{equation}
	\label{RLnp}
	\min_{\tau} \left\lbrace \dfrac{1}{N} \sum_{i=1}^N  \left[ ( Y_i - \mu(X_i)) - (D_i - p(X_i)) \tau(X_i) \right]^2 \right\rbrace
\end{equation}

Like for MCM, current implementations consider a linear working model for the IATE \cite{Nie2017,Zhao2017} and solve 

\begin{equation}
	\label{RLli}
	\hat{\beta}_{RL} = \argmin_{\beta} \left\lbrace \dfrac{1}{N} \sum_{i=1}^N \left[ ( Y_i - \mu(X_i)) -  X_i^{RL} \beta \right]^2 \right\rbrace,
\end{equation}

where $X_i^{RL}=(D_i - p(X_i))X_i$ can be considered as an alternative way to modify covariates. The estimated IATEs are then obtained by $\hat{\tau}_{RL}(x) = x\hat{\beta}_{RL}$. Similar to MCM, \ref{RLnp} can be rewritten as

\begin{equation}
	\label{RLgen}
	\min_{\tau} \left\lbrace \dfrac{1}{N} \sum_{i=1}^N  (D_i - p(X_i))^2 \left[ \dfrac{Y_i - \mu(X_i)}{D_i - p(X_i)} - \tau(X_i) \right]^2 \right\rbrace
\end{equation} 

and solved with any suitable method after estimating the nuisance parameters in a first step \cite<see also>{Schuler2018AEffects}.

\citeA{Nie2017} show that R-learning can perform as good as an oracle estimator that knows the real nuisance parameters in the special case of solving \ref{RLnp} with penalized kernel regression. This result requires that the estimators of the nuisance parameters need to be fourth root consistent in the semiparametric case.

\subsubsection{Summary of generic approaches to estimate IATEs}

One goal of this paper is to structure approaches that estimate IATEs coming from different literatures. One way to put the approaches above on more common ground is by noting that they can all be considered as solving a weighted minimization problem with modified outcomes:

\begin{equation}
	\label{general}
	\min_{\tau} \left\lbrace \dfrac{1}{N} \sum_{i=1}^N   w_i  \left[  Y_i^*  - \tau(X_i) \right]^2 \right\rbrace.
\end{equation}

Table \ref{tab:general-approach} summarizes the weights, $w_i$, and outcome modifications, $Y_i^*$, underlying the different approaches. This common representation is helpful to see and understand the differences of the methods. The two MOM methods require no additional weighting because $\tau(x) = E[Y_{i,IPW}^* \mid X_i = x] = E[Y_{i,DR}^* \mid X_i = x]$. For MCM, $\tau(x) = E[2 T_i Y_i \mid X_i = x] = E[2 T_i (Y_i - \mu(X_i)) \mid X_i = x]$ in the special case of 50:50 randomization of the treatment ($p(x)=0.5$). Any deviating assignment mechanism requires reweighting with IPW weights to control for the deviation from 50:50 randomization. The modified outcome of R-learning is equivalent to MCM with efficiency augmentation if $p(x)=0.5$. However, while the MCM modified outcome does not change with other propensity scores and preserves the interpretation as mean comparison under 50:50 randomization, $E[(Y_i - \mu(X_i)) / (D_i - p(X_i)) \mid X_i = x]$ is lacking such an intuitive interpretation.

\begin{table}[t!]
\centering \doublespacing
\begin{threeparttable}
  \caption{Summary of generic approaches to estimate IATEs}
    \begin{tabular}{lcc}
        \toprule
		Approach & $w_i$ & $Y_i^*$ \\
\midrule
		MOM IPW & 1 & $Y_{i,IPW}^*$  \\
	    MOM DR & 1 & $Y_{i,DR}^*$  \\
	    MCM & $ T_i \dfrac{D_i - p(X_i)}{4 p(X_i)(1-p(X_i))} $ & $ 2 T_i Y_i$ \\
        MCM with EA & $ T_i \dfrac{D_i - p(X_i)}{4 p(X_i)(1-p(X_i))} $ & $2 T_i (Y_i - \mu(X_i))$ \\
	    R-Learning & $(D_i - p(X_i))^2$ & $\dfrac{Y_i - \mu(X_i)}{D_i - p(X_i)}$ \\
    	\bottomrule
    \end{tabular}%
     \label{tab:general-approach}
\end{threeparttable}
\end{table}

\subsection{Causal Forest}\label{sec:cf}

Another strand of literature modifies machine learning algorithms based on regression trees \cite{breiman1984} to estimate IATEs. We focus on Causal Forest that is a special case of the Generalized Random Forest of \citeA{Athey2017a} as the most recent estimator in this line of research.\footnote{Previous works concerned with estimating IATEs by modifying tree-based methods are \citeA{Su2009}, \citeA{Athey2016} and \citeA{Wager2017}.} Causal Forests build on the idea of Random Forests \cite{Breiman2001} and boil down to taking the difference of two weighted means in our case of binary treatments,

\begin{equation}
	\label{CF}
	\hat{\tau}_{CF}(x) = \sum_{i=1}^N  D_i w_i^1(x) Y_i - \sum_{i=1}^N (1-D_i) w_i^0(x)  Y_i,
\end{equation}

where the weights $w_i^d(x)$ define an adaptive local neighbourhood around the covariate value of interest, $x$. These weights are obtained from the tailored splitting procedure that we describe in Section \ref{sec:imp-rf}. \citeA{Athey2017a} show that this estimator can be consistent and asymptotically normal for a fixed covariate space.

\section{Implementation of estimators \label{sec3}}

Section \ref{sec:ga} describes generic approaches to split the estimation of IATEs into several prediction problems. The machine learning literature offers a large variety of potential methods such that the investigation of every possible combination of approaches and machine learning methods is not feasible given our restrictions on computational costs. Thus, this study is restricted to two machine learning methods for the implementation of the prediction problems. They are chosen to be representative for more general approaches. First, we consider Random Forests \cite{Breiman2001} that serve as a popular representative for methods that attempt \textit{local} approximations of conditional mean functions. Second, we consider Lasso \cite{Tibshirani1996} as a method that attempts \textit{global} approximation of conditional mean functions.\footnote{By 'local' we mean that each point of the conditional mean function is approximated by the (weighted) average of neighbouring observations. By 'global' we mean the attempt to approximate the conditional mean by a flexible functional fitted to all data simultaneously.} Both methods are increasingly popular in econometrics and are used in methodological contributions as well as in applications. 

We consider the combinations of the generic approaches in Section \ref{sec:ga} with Random Forest and Lasso.\footnote{We consider only 'pure' combinations where all estimation steps are conducted with one of the two machine learning methods and neglect the possibilities to estimate, e.g., the nuisance parameters via Random Forests and the IATEs via Lasso. In principle, this is possible but not pursued due to computational constraints. For the same reason, we do not pursue ensemble methods that combine different estimators for the nuisance parameters or for IATE estimation \cite<see, e.g., >{Rolling2018CombiningEffects,Schuler2018AEffects}.} Following \citeA{Chernozhukov2018}, we apply cross-fitting to all approaches that require the estimation of nuisance parameters. This means that the nuisance parameters and the IATEs are estimated in different samples to avoid overfitting. 

Table \ref{List_est} summarizes all estimators under investigation. We are currently not aware of a Random Forest implementation that supports weighted minimization. Thus, we implement MCM and R-learning only with Lasso. We add further an \textit{infeasible benchmark} estimator that has access to the true ITEs and uses them as outcome in a standard prediction problem.

\begin{table}[t]
  \centering \onehalfspacing
  \caption{List of considered causal machine learning estimators}
    \begin{tabular}{lccc}
    \toprule
          & \textbf{Random Forest} & \textbf{Lasso} & \textbf{Cross-fitting} \\
    \midrule
    Infeasible benchmark &   x    &   x    &  \\
    Conditional mean regression & x     & x     &  \\
    MOM IPW & x     & x     & x \\
    MOM DR & x     & x     & x \\
    MCM &       & x     & x \\
    MCM with EA &       & x     & x \\
    R-learning &       & x     & x \\
    Causal Forest & x     &       &  \\
    Causal Forest with local centering & x     &       & x \\
    \bottomrule
    \end{tabular}%
  \label{List_est}%
\end{table}%

Like all machine learning methods, Random Forests and Lasso involve a variety of choices in the implementation. The following sections briefly explain Random Forests and Lasso, present the details of the implementation and explain the use of cross-fitting. The resulting estimators target the estimation of IATEs. Additionally, we consider the methods to estimate GATEs and ATE as a computationally cheap by-product of our analysis. To this end, we average the estimated IATEs, $\hat{\tau}(x)$, to GATEs by $\hat{\tau}(g) = N_g^{-1} \sum_{i=1}^{N} \mathds{1}[G_i = g] \hat{\tau}(X_i)$, with $N_g = \mathds{1}[G_i = g]$, and to the ATE by $\hat{\tau} = N^{-1}\sum_{i=1}^{N} \hat{\tau}(X_i)$.\footnote{Specialized estimators for GATE and ATE might outperform these aggregate estimators but this is beyond the scope of this paper \cite<see, e.g.,>{Lee2017,Chernozhukov2018}.}

\subsection{Random Forest} \label{sec:imp-rf}

The building block of Random Forests for conditional mean estimation are regression trees \cite{breiman1984}. Regression trees recursively partition the sample along covariates to minimize MSE of the outcome. This leads to the tree structure and the means in the final leaves are used as predictions \cite<for an introduction, see>{Hastie2009}. However, regression trees are unstable and exhibit a high variance. The Random Forest of \citeA{Breiman2001} addresses this issue by combining a large number of decorrelated regression trees. The decorrelation is achieved by growing each tree on a random subsample (generated either by bootstrapping or subsampling) and by randomly choosing the covariates for each split decision.

The standard regression and probability forests split the trees to minimize the MSE of the observed outcomes. Those trees can then be used to form predictions for a realization of $X_i$. These predictions are formed as a weighted average of the observed outcomes where the weights are larger, the more often the observed outcome shares a final leave with the realization of $X_i$. These kind of forests are required for the conditional mean regression and the modified outcome approaches. 

The Causal Forest of \citeA{Athey2017a} follows a similar structure. However, instead of splitting the sample according to observed outcomes, Causal Forests split the samples along the gradient of the mean difference with the pseudo outcomes

\begin{equation}
	\label{pseudo}
	\rho_i = (D_i - \bar{D}_P)(Y_i - \bar{Y}_P - (D_i - \bar{D}_P) \hat{\beta}_P) / Var_P(D_i),
\end{equation}

where $\bar{D}_P$ and $\bar{Y}_P$ are averages of treatment indicator and outcome, $\hat{\beta}_P$ is the mean difference and $Var_P(D_i)$ is the variance of the treatment in the parent node. This splitting rule is tailored to maximize heterogeneity and produces splits that are used to calculate the weights for the weighted mean comparison in Equation \ref{CF}. 

We also consider the Causal Forest with \textit{local centering}. This means that $D_i$ and $Y_i$ in Equation \ref{pseudo} are replaced by $D_i - p(X_i)$ and $Y_i - \mu(X_i)$, respectively. The nuisance parameters are again estimated in a first step and this partialling out should remove the confounding at a high level before building the Causal Forest. \citeA{Athey2017a} show that local centering can improve the performance of Causal Forests substantially in the presence of confounding.

We implement the regression forests for the conditional mean regression and the modified outcomes as well as the Causal Forests using the R package \texttt{grf} \cite{Athey2017a}. We provide the forests with 105 baseline covariates for the prediction and set the number of variables that are considered at each split to 70. The minimum leaf size is set to one and one forests consists of 1,000 trees. We build honest trees such that the building of the tree and the estimation of the parameters are conducted in separate samples \cite{Athey2016}. To this end, we split the sample randomly for each tree in three parts: 25\% are used to build the tree, 25\% to calculate the predictions, and 50\% are left out.

\subsection{Lasso}  \label{sec:imp-lasso}

The Lasso is a shrinkage estimator and can be considered as an OLS estimator with a penalty on the sum of the absolute coefficients. The standard least squares Lasso solves the following minimization problem

\begin{equation} \label{Lasso}
\min_{\beta} \left[ \sum_{i=1}^{N} w_i \left( Y_i -  X_i \beta \right)^2  \right] + \lambda \sum_{j=1}^{p} \left| \beta_j \right|,
\end{equation}

where $w_i$ are weights, $p$ is the number of covariates and $\lambda$ is a tuning parameter to be optimally chosen.\footnote{For the estimation of the propensity score, we use the equivalent logistic regression.} We obtain the standard OLS coefficients if the penalty term is equal to zero and we have at least as many observations as covariates. For a positive penalty term, at least some coefficients are shrunken towards zero to satisfy the constraint. The Lasso serves as a variable selector because some coefficients are set to zero if the penalty term is sufficiently increased. By incrementing the penalty term to a sufficiently large number, eventually all coefficients besides the constant are zero. The idea of this procedure is to shrink those variables with little or no predictive power to zero and to use the remaining shrunken coefficients for prediction. The degree of shrinkage should be chosen to balance the bias-variance trade-off and is the crucial tuning parameter of the Lasso.

We apply the R package \texttt{glmnet} to produce the predictions in the different approaches \cite{Friedman2010}. We provide the estimator a set of 1,749 potential covariates including second order interactions and up-to fourth order polynomials.\footnote{We exclude binary variables that represent less than 1\% of the observations. Furthermore, we keep only one variable of variable combinations that show correlations of larger magnitude than $\pm0.99$ to speed up computation.} The tuning parameter is selected via 10-fold cross-validation.

\subsection{Cross-fitting} \label{sec:cross-fit}

Some approaches require the estimation of the nuisance parameters in a first step. We follow \citeA{Chernozhukov2018} and apply cross-fitting to remove bias due to overfitting that is induced if nuisance and main parameters are estimated using the same observations. We implement a 50:50 version of their DML1 procedure. This means that we split the sample in two parts of the same size. In the first half, we estimate models for the nuisance parameters. We take these models to predict the nuisance parameters in the second half. These predicted nuisance parameters are then used in the estimation of the IATEs, $\hat{\tau}_1(x)$. We reverse the role of the two halves to obtain $\hat{\tau}_2(x)$. The estimates of the IATE are then calculated as $\hat{\tau}(x) = 1/2 (\hat{\tau}_1(x) + \hat{\tau}_2(x))$.

\subsection{Alternative estimation approaches}

Table \ref{List_est} above lists all estimators that we consider in this study. This list comprises not all alternatives, because we are not able to consider all estimators that have been proposed or would be possible combinations of the generic approaches and existing machine learning methods. We do not consider methods that are tailored for experimental studies \cite<e.g.,>{Imai2013,Grimmer2017,McFowland2018}. Furthermore, restrictions in computation power force us to commit to approaches where we can leverage synergies in the implementation as illustrated in Figure \ref{fig:arrow} of Appendix \ref{sec:app-impl}. Thus, we are not able to consider the X-learner of \citeA{Kunzel2017}, the three conditional outcome difference methods proposed by \citeA{Powers2018}, Orthogonal Random Forests \cite{Oprescu2018}, Penalized Causal Forests \cite{Lechner2018PenalizedEffects}, methods based on neural nets  \cite<e.g.,>{johansson2016learning,Shalit2016EstimatingAlgorithms,Schwab2018PerfectNetworks}, Bayesian approaches like those based on Bayesian additive regression trees (BART) \cite{Hill2011,Hahn2017} or Bayesian forests \cite{Taddy2016}, and potentially other approaches that we are currently not aware of. 

Further, the generic approaches discussed in Section \ref{sec:ga} could be implemented using different machine learning algorithms like Boosting, Elastic Nets, Neural Nets, Ridge or any other supervised machine learning algorithm that minimizes the required loss functions \cite<see for an overview>{Hastie2009}.

\section{Simulation set-up \label{sec4}}

\subsection{Previous Empirical Monte Carlo Study} \label{sec:literature}

The simulation study of \citeA{Wendling2018} is close in spirit to our approach, in the sense that their and our DGP relies as much as possible on real data. They compare eight conditional outcome difference estimators for binary outcomes, i.e., they focus on probability models. Their four DGPs are based on the covariates and the observational treatment assignment of four medical datasets. Thus, the IATEs and the resulting binary potential outcomes are the only components that need to be specified. The outcomes are simulated based on predictions of $\mu(0,x)$ and $\mu(1,x)$ from logistic neural networks \cite<for more details, see>{Wendling2018}.\footnote{\citeA{Nie2017} use a similar EMCS for binary outcomes to assess the performance of different implementations of R-learning. However, they  do not estimate the IATE but specify it to depend on two covariates.} This is a realistic approach in the medical context. However, it removes two important features from the true outcome generating process. First, the projection of the outcome on observable covariates removes the impact of unobservable variables. Second, the true error structure is lost by imposing a logistic error term. Our EMCS aims to preserve these features of the data at least for the non-treatment outcome. \citeA{Wendling2018} find that conditional mean regressions \cite<implemented with BART, see>{Hill2011} and causal boosting \cite{Powers2018} perform consistently well, while causal MARS \cite{Powers2018} and Causal Forests \cite{Athey2017a} are found to be competitive for complex IATE but perform poorly if the variance of the IATE is relatively low. We implement conditional mean regressions and Causal Forests, but omit causal boosting and MARS because of computational restrictions.

\subsection{Empirical Monte Carlo Study}

Similar to \citeA{Wendling2018} for the medical context, our study aims to approximate a real application in economic policy evaluation as close as possible. The idea of an EMCS is introduced by \citeA{Huber2013} and \citeA{lechner2013sensitivity}. It aims to take as many components of the DGP as possible from real data. We build this EMCS on 96,298 observations of Swiss administrative social security data that is already used in previous evaluation studies \cite{behncke2010caseworker,behncke2010unemployed,Huber2017}. In particular, the EMCS mimics an evaluation of job search programs as in \citeA{Knaus2017}. The interest is in the heterogeneous effects of such a program on employment over the 33 months after the program start.\footnote{Appendix \ref{sec:App_Dat} provides more details about the outcomes and the rest of the dataset.} 

Before we describe and motivate our EMCS approach, we list the general steps to evaluate estimators for IATEs, GATEs and ATEs in Table \ref{EMCS}. We leave out a validation sample (10,000 randomly drawn observations) to compare the estimated IATEs against the true ITEs, while previous EMCS compare in-sample estimates to true values of a known IATE. This modification is intended to focus on the out-of-sample predictive power of the estimated causal effects. The advantage of this procedure is that we can specify the ITEs as ground truth without knowing the IATE as we describe below. 

\begin{table}[t]
  \centering
  \caption{Empirical Monte Carlo Study}
    \begin{tabularx}{\textwidth}{rX}
        \toprule
         
		1. & Take the full sample and estimate the propensity score, $p^{full}(x)$, using the method and specification of choice. \\
		[0.2em]
		2. & Remove all treated and keep only the $N_{nt}$ non-treated observations. This means that $Y_i^0$ is observed for all members of the remaining subpopulation. \\
		[0.2em]
		3. & Specify the true ITEs, $\xi_i$. \\
		[0.2em]
		4. & Calculate the potential outcome under treatment as $\hat{Y}_i^1 = Y_i^0 + \xi_i$ for all observations. \\
		[0.2em]
		5. & Set aside a random validation sample of $N_v$ observations. Remove this validation sample from the main sample.  \\
		[0.2em]
		6. & Calculate any other parameters of interest in the validation sample as benchmark. Like GATEs as $\tau(g) = \left( \sum_{i=v}^{N_v} \mathds{1}[G_v = g] \right)^{-1} \sum_{v=1}^{N_v} \mathds{1}[G_v = g] \xi_v$ or ATEs as $\tau = N_v^{-1}\sum_{v=1}^{N_v} \xi_v$. \\
		[0.2em]
		7. & Draw a random sample of size $N_s$ from the remaining $N_{nt}-N_v$ observations. \\
		[0.2em]
		8. & Simulate pseudo treatment indicators $D_i \sim Bernoulli(p^{sim}(x))$, where $p^{sim}(x)$ is a potentially modified version of $p^{full}(x)$ to control the ratio of treated and controls or other features of the selection process. \\
		[0.2em]
		9. & Use the observation rule in \ref{OR} to create the observable outcome $Y_i$. \\
		[0.2em]
		10. & Use the $N_s$ observations to estimate $\tau(x)$ with all estimators of interest.  \\
		[0.2em]
		11. & Predict $\hat{\tau}(x)$ for all observations in the validation sample and use them to calculate $\hat{\tau}(g)$ as well as $\hat{\tau}$ for each estimator. \\
		[0.2em]
		12. & Repeat steps 7 to 11 $R$ times. \\
		[0.2em]
		13. & Calculate performance measures. \\

    	\bottomrule
    \end{tabularx}%
  \label{EMCS}%
\end{table}%

After removing the 10,000 observations of our validation sample, the remaining 78,844 observations form our 'population' from which we draw random subsamples of size 1,000 and 4,000 for estimation. We replicate this 2,000 times for the smaller and 500 times for the larger samples. The precision of the estimators and the computational costs increase with the sample size. Thus, we reduce the number of replications when we increase the sample size to restrict the latter. In case of $\sqrt{N}$-convergence, this will keep the simulation error approximately constant. Table \ref{tab:list_dgp} below shows the variants of the DGP for different $N_s$, $R$, $p(x)$ and $\xi_i$. Before, we explain the specification of the two latter functions.

\subsubsection{Propensity score} \label{sec:p_score}

The 'population' propensity score is estimated in the full sample with 7,454 treated and 88,844 controls. After this estimation step, all treated are removed from the sample. The specification of the propensity score is taken from \citeA{Huber2017} and estimated using a standard logistic regression. We manipulate the constant to create a 50:50 split into treated and non-treated in the simulated samples.\footnote{We remove the 342 observations with a modified propensity score below 5\% and above 95\%. We deviate at this point from the real dataset and make the problem better behaved than in reality in terms of common support \cite<see discussion in, e.g.,>{lechner2017practical}. We leave the investigation of performance in the presence of unbalanced ratios and insufficient common support for future studies and focus here on a relatively nice setting to start with.} Appendix \ref{sec:app-p_score} provides the details of the specification of the original propensity score and the distribution of the modified propensity score.

\subsubsection{Specification of ITE}

We are not able to observe the ITEs or any of its aggregates in a real world dataset. Therefore, we either need to estimate or to specify them. We choose the latter because estimation might favor similar estimators under investigation. Thus, our goal is to create a challenging synthetic ITE that uses components from real data. In observational studies, the estimators must be able to disentangle selection bias and effect heterogeneity. We make it hard for the estimators by using the 'population' propensity score $p^{HLM}(x)$ directly to calculate the ITEs. To this end, the propensity score is normalized and put into a sine function

\begin{equation} \label{small_omega}
	\omega(x) = sin \left(1.25 \pi \dfrac{p^{HLM}(x)}{max(p^{HLM}(x))} \right) + \varepsilon_i,
\end{equation}

where $\varepsilon_i$ is random noise. This highly non-linear function of the propensity score is standardized to have mean zero and variance one before it is scaled by the parameter $\alpha$:

\begin{equation} \label{omega}
	\Omega(x) = \alpha \dfrac{\omega(x) - \bar{\omega}}{SD(\omega(x))},
\end{equation}

where $\bar{\omega}$ is the mean of $\omega(x)$ and $SD(\omega(x))$ is its standard deviation. Finally, we force the ITEs to respect two features of our outcome variable. This means that they are rounded to the next integer and that they must respect that $\hat{Y}_i^1$ falls between zero and 33.\footnote{The histogram of the (observed) $Y_i^0$ is provided in Figure \ref{fig:hist-y0}. Given the censored and integer nature of the outcome, we considered also using Poisson Lasso to estimate the outcome regressions. However, the computation time compared to least squared Lasso is substantially longer, while the predictive performance is very similar for our outcomes. Thus, we chose the least squares version for the simulations.} Thus, the final ITEs take the form

\begin{equation} \label{xi}
\xi(x,y^0) =  \left\{ \begin{array}{l l}
    \lfloor \Omega(x) \rceil & \text{if $0 \leq  y^0 + \lfloor \Omega(x) \rceil \leq 33$}\\
    -y^0 & \text{if $y^0 + \lfloor \Omega(x) \rceil < 0$}\\
    33 - y^0 & \text{if $y^0 + \lfloor \Omega(x) \rceil > 33,$  }
  \end{array} \right.
\end{equation}

where $\lfloor \cdot \rceil$ indicates that we round to the nearest integer.

$\xi(x,y^0)$ is highly non-linear and complicated due to the logistic function, the sine function and the rounding. Additionally, enforcement of the censoring makes it dependent on $Y_i^0$ that is taken directly from the data and thus depends on the covariates in an unknown fashion. Thus, we know the true ITEs but we do not know the functional form of the true IATE. 

The true ITEs depend on the observables $X_i$ and additionally on some unobservables through $Y_i^0$.\footnote{These unobservables do not invalidate the CIA in our simulated samples, as $Y_i^0$ and thus the unobservables are not part of the population propensity score. The alternative to ensure a valid CIA in an EMCS is to keep the true treatment allocation structure and to specify the potential outcomes as function of the observables. This is the approach of \citeA{Wendling2018} that is discussed in \ref{sec:literature}.} This means that the estimators approximate $\xi(x,y^0)$ using observables and produce estimates $\hat{\tau}(x)$ of $\tau(x)$. The goal of the EMCS is to figure out which estimators approximate the ITE comparatively well in this arguably realistic setting. The relative performance of the estimators translate then directly into the ability to approximate the unknown IATEs because estimators that minimize the MSE of the ITE also minimize the MSE of the IATE \cite<see, e.g.,>{Kunzel2017}. Note that the aggregation of IATEs to GATEs and ATE in step 6 of Table \ref{EMCS} can be considered as true values because the influence of $Y_i^0$ is averaged out for them asymptotically. This implies that the MSE of GATEs and ATEs would be approximately zero if $\hat{\tau}(x) = \tau(x)$, while the MSE of ITEs might still be positive in this case.

We consider three different values of $\alpha$ in Equation \ref{omega} to vary the size of the ITEs: $\alpha = 0$ (ITE0), $\alpha = 2$ (ITE1) and $\alpha = 8$ (ITE2). Additionally, we create one specification without random noise and one with error term in Equation \ref{small_omega}, $\varepsilon_i = 0$ and  $\varepsilon_i \sim 1 - Poisson(1)$, respectively. Table \ref{tab:ITE-desc} reports basic descriptive statistics of the resulting potential outcomes, ITEs, and GATEs. ITE0 without random noise creates a benchmark scenario that is most likely to be informative about which estimators are prone to confuse effect heterogeneity with selectivity. ITE1 leads to a scenario with moderate variance of the resulting ITEs. Their standard deviation amounts to about 14\% of the non-treatment outcome. ITE2 produces bigger ITEs with a standard deviation of about 6, which is roughly 50\% of the standard deviation of the non-treatment outcome. Thus, they should be less difficult to detect. 

\begin{table}[t]
    \centering
\begin{threeparttable}
  \caption{Descriptive statistics of simulated outcomes and ITEs}
    \begin{tabular}{lccccc}
            \toprule
                      & \multicolumn{1}{l}{Mean} & \multicolumn{1}{l}{Std. Dev.} & \multicolumn{1}{l}{Skewness} & \multicolumn{1}{l}{Kurtosis} & \multicolumn{1}{l}{Percent censored} \\
                      \midrule
    \multicolumn{6}{l}{\textit{Without random noise ($\varepsilon_i = 0$):}} \\
    $Y^0$ in all DGPs & 16.1  & 12.8  & -0.1  & 1.4   & - \\
    $Y^1$ in ITE0 & 16.1  & 12.8  & -0.1  & 1.4   & - \\
    $Y^1$ in ITE1 & 16.3  & 12.6  & -0.1  & 1.4   & - \\
    $Y^1$ in ITE2 & 16.3  & 12.6  & 0.1   & 1.5   & - \\
    ITE0  & 0.0   & 0.0   & -  & -   & 0.0 \\
    ITE1  & 0.1   & 1.8   & -0.3  & 2.3   & 39.2 \\
    ITE2  & 0.2   & 6.4   & -0.4  & 2.5   & 43.7 \\
    GATE0  & 0.0   & 0.0   & -  & -   & - \\
    GATE1 & -0.4  & 1.8   & -0.1  & 2.0   & - \\
    GATE2 & -1.8  & 6.2   & -0.3  & 2.1   & - \\
    [0.4em]
    \multicolumn{6}{l}{\textit{With random noise ($\varepsilon_i \sim 1 - Poisson(1)$):}} \\
    $Y^1$ in ITE0 & 16.2  & 12.7  & -0.1  & 1.4   & - \\
    $Y^1$ in ITE1 & 16.3  & 12.6  & -0.1  & 1.4   & - \\
    $Y^1$ in ITE2 & 16.5  & 12.3  & 0.0   & 1.5   & - \\
    ITE0  & 0.1   & 0.9   & -1.2  & 5.1   & 26.6 \\
    ITE1  & 0.1   & 1.8   & -1.0  & 4.9   & 36.7 \\
    ITE2  & 0.3   & 6.3   & -1.0  & 4.8   & 41.1 \\
    GATE0 & 0.0   & 1.1   & -1.2  & 3.5   & - \\
    GATE1 & 0.0   & 1.7   & -0.7  & 3.2   & - \\
    GATE2 & 0.0   & 5.8   & -0.6  & 3.7   & - \\
            \bottomrule
    \end{tabular}%
    \begin{tablenotes} \item \textit{Notes:} Potential outcomes and ITEs are considered for all observations. GATEs are considered for the validation sample.    \end{tablenotes}  
  \label{tab:ITE-desc}%
\end{threeparttable}%
\end{table}
\doublespacing

The ITEs without and with random noise are created to be similar in their first two moments. However, they differ substantially in their variation that can be explained by observables. The influence of $Y_i^0$ on the ITEs is substantial because between 27\% and 44\% of the observations are censored for the non-zero ITEs. This explains why ITE1 and ITE2 without random noise are not deterministic either and therefore not perfectly predictable by $X_i$. Still, the out-of-sample $R^2$ of Random Forest and Lasso predictive regressions shown in Table \ref{tab:r-sq} document that we can explain between about 50\% and 70\% of the ITEs with our covariates. With random noise, this explainable part decreases to close to zero for ITE0 and 6.3\% for ITE1. We consider the latter to be a more realistic scenario because the individual component is expected to be relatively large.\footnote{For example, \citeA{Djebbari2008HeterogeneousPROGRESA} provide evidence that the ITEs in their applications show only little systematic variation.} Thus, we select ITE1 and ITE2 with random noise as our baseline DGPs additionally to the benchmark scenario ITE0 without random noise.

\begin{table}[t]     %
  \centering
\begin{threeparttable}
  \caption{List of DGPs}
    \begin{tabular}{lccccc}
        \toprule
           & \textbf{$N_s$} & $\alpha$ in \ref{omega} & Propensity score & \textbf{$R$} & $\varepsilon_i$ in \ref{small_omega} \\
          \midrule

\multicolumn{6}{l}{\textit{With selection and without random noise:}} \\
    ITE0*   & 1,000 & $\alpha = 0$  & $p^{HLM}(x)$ & 2,000 & $\varepsilon_i = 0$ \\
    ITE1   & 1,000 & $\alpha = 2$  & $p^{HLM}(x)$ & 2,000 & $\varepsilon_i = 0$ \\
    ITE2   & 1,000 & $\alpha = 8$  & $p^{HLM}(x)$ & 2,000 & $\varepsilon_i = 0$ \\
    ITE0*   & 4,000 & $\alpha = 0$  & $p^{HLM}(x)$ & 500 & $\varepsilon_i = 0$ \\
    ITE1   & 4,000 & $\alpha = 2$  & $p^{HLM}(x)$ & 500 & $\varepsilon_i = 0$ \\
    ITE2   & 4,000 & $\alpha = 8$  & $p^{HLM}(x)$ & 500 & $\varepsilon_i = 0$ \\
    
    \multicolumn{6}{l}{\textit{With selection and random noise:}}\\
    ITE0~~~~~~~~~~~   & 1,000 & $\alpha = 0$  & $p^{HLM}(x)$ & 2,000 & $\varepsilon_i \sim 1 - Poisson(1)$ \\
    ITE1*   & 1,000 & $\alpha = 2$  & $p^{HLM}(x)$ & 2,000 & $\varepsilon_i \sim 1 - Poisson(1)$ \\
    ITE2*   & 1,000 & $\alpha = 8$  & $p^{HLM}(x)$ & 2,000 & $\varepsilon_i \sim 1 - Poisson(1)$ \\
    ITE0   & 4,000 & $\alpha = 0$  & $p^{HLM}(x)$ & 500 & $\varepsilon_i \sim 1 - Poisson(1)$ \\
    ITE1*   & 4,000 & $\alpha = 2$  & $p^{HLM}(x)$ & 500 & $\varepsilon_i \sim 1 - Poisson(1)$ \\
    ITE2*   & 4,000 & $\alpha = 8$  & $p^{HLM}(x)$ & 500 & $\varepsilon_i \sim 1 - Poisson(1)$ \\
    
    \multicolumn{6}{l}{\textit{With random assignment and without random noise:}}  \\
    ITE0   & 1,000 & $\alpha = 0$  & 0.5 & 2,000 & $\varepsilon_i = 0$ \\
    ITE1   & 1,000 & $\alpha = 2$  & 0.5 & 2,000 & $\varepsilon_i = 0$ \\
    ITE2   & 1,000 & $\alpha = 8$  & 0.5 & 2,000 & $\varepsilon_i = 0$ \\
    ITE0   & 4,000 & $\alpha = 0$  & 0.5 & 500 & $\varepsilon_i = 0$ \\
    ITE1   & 4,000 & $\alpha = 2$  & 0.5 & 500 & $\varepsilon_i = 0$ \\
    ITE2   & 4,000 & $\alpha = 8$  & 0.5 & 500 & $\varepsilon_i = 0$ \\
    
    \multicolumn{6}{l}{\textit{With random assignment and random noise:}}  \\
    ITE0   & 1,000 & $\alpha = 0$  & 0.5 & 2,000 & $\varepsilon_i \sim 1 - Poisson(1)$ \\
    ITE1   & 1,000 & $\alpha = 2$  & 0.5 & 2,000 & $\varepsilon_i \sim 1 - Poisson(1)$ \\
    ITE2   & 1,000 & $\alpha = 8$  & 0.5 & 2,000 & $\varepsilon_i \sim 1 - Poisson(1)$ \\
    ITE0   & 4,000 & $\alpha = 0$  & 0.5 & 500 & $\varepsilon_i \sim 1 - Poisson(1)$ \\
    ITE1   & 4,000 & $\alpha = 2$  & 0.5 & 500 & $\varepsilon_i \sim 1 - Poisson(1)$ \\
    ITE2   & 4,000 & $\alpha = 8$  & 0.5 & 500 & $\varepsilon_i \sim 1 - Poisson(1)$ \\
    
        \bottomrule

    \end{tabular}%
    \begin{tablenotes} \item \textit{Notes:} Asteriks mark the baseline DGPs. \end{tablenotes}
\label{tab:list_dgp}
  \end{threeparttable}

\end{table}%

The first column of Table \ref{tab:ITE-desc} shows that we specify the mean of the ITEs, the ATE, close to zero.\footnote{Appendix \ref{sec:app-ite} shows in detail how the ITEs and potential outcomes are distributed, how the ITEs are related to the propensity score and $Y_i^0$, as well as an interpretation of the simulated selection behavior of caseworkers. Note that the lower standard deviations of $Y_i^1$ compared to $Y_i^0$ result from the censoring that moves mass away from the bounds (see Figures \ref{fig:cdf1} and \ref{fig:cdf2}).} Appendix \ref{sec:App_gate} describes how we aggregate the ITEs into 64 groups with sizes between 32 and 420 observations to specify the true GATEs. 

In summary, we consider six different scenarios defined by different choices for the scale of the ITEs and the random noise variables. Additionally to the DGP with selection into the treatment, we consider also the case of an experiment with 50:50 random assignment. These twelve different DGPs are considered for the sample sizes of 1,000 and 4,000 observations leading to a total number of 24 different settings. Table \ref{tab:list_dgp} summarizes all parameter settings in which the eleven estimators are compared.

\subsubsection{Performance measures} \label{sec:performance}

We consider three major performance measures: mean squared error (MSE), absolute bias ($|Bias|$) and standard deviation ($SD$) for the prediction of each observation $v$ in the validation sample:\footnote{The formulas are written for the ITE. The same measures are used for GATE and ATE.} 

\begin{equation} \label{MSE}
	MSE_v = \dfrac{1}{R} \sum_{r=1}^{R} \left[ \xi(x_v,y_v^0) - \hat{\tau}(x_v)_r \right]^2
\end{equation}

\begin{equation}
	|Bias_v| = | \underbrace{\dfrac{1}{R} \sum_{r=1}^{R}  \hat{\tau}(x_v)_r}_{\bar{\hat{\tau}}(x_v)_r} - \xi(x_v,y_v^0) |
\end{equation}

\begin{equation} \label{SD}
	SD_v =  \sqrt[]{\dfrac{1}{R} \sum_{r=1}^{R} \left[ \hat{\tau}(x_v)_r - \bar{\hat{\tau}}(x_v)_r \right]^2}
\end{equation}

Most simulation studies are interested in only few parameters such that the performance measure for each parameter can be reported and interpreted. However, in our case we have 10,000 parameters such that we need to summarize the performance over the whole validation sample by taking the averages $\overline{MSE}$, $|\overline{Bias}|$ and $\overline{SD}$.\footnote{For example, $\overline{MSE} = N_v^{-1} \sum_{v=1}^{N_v} MSE_v$.} Additionally, we apply the Jarque-Bera test (JB) to the distribution of predictions for each observation $v$ in the validation sample and report the fraction of observations for which normality is rejected at the 5\% confidence level.\footnote{Appendix \ref{sec:final_app} discusses and provides also alternative performance measures.}

\section{Results \label{sec5}}

\subsection{IATE estimation}

Table \ref{tab:results} shows the main performance measures for the three baseline DGPs.\footnote{The full tables with more performance measures are provided in Tables \ref{tab:app-ite0} for ITE0, \ref{tab:app-ite1-rn} for ITE1, and \ref{tab:app-ite2-rn} for ITE2.} At first, we compare estimators within similar approaches to identify the competitive versions. Afterwards, we compare the competitive versions over all approaches to identify those estimators that show an overall good performance and provide a general comparison of Random Forest and Lasso based methods.


\begin{table}[htp]
  \centering  \footnotesize
\begin{threeparttable}

  \caption{Simulation results of ITE estimation for baseline DGPs}
    \begin{tabular}{lccccccccc}
    \toprule
          & \multicolumn{4}{c}{1000 observations} &       & \multicolumn{4}{c}{4000 observations} \\
\cmidrule{2-5}\cmidrule{7-10}          & $\overline{MSE}$    & $|\overline{Bias}|$  & $\overline{SD}$    & $JB$    &       & $\overline{MSE}$    & $|\overline{Bias}|$  & $\overline{SD}$    & $JB$  \\
\cmidrule{1-10}          & \multicolumn{9}{c}{\textbf{ITE0 with selection and without random noise}} \\
\cmidrule{1-10}   
    \textit{Random Forest:} &       &       &       &       &       &       &       &       &  \\
    Infeasible & \multicolumn{9}{c}{No variation in dependent variable} \\
    Conditional mean regression & 3.69  & 0.62  & 1.78  & 6\%   &       & 2.79  & 1.33  & 1.49  & 4\% \\
    MOM IPW & 10.52 & 2.05  & 2.16  & 18\%  &       & 5.76  & 1.89  & 1.74  & 12\% \\
    MOM DR & \textbf{2.00} & 0.40  & 1.35  & 7\%   &       & 1.16  & 0.85  & 1.03  & 8\% \\
    Causal Forest & 3.52  & 0.75  & 1.69  & 12\%  &       & 2.31  & 1.21  & 1.29  & 6\% \\
    Causal Forest with local centering & 3.42  & 0.34  & 1.81  & 10\%  &       & 2.05  & 1.13  & 1.40  & 9\% \\
    \textit{Lasso:} &       &       &       &       &       &       &       &       &  \\
    Infeasible & \multicolumn{9}{c}{No variation in dependent variable} \\
    Conditional mean regression & 11.21 & 0.69  & 3.19  & 91\%  &       & 6.14  & 1.92  & 2.30  & 35\% \\
    MOM IPW & 11.31 & 1.09  & 2.99  & 100\% &       & 5.02  & 1.56  & 1.93  & 100\% \\
    MOM DR & 45.39 & 0.60  & 6.31  & 100\% &       & \textbf{0.51} & 0.51  & 0.62  & 35\% \\
    MCM   & 13.03 & 1.50  & 3.05  & 100\% &       & 5.79  & 1.65  & 1.94  & 100\% \\
    MCM with efficiency augmentation & 2.08  & 0.42  & 1.36  & 100\% &       & \textbf{0.48} & 0.49  & 0.62  & 97\% \\
    R-learning & \textbf{2.03} & 0.45  & 1.33  & 100\% &       & \textbf{0.47} & 0.49  & 0.61  & 97\% \\
    \midrule
          & \multicolumn{9}{c}{\textbf{ITE1 with selection and random noise}} \\
    \midrule
    \textit{Random Forest:} &       &       &       &       &       &       &       &       &  \\
    Infeasible & 2.98  & 1.29  & 0.15  & 71\%  &       & 2.93  & 1.30  & 0.11  & 26\% \\
    Conditional mean regression & 7.04  & 1.45  & 1.78  & 8\%   &       & 6.05  & 1.44  & 1.49  & 4\% \\
    MOM IPW & 12.92 & 2.26  & 2.20  & 16\%  &       & 8.42  & 1.76  & 1.77  & 11\% \\
    MOM DR & \textbf{5.08} & 1.36  & 1.33  & 8\%   &       & 4.17  & 1.32  & 1.01  & 9\% \\
    Causal Forest & 6.86  & 1.49  & 1.68  & 12\%  &       & 5.61  & 1.48  & 1.29  & 6\% \\
    Causal Forest with local centering & 6.50  & 1.35  & 1.79  & 12\%  &       & 5.10  & 1.32  & 1.39  & 10\% \\
    \textit{Lasso:} &       &       &       &       &       &       &       &       &  \\
    Infeasible & 3.00  & 1.28  & 0.21  & 100\% &       & 2.93  & 1.28  & 0.16  & 83\% \\
    Conditional mean regression & 14.26 & 1.46  & 3.16  & 90\%  &       & 9.20  & 1.43  & 2.30  & 36\% \\
    MOM IPW & 15.69 & 1.56  & 3.12  & 100\% &       & 8.03  & 1.46  & 2.01  & 100\% \\
    MOM DR & 48.76 & 1.40  & 6.32  & 100\% &       & \textbf{3.66} & 1.34  & 0.64  & 96\% \\
    MCM   & 15.31 & 1.72  & 3.10  & 100\% &       & 8.14  & 1.51  & 1.96  & 100\% \\
    MCM with efficiency augmentation & 5.27  & 1.37  & 1.36  & 100\% &       & \textbf{3.62} & 1.33  & 0.63  & 98\% \\
    R-learning & 5.16  & 1.38  & 1.29  & 100\% &       & \textbf{3.65} & 1.34  & 0.63  & 98\% \\
    \midrule
          & \multicolumn{9}{c}{\textbf{ITE2 with selection and random noise}} \\
    \midrule
    \textit{Random Forest:} &       &       &       &       &       &       &       &       &  \\
    Infeasible & 38.46 & 4.43  & 0.52  & 66\%  &       & 37.86 & 4.43  & 0.41  & 31\% \\
    Conditional mean regression & 43.74 & 4.58  & 1.74  & 8\%   &       & 42.26 & 4.54  & 1.46  & 5\% \\
    MOM IPW & 46.83 & 4.83  & 2.23  & 16\%  &       & 42.69 & 4.54  & 1.80  & 11\% \\
    MOM DR & \textbf{41.45} & 4.50  & 1.32  & 10\%  &       & \textbf{40.03} & 4.45  & 1.03  & 11\% \\
    Causal Forest & 43.87 & 4.61  & 1.66  & 12\%  &       & 42.34 & 4.58  & 1.29  & 7\% \\
    Causal Forest with local centering & 42.84 & 4.50  & 1.78  & 12\%  &       & 41.05 & 4.46  & 1.40  & 9\% \\
    \textit{Lasso:} &       &       &       &       &       &       &       &       &  \\
    Infeasible & 38.66 & 4.42  & 0.71  & 100\% &       & 37.84 & 4.40  & 0.53  & 84\% \\
    Conditional mean regression & 50.11 & 4.52  & 3.15  & 92\%  &       & 44.31 & 4.46  & 2.33  & 34\% \\
    MOM IPW & 49.82 & 4.50  & 3.20  & 100\% &       & 43.21 & 4.43  & 2.17  & 97\% \\
    MOM DR & 537.16 & 4.55  & 5.04  & 100\% &       & 40.11 & 4.48  & 0.76  & 97\% \\
    MCM   & 49.25 & 4.47  & 3.18  & 100\% &       & 42.63 & 4.41  & 2.07  & 100\% \\
    MCM with efficiency augmentation & 41.99 & 4.51  & 1.41  & 100\% &       & \textbf{40.04} & 4.47  & 0.75  & 99\% \\
    R-learning & 42.13 & 4.54  & 1.35  & 100\% &       & 40.25 & 4.49  & 0.74  & 98\% \\
    \bottomrule
    \end{tabular}%
     \begin{tablenotes} \item \textit{Notes:} $\overline{MSE}$ shows the mean MSE of all 10,000 observations in the validation sample, $|\overline{Bias}|$ denotes the mean absolute bias, $\overline{SD}$ the mean standard deviation, and $JB$ the fraction of observations for which the Jarque-Bera test is rejected at the 5\% level. Bold numbers indicate the best performing estimators in terms of $\overline{MSE}$ and estimators within two standard (simulation) errors of the lowest $\overline{MSE}$.   \end{tablenotes}  
    
  \label{tab:results}%
\end{threeparttable}%
\end{table}
\doublespacing

\subsubsection{Conditional mean regressions}

The Random Forest version of conditional mean regressions clearly outperforms the Lasso version in terms of mean MSE. The differences are particularly striking in the smaller sample estimation of ITE0 where the mean MSE of the Lasso version is more than three times larger compared to the Random Forest version. The substantially worse performance of the Lasso version is consistently observed over all baseline DGPs and sample sizes. This is is mostly driven by a substantially lower mean SD of Random Forest based conditional mean regressions that is thus the dominant choice within the two considered versions of conditional mean regressions.

\subsubsection{Modified outcome methods}

The ranking of the MOM estimators depends on the sample size. Table \ref{tab:results} shows that Random Forests are superior to the Lasso versions in the smaller samples. Especially the DR modification with Random Forest performs well due to relatively low mean SD. In contrast, the Lasso equivalent is by far the worst estimator in the smaller samples. It has up to twice as large mean SD compared to the next worst estimator and shows consequently a very high mean MSE. One potential reason is that only the DR estimators require the estimation of $\mu(d,x)$ as a nuisance parameter. These predictions are then based on only 250 observations when using cross-fitting, while $\mu(x)$ and $p(x)$ are based on 500 observations. The instability of the Lasso as outcome predictor in small samples seems to spillover to the IATE estimation.\footnote{\citeA{Chernozhukov2018} observe a similar problem of global approximations for the estimation of average effects. See also \citeA{Waernbaum2017} for conditions under which a poor approximation of the outcome leads to worse performance of DR estimators compared to IPW. Similarly, \citeA{Kang2007} demonstrate that double robust methods can perform poorly if both nuisance parameters are misspecified.} The results for the larger sample size indicate that the poor performance is a small sample issue. The DR modification with Lasso outperforms the other versions of MOM in ITE0 and ITE1 and is also close to its Random Forest equivalent for ITE2. The good performance is mainly driven by relatively low mean SDs.

As expected from the results of \citeA{Powers2018}, the IPW modification has relatively high mean SD and is therefore not competitive. This is despite the fact that our DGP does not lead to extreme propensity scores and creates thus a relatively favorable setting for IPW. Therefore, the DR modification seems to be in general the dominant choice as long as the Lasso version is not used in small samples.

\subsubsection{MCM and R-learning}

The results of the three estimators with modified covariates are similar to the results for the MOM. The MCM is clearly outperformed by its efficiency augmented version and R-learning that both use the outcome regression additionally to the propensity score as a nuisance parameter. For MCM, the efficiency augmentation more than halves the mean SD in all baseline DGPs. Thus, the additional computational effort is fruitful when using MCM. For all DGPs, efficiency augmented MCM and R-learning perform very similar along all dimensions. This finding is in line with the synthetic simulation in Appendix D of \citeA{Chen2017} who find also very similar results for those two.

\subsubsection{Causal Forests}

The Causal Forest is specialized to maximize heterogeneity in experimental settings but it is not build to explicitly account for selection. Thus, it is prone to choose splits that do not sufficiently remove selection bias. However, Causal Forests with local centering address this problem by partialling out the selection effects in a first step. They are specialized to maximize effect heterogeneity and to account for selection bias. Consequently, they uniformly perform better than Causal Forests. This is driven by a relatively low mean absolute bias, but higher mean SD partly offsets this advantage. However, the differences between the two Causal Forest versions are moderate but the version with local centering is the dominant choice if the goal is to minimize mean MSE. However, the improvement comes at the cost of estimating additionally two nuisance parameters before estimating the Causal Forest.\footnote{Together with the conditional mean regression based on Random Forests, the Causal Forest is thus attractive if computation time is a concern. Appendix \ref{sec:comp-time} shows that both require very similar computation time and are the fastest Random Forest based estimators under consideration.}

\subsubsection{Overall comparison} \label{sec:comp}

The results in Table \ref{tab:results} show that no estimator is uniformly superior for all sample sizes and DGPs. However, we can categorize the estimators into those that show a relatively good performance in all settings, the volatile ones with outstanding performance only in particular settings, and those that are never competitive. 

The first category comprises Random Forest MOM DR, MCM with efficiency augmentation, R-learning and Causal Forest with local centering. These four estimators are in a similar range over all DGPs and sample sizes and belong consistently to the five best estimators. Thus, they seem to be reasonable choices to estimate IATEs. Causal Forest with local centering is the only one of those four that shows never the best performance in terms of mean MSE. This is driven by a larger mean SD that works against the very competitive mean absolute bias. The feature that unifies all four best performing estimators is that they use propensity score and outcome regressions as nuisance parameters in the estimation process.

The MOM DR with Lasso belongs to the second category because it is very competitive for larger samples but the worst choice in smaller samples. Thus, it remains a risky choice for applications because the critical sample size for good performance may depend on the particular dataset.

Finally, conditional mean regressions, MOM IPW with Lasso and Causal Forest should not be considered in settings like ours if minimizing MSE has a high priority. However, if computational constraints are binding, conditional mean regressions with Random Forests and Causal Forests can be attractive options.

\subsubsection{Random Forest vs. Lasso}

A direct comparison of Random Forest and Lasso is possible for conditional mean regressions and MOM. For the smaller sample size, Random Forest clearly outperforms the Lasso based versions. This is driven by the substantially lower mean SD of Random Forest based estimators. The reason is that the global approximations of Lasso are rather instable for small samples. This instability is reduced for larger samples and the Lasso based MOM performs better than the Random Forest equivalents for ITE0 and ITE1. 

This dependence on the sample size is not observed for the Lasso specific estimators MCM with efficiency augmentation and R-learning. Both show competitive performance regardless of the sample size. This is particularly surprising given the highly non-linear ITEs. However, all Lasso based methods are far from being normally distributed. For at least 30\% of the validation observations, the JB test rejects normality. For many estimators it is even rejected for all observations in the validation sample, while we would expect only a fraction of 5\% to be rejected under normality. Columns 9 in the Tables of Appendix \ref{sec:final_app} show that this is due to excess kurtosis, which indicates that the Lasso based methods are prone to produce outliers. It is mitigated for the sample size of 4,000 but still the JB test is rejected for a large majority. This reflects the theoretical results of \citeA{Leeb2005ModelFiction,Leeb2008SparseEstimator} that shrinkage estimators like Lasso exhibit non-normal finite sample distributions. 

In contrast, all Random Forest based estimators appear to be approximately normally distributed. This is also reflected by a mean skewness close to zero and a kurtosis close to three. Decently performing Random Forest based estimators might be therefore preferable to slightly better performing Lasso based estimators. The former produce less outliers and seem therefore more reliable and robust in empirical applications as well as more amendable to statistical inference.

\subsection{GATE and ATE estimation}

Table \ref{tab:results-gate} shows the main performance measures of GATE estimation for the three baseline DGPs.\footnote{The full tables with more performance measures are provided in Tables \ref{tab:app-gate0-base} for ITE0, \ref{tab:app-gate1-rn} for ITE1, and \ref{tab:app-gate2-rn} for ITE2. The results for all DGPs are provided in Appendix \ref{sec:app-gate}.} We observe similar patterns as for the IATE estimation and the categorization of estimators in Section \ref{sec:comp} remains by and large the same. The four estimators that show a consistently good performance for IATEs are also good choices for the estimation of GATEs.


\begin{table}[htp]
 \centering
\begin{threeparttable}
  \footnotesize
  \caption{Simulation results of GATE estimation for baseline DGPs}
    \begin{tabular}{lccccccccc}
    \toprule
          & \multicolumn{4}{c}{1000 observations} &       & \multicolumn{4}{c}{4000 observations} \\
\cmidrule{2-5}\cmidrule{7-10}          & $\overline{MSE}$    & $|\overline{Bias}|$  & $\overline{SD}$    & $JB$    &       & $\overline{MSE}$    & $|\overline{Bias}|$  & $\overline{SD}$    & $JB$  \\
    \midrule
           \multicolumn{10}{r}{\textbf{GATEs from ITE0 with selection and without random noise}} \\
    \midrule
    \textit{Random Forest:} &       &       &       &       &       &       &       &       &  \\
    Conditional mean regression & 1.77  & 0.55  & 1.19  & 22\%  &       & 1.07  & 0.49  & 0.86  & 8\% \\
    MOM IPW & 4.44  & 1.59  & 1.16  & 47\%  &       & 1.12  & 0.67  & 0.66  & 6\% \\
    MOM DR & \textbf{0.87} & 0.38  & 0.85  & 20\%  &       & 0.30  & 0.25  & 0.48  & 14\% \\
    Causal Forest & 1.44  & 0.70  & 0.96  & 17\%  &       & 0.74  & 0.64  & 0.53  & 0\% \\
    Causal Forest with local centering & 1.08  & 0.33  & 0.99  & 8\%   &       & 0.35  & 0.22  & 0.54  & 3\% \\
    \textit{Lasso:} &       &       &       &       &       &       &       &       &  \\
    Conditional mean regression & 3.35  & 0.55  & 1.70  & 34\%  &       & 1.45  & 0.47  & 1.06  & 6\% \\
    MOM IPW & 3.15  & 0.78  & 1.50  & 100\% &       & 1.19  & 0.53  & 0.87  & 86\% \\
    MOM DR & 38.85 & 0.59  & 6.20  & 100\% &       & 0.30  & 0.31  & 0.45  & 38\% \\
    MCM   & 4.56  & 1.20  & 1.62  & 100\% &       & 1.65  & 0.75  & 0.92  & 94\% \\
    MCM with efficiency augmentation & 1.04  & 0.41  & 0.93  & 66\%  &       & \textbf{0.27} & 0.26  & 0.45  & 55\% \\
    R-learning & 1.04  & 0.44  & 0.92  & 73\%  &       & \textbf{0.27} & 0.28  & 0.44  & 25\% \\
    \midrule
          & \multicolumn{9}{c}{\textbf{GATEs from ITE1 with selection and random noise}} \\
    \midrule
    \textit{Random Forest:} &       &       &       &       &       &       &       &       &  \\
    Conditional mean regression & 2.30  & 0.85  & 1.18  & 20\%  &       & 1.53  & 0.76  & 0.86  & 6\% \\
    MOM IPW & 3.84  & 1.41  & 1.17  & 41\%  &       & 0.99  & 0.54  & 0.67  & 11\% \\
    MOM DR & \textbf{1.16} & 0.59  & 0.83  & 20\%  &       & \textbf{0.49} & 0.42  & 0.48  & 17\% \\
    Causal Forest & 2.04  & 1.01  & 0.95  & 19\%  &       & 1.26  & 0.93  & 0.53  & 2\% \\
    Causal Forest with local centering & 1.38  & 0.56  & 0.97  & 17\%  &       & 0.58  & 0.44  & 0.54  & 5\% \\
    \textit{Lasso:} &       &       &       &       &       &       &       &       &  \\
    Conditional mean regression & 3.68  & 0.78  & 1.69  & 41\%  &       & 1.66  & 0.60  & 1.06  & 14\% \\
    MOM IPW & 3.03  & 0.65  & 1.53  & 100\% &       & 1.17  & 0.47  & 0.89  & 77\% \\
    MOM DR & 39.33 & 0.79  & 6.20  & 100\% &       & 0.59  & 0.50  & 0.46  & 42\% \\
    MCM   & 3.98  & 0.93  & 1.65  & 97\%  &       & 1.31  & 0.52  & 0.93  & 91\% \\
    MCM with efficiency augmentation & 1.40  & 0.62  & 0.93  & 80\%  &       & 0.55  & 0.47  & 0.45  & 39\% \\
    R-learning & 1.45  & 0.68  & 0.91  & 75\%  &       & 0.61  & 0.51  & 0.45  & 48\% \\
    \midrule
          & \multicolumn{9}{c}{\textbf{GATEs from ITE2 with selection and random noise}} \\
    \midrule
    \textit{Random Forest:} &       &       &       &       &       &       &       &       &  \\
    Conditional mean regression & 5.28  & 1.57  & 1.15  & 20\%  &       & 3.97  & 1.44  & 0.85  & 22\% \\
    MOM IPW & 3.75  & 1.25  & 1.18  & 41\%  &       & 1.72  & 0.95  & 0.68  & 14\% \\
    MOM DR & \textbf{3.50} & 1.29  & 0.84  & 23\%  &       & 2.19  & 1.08  & 0.49  & 23\% \\
    Causal Forest & 5.33  & 1.71  & 0.94  & 25\%  &       & 4.17  & 1.60  & 0.53  & 8\% \\
    Causal Forest with local centering & 3.74  & 1.28  & 0.98  & 11\%  &       & 2.43  & 1.12  & 0.55  & 9\% \\
    \textit{Lasso:} &       &       &       &       &       &       &       &       &  \\
    Conditional mean regression & 5.73  & 1.34  & 1.71  & 42\%  &       & 2.59  & 0.95  & 1.09  & 11\% \\
    MOM IPW & 4.00  & 1.01  & 1.59  & 100\% &       & 1.92  & 0.82  & 0.96  & 45\% \\
    MOM DR & 30.36 & 1.43  & 4.71  & 100\% &       & 2.71  & 1.23  & 0.52  & 69\% \\
    MCM   & 3.65  & 0.66  & 1.67  & 100\% &       & \textbf{1.58} & 0.61  & 0.97  & 81\% \\
    MCM with efficiency augmentation & 3.94  & 1.35  & 0.95  & 72\%  &       & 2.63  & 1.22  & 0.51  & 67\% \\
    R-learning & 4.27  & 1.43  & 0.93  & 72\%  &       & 2.95  & 1.29  & 0.50  & 50\% \\
    \bottomrule
    \end{tabular}%
     \begin{tablenotes} \item \textit{Notes:} $\overline{MSE}$ shows the mean MSE of all 10,000 observations in the validation sample, $|\overline{Bias}|$ denotes the mean absolute bias, $\overline{SD}$ the mean standard deviation, and $JB$ the fraction of observations for which the Jarque-Bera test is rejected at the 5\% level. Bold numbers indicate the best performing estimators in terms of $\overline{MSE}$ and estimators within two standard (simulation) errors of the lowest $\overline{MSE}$. \end{tablenotes}  
  \label{tab:results-gate}%
\end{threeparttable}%
\end{table}
\doublespacing 

For GATE estimation, we observe a new candidate with outstanding performance in a particular setting. MCM performs remarkably well for ITE2 showing the second lowest mean MSE. The mean absolute bias of MCM is already competitive for the estimation of IATEs of ITE2 in Table \ref{tab:results}. However, the mean SD is more than twice as large compared to the best estimators, which prevents a competitive performance in terms of mean MSE. The averaging of these noisy but relatively unbiased estimators seems to produce a competitive estimator for the higher aggregation level. Still, MCM performs poorly for the other DGPs. 

The averaging improves also the performance of locally centered Causal Forests relative to its uncentered version. The results for the estimation of IATEs show that the advantage in terms of mean absolute bias is partly offset by a higher variability. The aggregation step reduces this difference such that the lower bias translates also into a substantially lower mean MSE.

The aggregation leads further to a substantial reduction in the excess kurtosis of all Lasso estimators (see Tables of Appendix \ref{sec:app-gate}). However, the JB test is still rejected for most observations. Note that we observe for all estimators a substantial amount of bias although the influence of $Y_i^0$ and the irreducible noise is averaged out to a large extent. This indicates that the estimators are not able to completely remove the selection bias, which is particularly problematic if we would be interested in statistical inference. The results for ITE0 without noise and with random assignment in Appendix \ref{sec:app-gate} provide evidence in this direction.

The results for the estimation of ATEs in Appendix \ref{sec:app-ate} are mostly in line with those for GATEs. Again, MCM is highly competitive and provides the best performing estimators for ITE2. Also the benefits of averaging the locally centered Causal Forest are observed. The bias is halved compared to their uncentered version while both show similar SDs. 

The skewness and kurtosis show that the ATE estimators are mostly normally distributed with mean skewness close to zero and mean kurtosis close to three. The obvious exception is MOM DR with Lasso for which also averaging the IATEs does not mitigate the bad performance due to extreme outliers.

Finally, we note that the comparison of the mean MSE for the two sample sizes indicates that GATEs and ATEs estimators show a substantially faster convergence rate compared to the respective IATE estimators. This indicates that the additional averaging of noisily estimated IATEs results in faster convergence and the ATEs may be estimable with close to parametric rates. However, we do not overemphasize this finding as it is only based on two sample sizes.

\subsection{Alternative DGPs} \label{sec:alternative}

The discussions in the previous section focus on the results of the three baseline DGPs. This section summarizes the major insights from the alternative DGPs. Their results are provided in Appendices \ref{sec:app-ite-base} to \ref{sec:app-ate-rarn} where we also discuss details and peculiarities of the specific DGPs and aggregation levels. In general, the four estimators that are identified as the best performing for the baseline DGPs belong also to the best performing ones in the alternative DGPs.

For the estimation of \textit{IATEs}, we observe new candidates that are only successful in particular DGPs with selection into treatment. For example, Random Forest MOM IPW performs very good for ITE2 without noise. However, these and other peculiarities discussed in the Appendix hold only for either mean MSE or median MSE, while the four best performing candidates are usually competitive in both measures. Additionally, we assess whether our findings stay robust if we ignore the natural bounds of our outcome variable when creating the DGP. The results in Appendix \ref{sec:app-ite-nc} show that this is the case in a DGP that allows treated outcomes outside the natural bounds of the original outcome when defining the ITEs.

We also consider all previously discussed DGPs with \textit{random treatment assignment}. This means that the estimation problem becomes easier because selection bias is no longer a concern. By and large the results are in line with the respective results for the DGPs with selection into treatment, especially the conclusions about the best performing estimators are not changed. As expected, the mean MSEs for the DGP with random assignment are lower for most of the estimators and thus closer to the infeasible benchmark. This is always driven by a lower mean absolute bias while the mean SDs are very similar to the equivalent DGPs with selection. This suggests that the methods are not able to completely remove the selection bias.

Two other differences to the baseline DGPs are noteworthy. First, MOM DR with Lasso shows competitive performance already in small samples. This indicates that the bad performance is related to large errors made in the outcome \textit{and} the selection equation in small samples, which is in line with the simulation evidence of \citeA{Kang2007} for DR ATE estimators. Second, Causal Forest and its locally centered version show a nearly identical performance. This illustrates that local centering is only beneficial when there is selection into treatment.

The results for \textit{GATE and ATE} estimators confirm the observation in the baseline DGPs that IATE estimators with low bias but high variance can provide competitive estimators if they are averaged to higher aggregation levels. In particular, averaging MCM IATEs shows in many alternative DGPs a relatively good performance. However, MCM performs worst in some other DGPs in a non-systematic way. Thus, the results show that noise can be averaged out for these higher aggregation levels but there is no guarantee for this. The estimators that are already successful for the IATEs remain the most reliable choices for GATEs and ATEs.

In general, we find that the differences between the estimators become smaller, the more the IATEs are aggregated. Especially, the SDs become more similar by averaging IATEs such that the differences between the estimators are mainly driven by bias.

\section{Conclusion}

This is the first comprehensive simulation study in economics that investigates the finite sample performance of a large number of different causal machine learning estimators. We rely on arguably realistic DGPs that have potentially more external validity than the mostly synthetic DGPs considered so far in the limited simulation literature for these estimators. We consider DGPs with and without selection into treatment. Our main goal is to estimate individualized average treatment effects. Additionally, we report the performance of estimators that aggregate individualized average treatment effects to an intermediate and the population level.  

We do not find any single causal machine learning estimator that consistently outperforms all other estimators. However, we do find a group of four estimators that show competitive performance for all DGPs. This group includes the Causal Forest with local centering, Random Forest based MOM DR, MCM with efficiency augmentation, and R-learning. These estimators explicitly use both, the outcome and the treatment equations in a multiple step procedure. The estimators that use the Lasso have heavy tails in the smaller samples. 

The best performing estimators for the individualized average treatment effects produce also the most reliable estimators for higher aggregation levels. However, in some settings also noisily estimated individualized average treatment effects with low bias produce competitive estimators for higher aggregates because the noise is averaged out.

Despite relying as much as possible on arguably realistic DGPs, the external validity of every simulation study is uncertain. Future research will show if our findings hold in other empirical settings. Furthermore, it may be possible to improve the performance of each method with more tailored implementations. Finally, we focus in this study on the finite sample performance of point estimates and leave the investigation of inference procedures to future research.

\newpage

\bibliographystyle{apacite}
\renewcommand{\APACrefYearMonthDay}[3]{\APACrefYear{#1}}
\bibliography{references}

\newpage
\begin{appendices}
\counterwithin{figure}{section}
\counterwithin{table}{section}
 
\huge \noindent \textbf{Appendices}

\normalsize
 
\section{Data \label{sec:App_Dat}} 

\subsection{Dataset}

The data we use includes all individuals who are registered as unemployed at Swiss regional employment agencies in the year 2003. The data contains rich information from different unemployment insurance databases (AVAM/ASAL) and social security records (AHV). This is the standard data used for many Swiss ALMP evaluations \cite<e.g.,>{gerfin2002microeconometric,Lalive2008TheUnemployment,lechner2007value}. We observe (among others) residence status, qualification, education, language skills, employment history, profession, job position, industry of last job, and desired occupation and industry. The administrative data is linked with regional labour market characteristics, such as the population size of municipalities and the cantonal unemployment rate. The availability of extensive caseworker information and their subjective assessment of the employability of their clients distinguishes our data. Swiss caseworkers employed in the period 2003-2004 were surveyed based on a written questionnaire in December 2004 \cite<see>{behncke2010caseworker,behncke2010unemployed}. The questionnaire contained questions about aims and strategies of the caseworker and the regional employment agency. 

In total, 238,902 persons registered as being unemployed in 2003. We only consider the first unemployment registration per individual in 2003. Each registered unemployed person is assigned to a caseworker. In most cases, the same caseworker is responsible for the entire unemployment spell of her/his client. If this is not the case, we focus on the first caseworker to avoid concerns about (rare) endogenous caseworker changes \cite<see>{behncke2010caseworker}. We only consider unemployed aged between 24 and 55 years who receive unemployment insurance benefits. We omitted unemployed persons who apply for disability insurance benefits, when the responsible caseworker is not clearly defined, or when their caseworkers did not answer the questionnaire (the response rate is 84\%). We drop unemployed foreigners with a residence permit that is valid for less than a year. Finally, we drop unemployed persons from five regional employment agencies that are not comparable to the other regional employment agencies. This sample is identical to the data used in \citeA{Huber2017} and contains 100,120 unemployed persons. We drop further 3,822 observations that participated in alternative treatments. This leaves us with 96,298 to estimate the propensity score. After dropping 342 observations with propensity score below 5\% and above 95\% and dropping the treated, we are left with 88,844 observations for the simulation. 10,000 are used as validation sample and 78,844 to draw the simulation samples.

\subsection{Descriptive statistics}

This Appendix provides descriptive statistics of the dataset that is used to build the EMCS. Table \ref{tab:Desc_stat} shows the mean and the standard deviation of the outcome, the estimated propensity score, and the variables that are used to estimate the propensity score. Other variables and transformations that are part of the 1,749 covariates for the heterogeneity analysis are omitted. The last column reports standardized differences to illustrate covariate imbalances between treated and controls. Standardized differences normalize the absolute mean difference between two groups by the square root of their mean variance:
\begin{equation} \label{eq:SD}
\text{Std. Diff.} = \dfrac{\mid \bar{X_1} -\bar{X_0}  \mid}{\sqrt{ 1/2(Var(X_1)+Var(X_0))}} \cdot 100
\end{equation}

We observe that the estimated propensity score is highly imbalanced with a standardized difference of 77, while already a value of 20 is considered as indication for large imbalances \cite{Rosenbaum1985ConstructingScore}. This shows that we operate in a setting of high selectivity. The imbalances are mainly driven by differences in the language regions, the previous labor market history and employability of the unemployed.

\begin{figure}[ht]
\caption{Histogram of observed $Y^0$}
\centering
\includegraphics[width=0.75\textwidth]{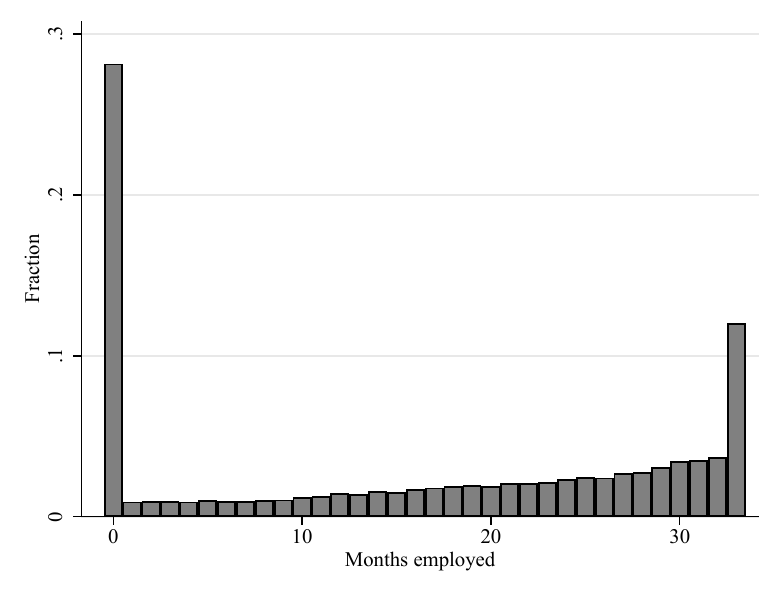}
\captionsetup{width=.75\textwidth}
\caption*{\footnotesize \textit{Notes:} Histogram of the non-treated outcomes counting the months of employment in the 33 months after the start of the job search program.}
\label{fig:hist-y0}
\end{figure}

Figure \ref{fig:hist-y0} shows the distribution of the observed non-treated outcomes that are used in the EMCS. We observe mass points at the bounds of the outcome variable. Nearly 30\% of the unemployed find no job at all in the 33 months after the start of the job search program and roughly 10\% find a job right away and stay employed for the whole period. 

\begin{figure}[ht]
\caption{Heatmap of baseline covariates}
\centering
\includegraphics[width=0.75\textwidth]{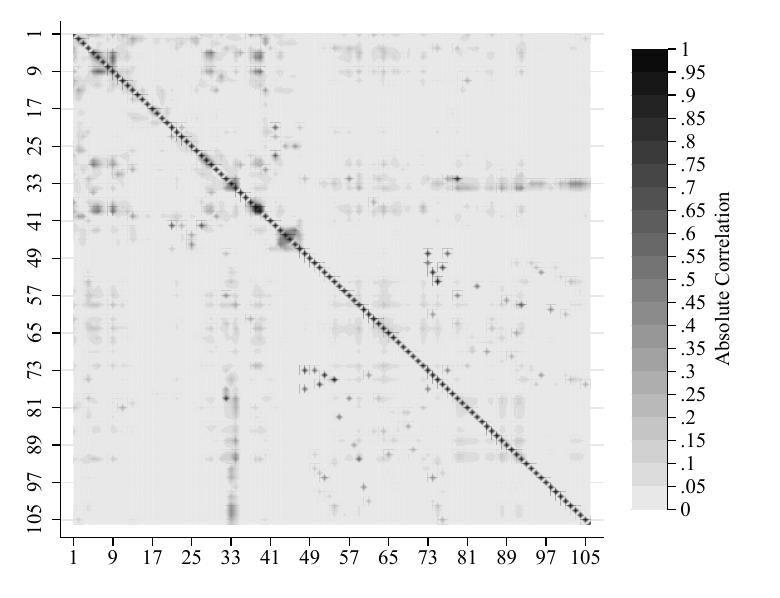}
\captionsetup{width=.75\textwidth}
\caption*{\footnotesize \textit{Notes:} The heatmap visualizes the absolute correlations of the 105 covariates that are used in the heterogeneity analysis.}
\label{fig:hmap}
\end{figure}

Finally, the absolute correlations of the 105 available covariates are visualized as a so-called heatmap in Figure \ref{fig:hmap}. The dark spots indicate that some of the covariates are highly correlated.

\begingroup
\singlespacing

\begin{ThreePartTable}
\begin{TableNotes}
\item \textit{Notes:} SD means standard deviation. Std. Diff. stands for standardized differences and are calcualted according to Equation \ref{eq:SD}.
\end{TableNotes}

\begin{center}  \footnotesize 
    \begin{longtable}{lccccccc}  
\caption{Descriptive statistics} \label{tab:Desc_stat} \\

    \toprule
          & \multicolumn{2}{c}{Treated} &       & \multicolumn{2}{c}{Controls} &       &  \\
\cmidrule{2-3}\cmidrule{5-6}          & Mean  & SD    &       & Mean  & SD    &       & Std. Diff. \\
    \midrule
\endfirsthead

\multicolumn{8}{c}%
{{ \tablename\ \thetable{} -- continued from previous page}} \\
    \toprule
          & \multicolumn{2}{c}{Treated} &       & \multicolumn{2}{c}{Controls} &       &  \\
\cmidrule{2-3}\cmidrule{5-6}          & Mean  & SD    &       & Mean  & SD    &       & Std. Diff. \\
    \midrule
\endhead

\hline \multicolumn{8}{r}{{Continued on next page}} \\ \hline
\endfoot

\bottomrule
\insertTableNotes  
\endlastfoot

    Cumulated months of employment: 33 months & 16.8  & 12.8  &       & 16.1  & 12.8  &       & 4.8 \\
    \midrule
    Propensity score in population & 0.11  & 0.06  &       & 0.07  & 0.05  &       & 77.0 \\
    \midrule
    \multicolumn{8}{c}{Covariates: Characteristics of unemployed persons} \\
    \midrule
    Female & 0.46  & 0.50  &       & 0.44  & 0.50  &       & 3.4 \\
              *French & 0.03  & 0.16  &       & 0.11  & 0.31  &       & 32.1 \\
              *Italian & 0.01  & 0.10  &       & 0.04  & 0.18  &       & 16.1 \\
    Mother tongue other than German, French, Italian & 0.26  & 0.44  &       & 0.32  & 0.47  &       & 13.6 \\
              *French & 0.02  & 0.14  &       & 0.07  & 0.26  &       & 26.1 \\
              *Italian & 0.01  & 0.08  &       & 0.02  & 0.15  &       & 13.4 \\
    Unskilled & 0.21  & 0.40  &       & 0.23  & 0.42  &       & 5.5 \\
              *French & 0.02  & 0.15  &       & 0.05  & 0.22  &       & 13.7 \\
              *Italian & 0.01  & 0.10  &       & 0.03  & 0.16  &       & 12.4 \\
    Qualification: semiskilled & 0.13  & 0.34  &       & 0.16  & 0.37  &       & 7.8 \\
              *French & 0.01  & 0.11  &       & 0.04  & 0.21  &       & 18.9 \\
              *Italian & 0.002 & 0.05  &       & 0.01  & 0.08  &       & 6.7 \\
    Qualification: skilled without degree & 0.03  & 0.17  &       & 0.05  & 0.21  &       & 8.3 \\
              *French & 0.002 & 0.05  &       & 0.02  & 0.13  &       & 15.8 \\
              *Italian & 0.002 & 0.04  &       & 0.01  & 0.08  &       & 6.3 \\
    \# of unemp. spells in last 2 years & 0.49  & 1.09  &       & 0.58  & 1.21  &       & 7.6 \\
              *French & 0.05  & 0.40  &       & 0.17  & 0.72  &       & 20.0 \\
              *Italian & 0.02  & 0.26  &       & 0.06  & 0.42  &       & 10.6 \\
    Fraction of months emp. in last 2 years & 0.84  & 0.22  &       & 0.79  & 0.25  &       & 18.2 \\
              *French & 0.05  & 0.20  &       & 0.18  & 0.35  &       & 47.7 \\
              *Italian & 0.02  & 0.12  &       & 0.06  & 0.21  &       & 23.2 \\
    Employability rating low & 0.11  & 0.32  &       & 0.14  & 0.34  &       & 6.9 \\
              *French & 0.01  & 0.08  &       & 0.02  & 0.14  &       & 12.4 \\
              *Italian & 0.004 & 0.06  &       & 0.01  & 0.10  &       & 6.8 \\
    Employability rating medium & 0.76  & 0.43  &       & 0.74  & 0.44  &       & 3.7 \\
              *French & 0.05  & 0.22  &       & 0.19  & 0.39  &       & 43.6 \\
              *Italian & 0.01  & 0.12  &       & 0.04  & 0.20  &       & 17.6 \\
    Education: above vocational training & 0.45  & 0.50  &       & 0.44  & 0.50  &       & 3.0 \\
              *French & 0.03  & 0.17  &       & 0.10  & 0.30  &       & 28.6 \\
              *Italian & 0.01  & 0.08  &       & 0.02  & 0.15  &       & 13.6 \\
    Education: tertiary track  & 0.23  & 0.42  &       & 0.24  & 0.43  &       & 0.6 \\
              *French & 0.02  & 0.13  &       & 0.09  & 0.29  &       & 31.9 \\
              *Italian & 0.003 & 0.06  &       & 0.02  & 0.12  &       & 13.0 \\
    Vocational training degree & 0.28  & 0.45  &       & 0.23  & 0.42  &       & 11.7 \\
              *French & 0.002 & 0.05  &       & 0.01  & 0.11  &       & 12.1 \\
              *Italian & 0.01  & 0.11  &       & 0.04  & 0.19  &       & 17.2 \\
    Age in 10 year & 3.70  & 0.87  &       & 3.66  & 0.87  &       & 5.2 \\
    Age squared / 10,000 & 0.14  & 0.07  &       & 0.14  & 0.07  &       & 5.0 \\
    Married & 0.45  & 0.50  &       & 0.49  & 0.50  &       & 8.2 \\
    Foreigner with B permit & 0.10  & 0.29  &       & 0.14  & 0.35  &       & 13.3 \\
    Foreigner with C permit & 0.21  & 0.41  &       & 0.25  & 0.43  &       & 9.2 \\
    Lives in big city & 0.14  & 0.35  &       & 0.17  & 0.37  &       & 7.6 \\
    Lives in medium sized city & 0.17  & 0.37  &       & 0.13  & 0.34  &       & 9.9 \\
    Past income (in CHF 10,000) & 0.46  & 0.20  &       & 0.42  & 0.21  &       & 20.5 \\
    Number of employment spells in last 5 years & 0.10  & 0.13  &       & 0.12  & 0.15  &       & 15.1 \\
    Previous job in primary sector & 0.06  & 0.24  &       & 0.09  & 0.29  &       & 12.3 \\
    Previous job in secondary sector & 0.15  & 0.35  &       & 0.13  & 0.34  &       & 3.6 \\
    Previous job in tertiary sector & 0.64  & 0.48  &       & 0.58  & 0.49  &       & 11.5 \\
    Foreigner with mother tongue in canton's language & 0.12  & 0.32  &       & 0.11  & 0.32  &       & 1.6 \\
    Previous job self-employed & 0.003 & 0.06  &       & 0.01  & 0.08  &       & 4.5 \\
    Previous job manager & 0.08  & 0.28  &       & 0.07  & 0.26  &       & 4.1 \\
    Previous job skilled worker & 0.65  & 0.48  &       & 0.61  & 0.49  &       & 8.4 \\
    Previous job unskilled worker & 0.24  & 0.43  &       & 0.29  & 0.45  &       & 10.7 \\
    \midrule
    \multicolumn{8}{c}{Covariates: Allocation of unemployed to caseworkers} \\
    \midrule
    By industry & 0.68  & 0.47  &       & 0.54  & 0.50  &       & 30.2 \\
              *French & 0.03  & 0.18  &       & 0.10  & 0.29  &       & 24.9 \\
              *Italian & 0.01  & 0.10  &       & 0.03  & 0.18  &       & 16.5 \\
    By occupation & 0.59  & 0.49  &       & 0.56  & 0.50  &       & 5.2 \\
              *French & 0.05  & 0.21  &       & 0.17  & 0.37  &       & 39.7 \\
              *Italian & 0.01  & 0.11  &       & 0.04  & 0.21  &       & 20.3 \\
    By age & 0.03  & 0.18  &       & 0.03  & 0.18  &       & 0.6 \\
    By employability & 0.06  & 0.23  &       & 0.07  & 0.25  &       & 4.1 \\
    By region & 0.09  & 0.28  &       & 0.12  & 0.32  &       & 10.5 \\
    Other & 0.07  & 0.25  &       & 0.07  & 0.26  &       & 2.9 \\
    \midrule
    \multicolumn{8}{c}{Covariates: Caseworker characteristics} \\
    \midrule
    Age in years & 44.68 & 11.54 &       & 44.35 & 11.60 &       & 2.8 \\
              *French & 2.89  & 11.60 &       & 11.04 & 20.20 &       & 49.5 \\
              *Italian & 0.95  & 6.44  &       & 3.24  & 11.61 &       & 24.3 \\
    Female & 0.47  & 0.50  &       & 0.41  & 0.49  &       & 11.5 \\
              *French & 0.02  & 0.13  &       & 0.09  & 0.28  &       & 32.2 \\
              *Italian & 0.01  & 0.10  &       & 0.02  & 0.15  &       & 9.8 \\
    Tenure (in years) & 5.63  & 3.14  &       & 5.83  & 3.31  &       & 6.2 \\
              *French & 0.37  & 1.59  &       & 1.54  & 3.03  &       & 48.2 \\
              *Italian & 0.18  & 1.26  &       & 0.57  & 2.23  &       & 21.9 \\
    Own unemp. experience & 0.63  & 0.48  &       & 0.63  & 0.48  &       & 0.1 \\
              *French & 0.04  & 0.20  &       & 0.16  & 0.37  &       & 40.7 \\
              *Italian & 0.02  & 0.13  &       & 0.05  & 0.21  &       & 17.1 \\
    Indicator for missing caseworker characteristics & 0.04  & 0.20  &       & 0.04  & 0.20  &       & 0.7 \\
    \midrule
    \multicolumn{8}{c}{Covariates: Local labour market characteristics} \\
    \midrule
    French speaking REA & 0.06  & 0.24  &       & 0.24  & 0.43  &       & 52.2 \\
    Italian speaking REA & 0.02  & 0.15  &       & 0.08  & 0.26  &       & 24.9 \\
    Cantonal unemployment rate (in \%) & 3.55  & 0.75  &       & 3.74  & 0.86  &       & 24.5 \\
              *French & 0.23  & 0.93  &       & 1.02  & 1.84  &       & 53.9 \\
              *Italian & 0.10  & 0.64  &       & 0.32  & 1.13  &       & 24.4 \\
    Cantonal GDP per capita (in CHF 10,000) & 0.51  & 0.09  &       & 0.49  & 0.09  &       & 12.0 \\
    \midrule
    Number of observations & \multicolumn{2}{c}{7,545} & & \multicolumn{2}{c}{88,844} & & \\

\end{longtable}
\end{center}
\end{ThreePartTable}
\endgroup

\FloatBarrier
\clearpage

\section{DGP} \label{sec:DGP}

\subsection{Propensity score} \label{sec:app-p_score}

\begin{figure}[!h]
\caption{Propensity scores by treatment status}
\centering
\includegraphics[width=0.8\textwidth]{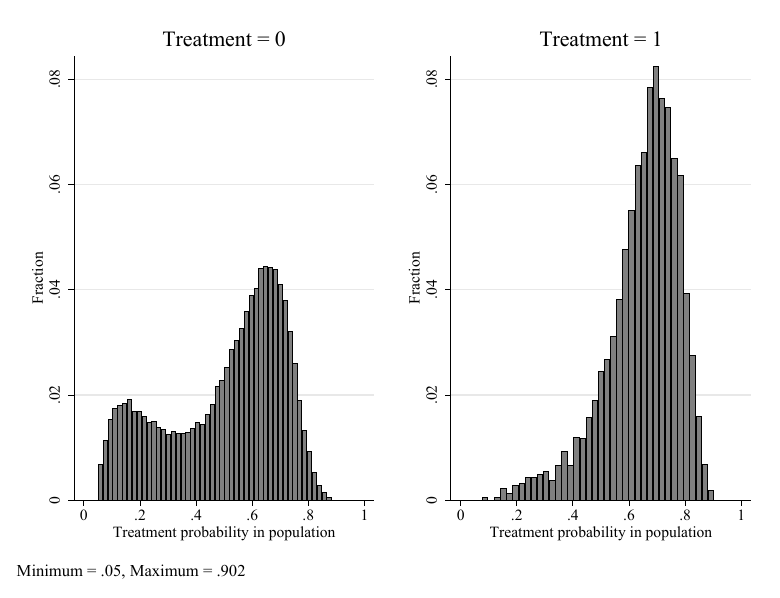}
\captionsetup{width=.8\textwidth}
\caption*{\footnotesize \textit{Notes:} Histogram of the manipulated propensity score for the non-treated (left) and the treated (right).}
\label{fig:p-score}
\end{figure}

Table \ref{tab:p_score} shows the estimated propensity score using the specification of \citeA{Huber2017}. Many variables have sizeable and statistically significant coefficients that create the selection into the treatment. Figure \ref{fig:p-score} shows the distribution of the propensity score after the manipulations described in Section \ref{sec:p_score}. Figure \ref{fig:p-score} shows that our setting does not creates problems due to extreme propensity scores or no overlap.

\begingroup

\begin{ThreePartTable}
\begin{TableNotes}
\item \textit{Notes:} Coefficients and average marginal effects of the propensity score based on the specification of \citeA{Huber2017}. Standard errors are in parentheses. ***/**/* indicate statistical significance at the 1\%/5\%/10\%-level.
\end{TableNotes}

\singlespacing
 \footnotesize 
\begin{longtable}{lcc} 
\caption{Propensity score} \label{tab:p_score} \\

    \toprule
          & (1)   & (2) \\
\cmidrule{2-3}          & Coefficients & Marginal effects \\
    \midrule
\endfirsthead

\multicolumn{3}{c}%
{{ \tablename\ \thetable{} -- continued from previous page}} \\
    \toprule
          & (1)   & (2) \\
\cmidrule{2-3}          & Coefficients & Marginal effects \\
    \midrule
\endhead

\hline \multicolumn{3}{r}{{Continued on next page}} \\ \hline
\endfoot

\bottomrule
\insertTableNotes  
\endlastfoot

\multicolumn{3}{c}{{Characteristics of unemployed persons}} \\
\midrule

    Female & 0.140*** & 0.00683*** \\
          & (0.0294) & (0.00145) \\
         ~~~~*French & -0.0813 & -0.00383 \\
          & (0.105) & (0.00481) \\
         ~~~~*Italian & -0.0562 & -0.00266 \\
          & (0.167) & (0.00770) \\
    Mother tongue other than German, French, Italian & -0.239*** & -0.0112*** \\
          & (0.0393) & (0.00177) \\
         ~~~~*French & -0.0312 & -0.00150 \\
          & (0.119) & (0.00561) \\
         ~~~~*Italian & 0.0151 & 0.000735 \\
          & (0.203) & (0.00995) \\
    Unskilled & 0.115*** & 0.00576** \\
          & (0.0447) & (0.00230) \\
         ~~~~*French & 1.019*** & 0.0756*** \\
          & (0.130) & (0.0136) \\
         ~~~~*Italian & 0.441** & 0.0259* \\
          & (0.206) & (0.0144) \\
    Qualification: semiskilled & -0.134*** & -0.00622*** \\
          & (0.0436) & (0.00195) \\
         ~~~~*French & 0.893*** & 0.0631*** \\
          & (0.144) & (0.0139) \\
         ~~~~*Italian & 0.693** & 0.0459* \\
          & (0.289) & (0.0251) \\
    Qualification: skilled without degree & 0.0718 & 0.00358 \\
          & (0.0828) & (0.00426) \\
         ~~~~*French & -0.0561 & -0.00265 \\
          & (0.288) & (0.0133) \\
         ~~~~*Italian & 0.313 & 0.0175 \\
          & (0.318) & (0.0202) \\
    \# of unemp. spells in last 2 years & 0.0105 & 0.000511 \\
          & (0.0131) & (0.000634) \\
         ~~~~*French & 0.0482 & 0.00234 \\
          & (0.0364) & (0.00176) \\
         ~~~~*Italian & 0.105* & 0.00510* \\
          & (0.0558) & (0.00270) \\
    Fraction of months emp. in last 2 years & 0.225*** & 0.0109*** \\
          & (0.0640) & (0.00310) \\
         ~~~~*French & -0.108 & -0.00522 \\
          & (0.199) & (0.00967) \\
         ~~~~*Italian & -0.0512 & -0.00248 \\
          & (0.334) & (0.0162) \\
    Employability rating low & -0.641*** & -0.0255*** \\
          & (0.0562) & (0.00184) \\
         ~~~~*French & 1.323*** & 0.115*** \\
          & (0.248) & (0.0335) \\
         ~~~~*Italian & 1.335*** & 0.118*** \\
          & (0.291) & (0.0404) \\
    Employability rating medium & -0.310*** & -0.0161*** \\
          & (0.0414) & (0.00232) \\
         ~~~~*French & 0.879*** & 0.0558*** \\
          & (0.194) & (0.0157) \\
         ~~~~*Italian & 0.773*** & 0.0519*** \\
          & (0.224) & (0.0200) \\
    Education: above vocational training & 0.0594* & 0.00289* \\
          & (0.0311) & (0.00152) \\
         ~~~~*French & 0.115 & 0.00581 \\
          & (0.139) & (0.00733) \\
         ~~~~*Italian & 0.131 & 0.00674 \\
          & (0.198) & (0.0107) \\
    Education: tertiary track  & 0.167*** & 0.00843*** \\
          & (0.0380) & (0.00200) \\
         ~~~~*French & -0.285* & -0.0125** \\
          & (0.151) & (0.00592) \\
         ~~~~*Italian & -0.281 & -0.0121 \\
          & (0.274) & (0.0104) \\
    Vocational training degree & 0.120*** & 0.00601*** \\
          & (0.0308) & (0.00158) \\
         ~~~~*French & -0.332 & -0.0139 \\
          & (0.267) & (0.00963) \\
         ~~~~*Italian & 0.125 & 0.00638 \\
          & (0.180) & (0.00965) \\
    Age in 10 year & 0.0955 & 0.00463 \\
          & (0.136) & (0.00662) \\
    Age squared / 10,000 & -0.814 & -0.0395 \\
          & (1.755) & (0.0851) \\
    Married & -0.0235 & -0.00114 \\
          & (0.0292) & (0.00141) \\
    Foreigner with B permit & -0.187*** & -0.00852*** \\
          & (0.0511) & (0.00219) \\
    Foreigner with C permit & -0.0815** & -0.00388** \\
          & (0.0376) & (0.00176) \\
    Lives in big city & -0.138*** & -0.00640*** \\
          & (0.0414) & (0.00185) \\
    Lives in medium sized city & 0.223*** & 0.0117*** \\
          & (0.0361) & (0.00203) \\
    Past income (in CHF 10,000) & 0.754*** & 0.0366*** \\
          & (0.0763) & (0.00371) \\
    Number of employment spells in last 5 years & -0.649*** & -0.0314*** \\
          & (0.105) & (0.00511) \\
    Previous job in primary sector & -0.263*** & -0.0116*** \\
          & (0.0619) & (0.00247) \\
    Previous job in secondary sector & 0.198*** & 0.0103*** \\
          & (0.0481) & (0.00266) \\
    Previous job in tertiary sector & 0.113*** & 0.00545*** \\
          & (0.0373) & (0.00178) \\
    Foreigner with mother tongue in canton's language & 0.302*** & 0.0163*** \\
          & (0.0430) & (0.00257) \\
    Previous job self-employed & -0.884*** & -0.0295*** \\
          & (0.236) & (0.00509) \\
    Previous job manager & -0.480*** & -0.0195*** \\
          & (0.0970) & (0.00326) \\
    Previous job skilled worker & -0.306*** & -0.0153*** \\
          & (0.0841) & (0.00436) \\
    Previous job unskilled worker & -0.309*** & -0.0141*** \\
          & (0.0896) & (0.00388) \\
                    \midrule
          \multicolumn{3}{c}{{Allocation of unemployed to caseworkers}} \\
\midrule     
    By industry & 0.349*** & 0.0167*** \\
          & (0.0296) & (0.00142) \\
         ~~~~*French & 0.114 & 0.00579 \\
          & (0.107) & (0.00563) \\
         ~~~~*Italian & -0.529*** & -0.0207*** \\
          & (0.175) & (0.00540) \\
    By occupation & 0.201*** & 0.00963*** \\
          & (0.0280) & (0.00134) \\
         ~~~~*French & 0.487*** & 0.0275*** \\
          & (0.122) & (0.00792) \\
         ~~~~*Italian & -0.463*** & -0.0186*** \\
          & (0.171) & (0.00566) \\
    By age & -0.0511 & -0.00243 \\
          & (0.0720) & (0.00334) \\
    By employability & -0.362*** & -0.0153*** \\
          & (0.0549) & (0.00201) \\
    By region & -0.349*** & -0.0151*** \\
          & (0.0451) & (0.00173) \\
    Other & -0.260*** & -0.0114*** \\
          & (0.0512) & (0.00204) \\
      \midrule
          \multicolumn{3}{c}{{Caseworker characteristics}} \\
\midrule   
    Age (in 10 years) & -0.000912 & -4.42e-05 \\
          & (0.00136) & (6.60e-05) \\
         ~~~~*French & 0.0217*** & 0.00105*** \\
          & (0.00547) & (0.000264) \\
         ~~~~*Italian & 0.00413 & 0.000200 \\
          & (0.0109) & (0.000528) \\       
    Female & 0.205*** & 0.0101*** \\
          & (0.0276) & (0.00139) \\
         ~~~~*French & -0.272** & -0.0119** \\
          & (0.118) & (0.00466) \\
         ~~~~*Italian & 0.469*** & 0.0279** \\
          & (0.172) & (0.0123) \\
    Tenure (in years) & 0.0266*** & 0.00129*** \\
          & (0.00425) & (0.000206) \\
         ~~~~*French & -0.0749*** & -0.00363*** \\
          & (0.0196) & (0.000946) \\
         ~~~~*Italian & 0.00549 & 0.000266 \\
          & (0.0276) & (0.00134) \\
    Own unemp. experience & 0.0251 & 0.00121 \\
          & (0.0281) & (0.00135) \\
         ~~~~*French & -0.00642 & -0.000311 \\
          & (0.115) & (0.00556) \\
         ~~~~*Italian & 0.627*** & 0.0395** \\
          & (0.197) & (0.0157) \\
    Indicator for missing caseworker characteristics & 0.0950 & 0.00479 \\
          & (0.0759) & (0.00398) \\

                    \midrule
          \multicolumn{3}{c}{{Local labour market characteristics}} \\
\midrule    
    French speaking REA & -1.346*** & -0.0497*** \\
          & (0.496) & (0.0143) \\
    Italian speaking REA & -3.548*** & -0.0628*** \\
          & (1.035) & (0.00636) \\
    Cantonal unemployment rate (in \%) & 0.215*** & 0.0104*** \\
          & (0.0256) & (0.00124) \\
         ~~~~*French & -0.690*** & -0.0334*** \\
          & (0.0666) & (0.00314) \\
         ~~~~*Italian & -0.00787 & -0.000381 \\
          & (0.179) & (0.00869) \\
    Cantonal GDP per capita (in CHF 10,000) & -3.805*** & -0.184*** \\
          & (0.232) & (0.0113) \\
    Constant & -1.700*** &  \\
          & (0.290) &  \\
          &       &  \\
         \midrule
    Number of observations & 96,298 & 96,298 \\

\end{longtable}

\end{ThreePartTable}

\endgroup

\clearpage

\subsection{Description of ITEs} \label{sec:app-ite}

\begin{figure}[!htb]
\caption{Histogram of ITE1}
\centering
\includegraphics[width=0.75\textwidth]{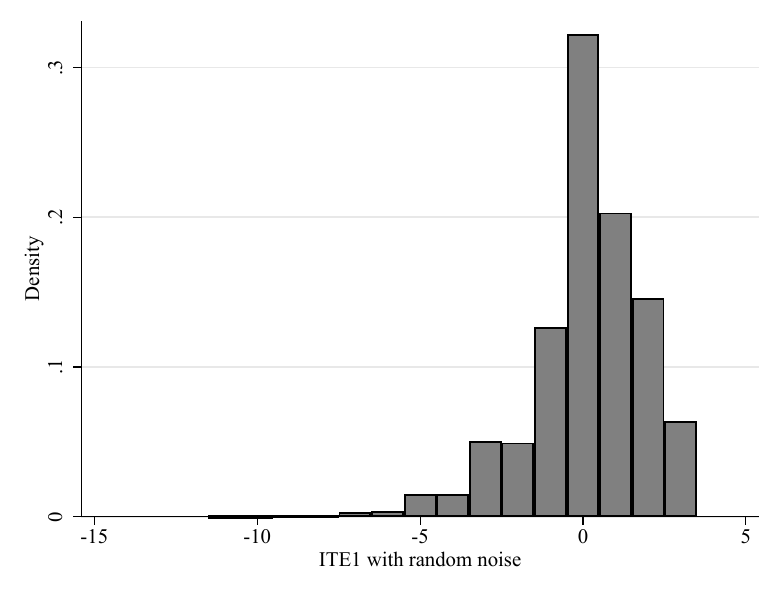}
\footnotesize \flushleft 
\label{fig:ite1}

\caption{Histogram of ITE2}
\centering
\includegraphics[width=0.75\textwidth]{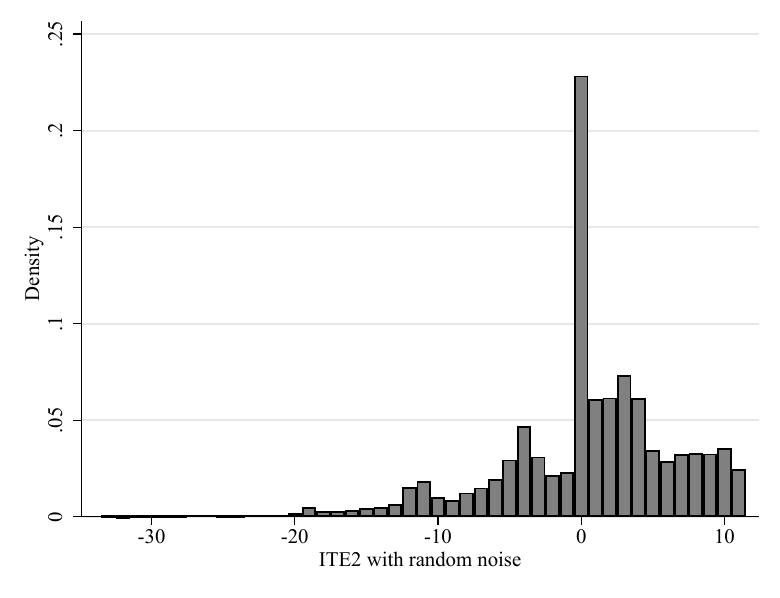}
\footnotesize \flushleft 
\label{fig:ite2}
\end{figure}

This Appendix complements the basic statistics of the baseline DGP provided in Table \ref{tab:ITE-desc} in the main text. Figures \ref{fig:ite1} and \ref{fig:ite2} show the histograms of the baseline ITE1 and ITE2 with random noise, respectively. We observe a bunching of ITEs at zero because we force the observations to respect the bounds of the outcome variable (see Figure \ref{fig:hist-y0}).

\begin{figure}[!htb]
\caption{Cumulative density functions of the potential outcomes of ITE1}
\centering
\includegraphics[width=0.75\textwidth]{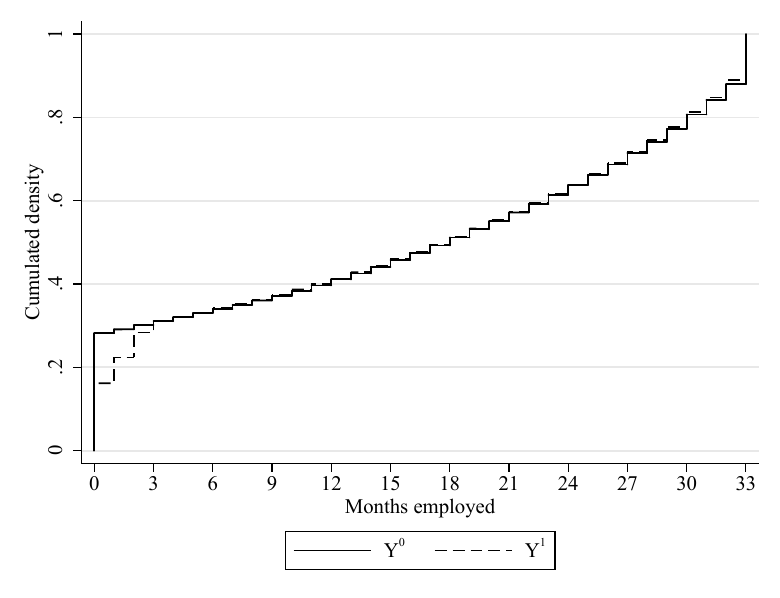}
\label{fig:cdf1}

\caption{Cumulative density functions of the potential outcomes of ITE2}
\centering
\includegraphics[width=0.75\textwidth]{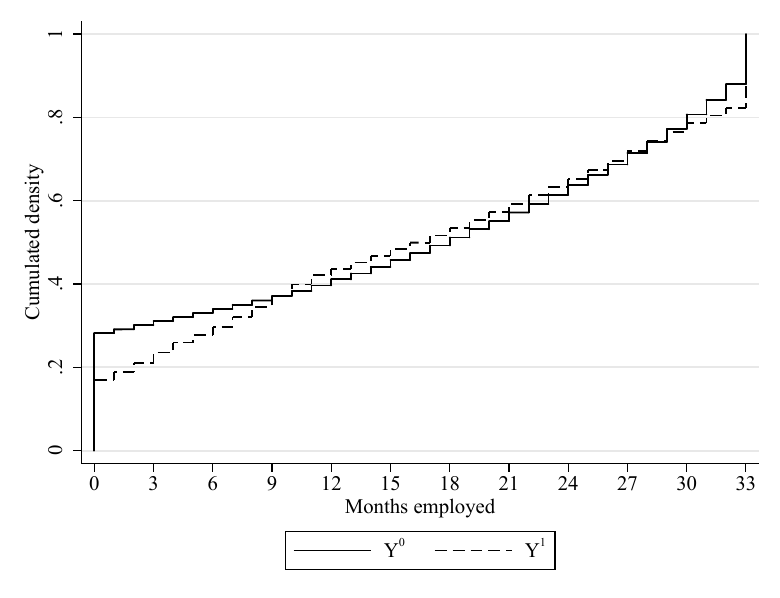}
\label{fig:cdf2}
\end{figure}

Figures \ref{fig:cdf1} and \ref{fig:cdf2} compare the cumulative density functions of $Y_i^0$ and $Y_i^1$ in the population for the baseline ITE1 and ITE2, respectively. Adding the ITEs to $Y_i^0$ while respecting the bounds results in more mass of $Y_i^1$ away from the bounds.\footnote{This is also the reason for the lower standard deviations of the $Y_i^1$s in Table \ref{tab:ITE-desc}.} Further, we observe that the impact of ITE1 on the distribution of the potential outcomes is moderate, while the larger ITE2 changes the potential outcome distribution substantially.

\begin{figure}[!htb]
\caption{Relation of propensity score and potential outcomes}
\centering
\includegraphics[width=0.75\textwidth]{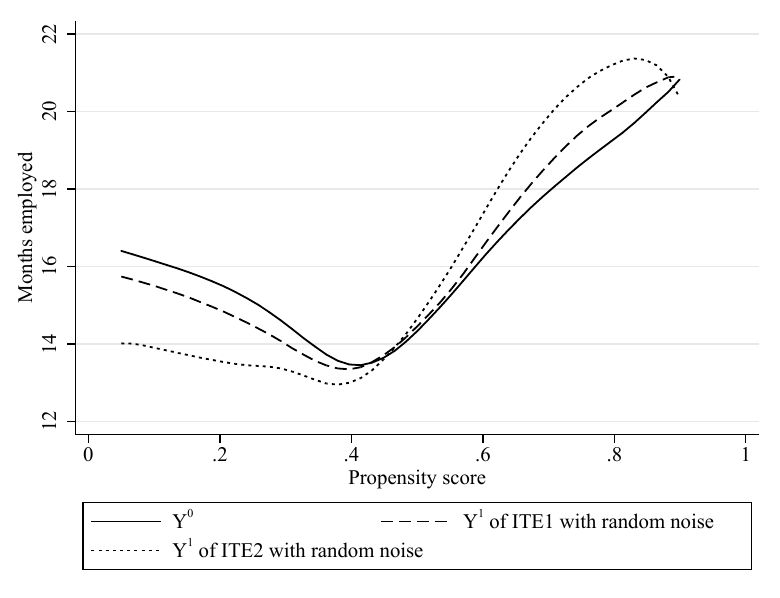}
\captionsetup{width=.75\textwidth}
\caption*{\footnotesize \textit{Notes:} Local constant regression with Epanechnikov kernel and Silverman's bandwidth rule. }
\label{fig:p_out}
\end{figure}

Figure \ref{fig:p_out} shows the relation of the propensity score with the ITEs. The 'simulated' caseworkers in our setting are rather successful in identifying unemployed with gains from the program. Those unemployed with a probability of lower than 50\% of being send to the program have on average also negative IATEs, which is evident from the dashed lines of the treated potential outcomes being below the solid line of the non-treated potential outcomes. In contrast, the unemployed that participate with a probability of more than 50\% benefit on average from the job search program. The simulated assignment mechanism is therefore favorable for most unemployed. However, we build in one feature that is often observed in applications, namely 'cream-skimming' \cite<see, e.g.,>{bell2002screening}. This means that unemployed persons with good labour market opportunities (a high $Y_i^0$) have a greater probability to participate in a JSP. However, the effect of the program is not necessarily positive for participants with good labour market opportunities because these participants would have good labour market opportunities even in the absence of training and just suffer from the lock-in effect \cite<see e.g.,>{Card2017WhatEvaluations}. The downward sloping average potential outcomes for very high propensity scores reflect this empirical observations.

Finally, Table \ref{tab:r-sq} provides an idea about how much of the variation in the potential outcomes, ITEs and the treatment can be explained by the observable characteristics at our disposal. Potential outcomes are relatively hard to approximate. The potential outcome of non-treated that is taken from real data shows an out-of-sample $R^2$ of at most about 10\%. The ITEs without noise are easier to explain with up to about 70\% explainable variation of ITE2. As we add these systematically varying ITE to the non-treated outcomes to obtain the treated outcomes, the different $Y_i^1$ are easier to approximate compared to $Y_i^0$ with explained variation of up to 30\%. The picture is changed if we add random noise to create the baseline ITE1 and ITE2. This makes the ITEs and the respective $Y_i^1$s much harder to approximate with $R^2$ of at most 11\%. The propensity score is the component that is most easy to approximate. The Lasso is even able to explain 80\% of the variation. The reason is that the Lasso has access to all relevant variables and uses the true link function. However, it may not recover exactly the true model due to the shrunken coefficiencts.

\begin{table}[!htb]
  \centering 
\begin{threeparttable}

  \caption{Out-of-sample R-squared of different components in \%}
    \begin{tabular}{lccccc} 
    \toprule
    Sample size: & \multicolumn{2}{c}{1000} &       & \multicolumn{2}{c}{4000} \\
\cmidrule{2-3}\cmidrule{5-6}    Machine learner: & \multicolumn{1}{c}{Random Forest} & \multicolumn{1}{c}{Lasso} &       & \multicolumn{1}{c}{Random Forest} & \multicolumn{1}{c}{Lasso} \\
    \midrule
    \multicolumn{6}{l}{\textit{Without random noise ($\varepsilon_i = 0$):}} \\
    $Y^0$    & 7.6   & 6.5   &       & 10.6  & 10.7 \\
    $Y^1$ ITE1 & 9.4   & 8.2   &       & 12.5  & 13.1 \\
    $Y^1$ ITE2 & 16.5  & 22.0  &       & 25.2  & 30.5 \\
    ITE1  & 56.8  & 62.3  &       & 62.1  & 67.7 \\
    ITE2  & 55.0  & 59.9  &       & 60.2  & 65.4 \\
    Propensity score & 56.8  & 62.9  &       & 75.2  & 80.3 \\
    \multicolumn{6}{l}{\textit{With random noise ($\varepsilon_i \sim 1 - Poisson(1)$):}} \\
    $Y^0$     & 7.6   & 6.5   &       & 10.6  & 10.7 \\
    $Y^1$ ITE0 & 8.1   & 6.6   &       & 10.9  & 10.7 \\
    $Y^1$ ITE1 & 8.4   & 6.9   &       & 11.2  & 11.1 \\
    $Y^1$ ITE2 & 7.1   & 5.8   &       & 9.7   & 10.3 \\
    ITE0  & 0.04  & 0.02  &       & 0.3   & 0.1 \\
    ITE1  & 4.8   & 4.2   &       & 6.3   & 6.3 \\
    ITE2  & 3.2   & 2.7   &       & 4.7   & 4.8 \\
    Propensity score & 56.8  & 62.9  &       & 75.2  & 80.3 \\
    \bottomrule
    \end{tabular}%
    \begin{tablenotes} \item \textit{Notes:} Table shows the average out-of-sample R-squared in the validation sample over all replications. \end{tablenotes}  
  \label{tab:r-sq}%
  \end{threeparttable}
\end{table}%

\FloatBarrier

\subsection{Specification and description of GATEs} \label{sec:App_gate}

\begin{figure}[!htb]
\caption{Histogram of group sizes used to calculate true GATEs}
\centering
\includegraphics[width=0.75\textwidth]{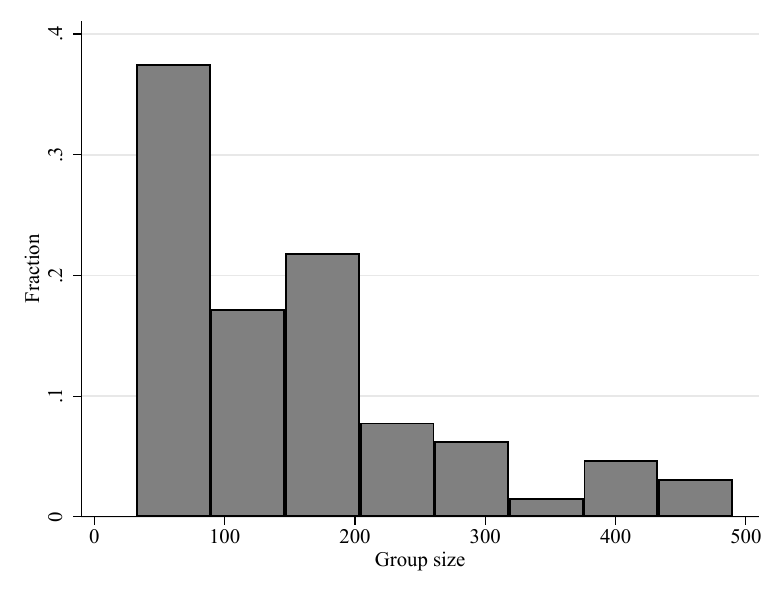}
\label{fig:hist-gate-gr-size}
\end{figure}

\begin{figure}[!htb]
\caption{Histogram of the GATEs of ITE1}
\centering
\includegraphics[width=0.75\textwidth]{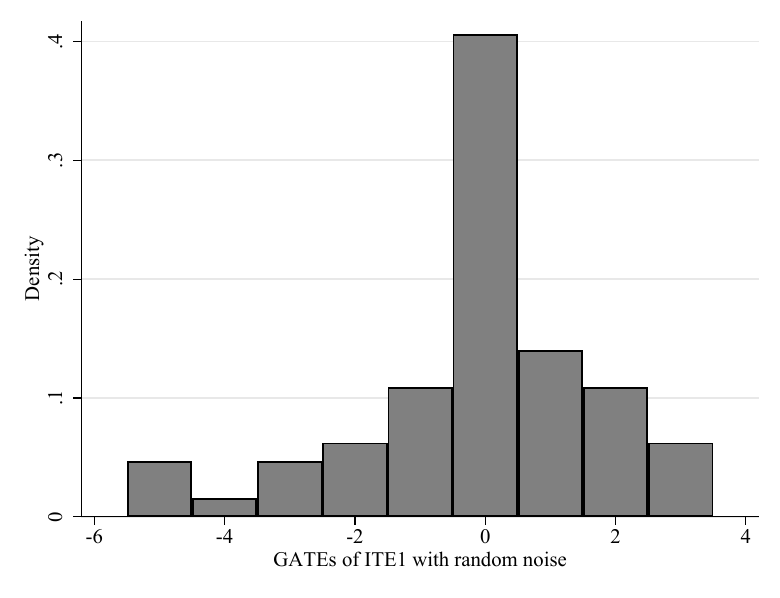}
\label{fig:hist-gate1}

\caption{Histogram of the GATEs of ITE2}
\centering
\includegraphics[width=0.75\textwidth]{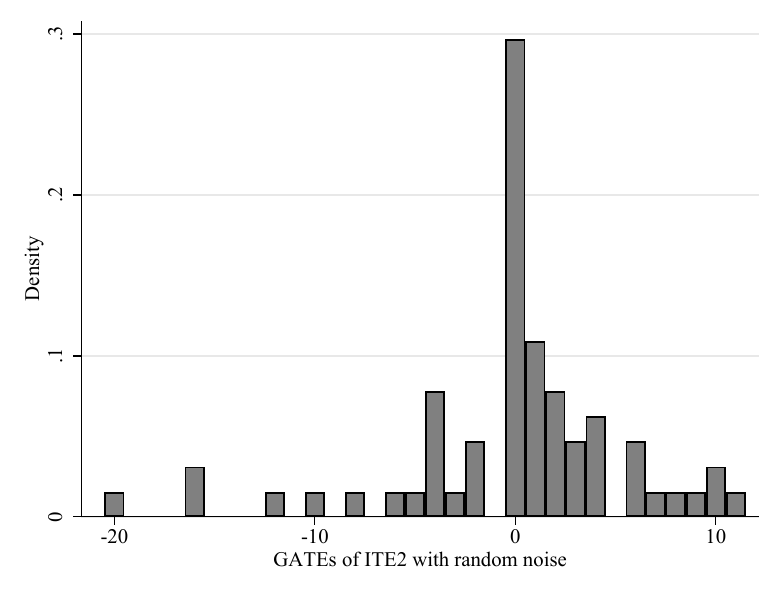}
\label{fig:hist-gate2}
\end{figure}

We want to create a setting where we need to summarize the heterogeneity over groups that are of interest to the policy maker and might be used by caseworkers to assign program participation. We consider the scenario where we categorize the unemployed by six characteristics: employability, gender, age, qualification, foreigner and language region. This splits the 10,000 validation observations into 64 groups of sizes 32 to 420 (histogram in Figure \ref{fig:hist-gate-gr-size}) by using the following combinations: employability (three categories) x female (binary) x foreigner (binary) x some qualification degree (binary). The biggest group with medium employability is additionally interacted with three age groups ($<30,~30-40,~>40$) and the German speaking cantons of Switzerland (binary). Figures \ref{fig:hist-gate1} and \ref{fig:hist-gate2} show the histogram of the distribution of the resulting GATEs of the baseline ITE1 and ITE2.

\FloatBarrier
\clearpage

\section{Implementation} \label{sec:app-impl}

\begin{figure}[!htb]
\caption{Overview of inputs and estimators}
\centering
\includegraphics[width=0.75\textwidth]{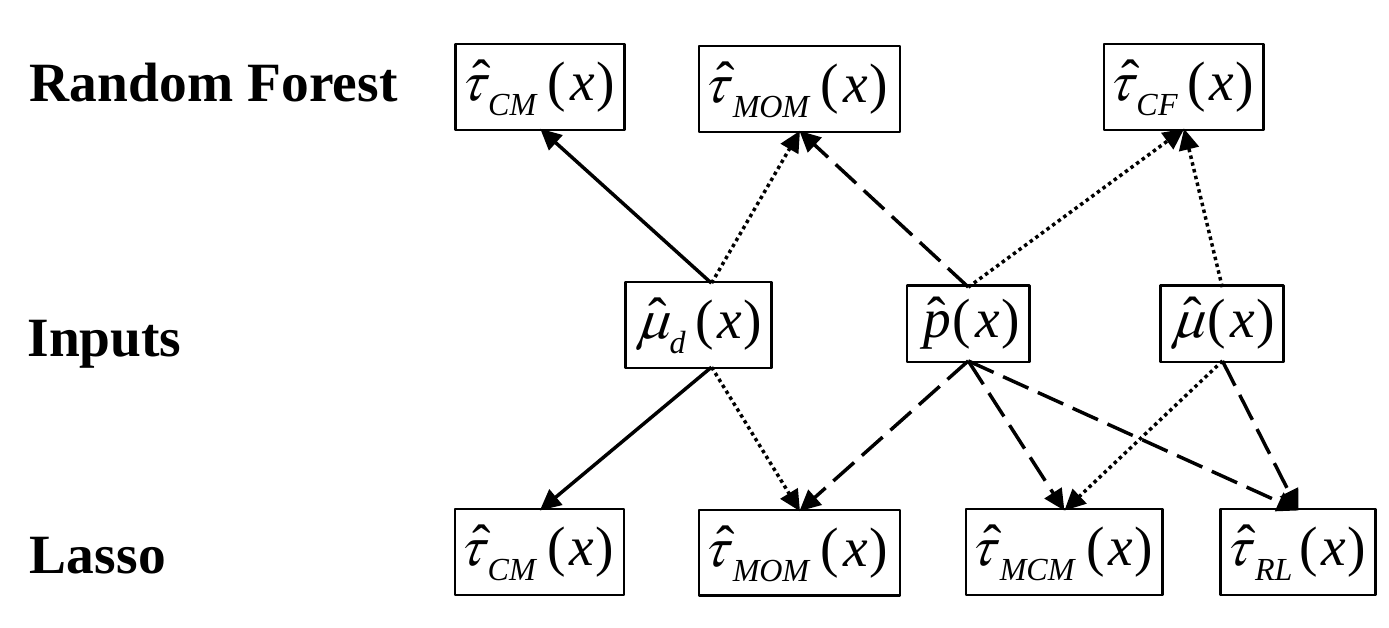}
\captionsetup{width=0.75\textwidth}
\caption*{\textit{Notes:} Solid lines indicate that this is the only input. Dashed lines represent necessary nuisance parameters. Dotted lines indicate optional nuisance parameters. $\tau_{MOM}(x)$ summarizes IPW and DR versions of MOM approaches.}
\label{fig:arrow}
\end{figure}

This Appendix provides a brief description of the implementation steps of the compared estimators.\footnote{We do not repeat the tuning parameter choices at each prediction step because they follow always the procedures that are described in Section \ref{sec:imp-rf} for Random Forest and Section \ref{sec:imp-lasso} for Lasso. Similarly, we omit the cross-fitting steps because the basic principle is already explained in Section \ref{sec:cross-fit}.} Before, Figure \ref{fig:arrow} provides a graphical summary how different parameters enter the estimation as either the only inputs, as necessary, or as optional nuisance parameters. It shows also how we can use one input for multiple estimators such that the estimation of IATEs requires at most one more step after having $\hat{\mu}_d(x)$, $\hat{p}(x)$ or $\hat{\mu}(x)$. These synergies are important to keep computational time under control.

\singlespacing
\begin{table}[H]
  \centering
  \caption{Conditional mean regression}
    \begin{tabularx}{\textwidth}{rX}
        \toprule
		1. & Regress $Y_i$ on $X_i$ in the non-treated sample to obtain a prediction model for $\hat{\mu}(0,x)$. \\
		[0.2em]
		2. & Regress $Y_i$ on $X_i$ in the treated sample to obtain a prediction model for $\hat{\mu}(1,x)$. \\
		[0.2em]
		3. & Calculate $\hat{\tau}_{CMR}(x) = \hat{\mu}(1,x) - \hat{\mu}(0,x)$ for each observation in the validation sample. \\
	 	\bottomrule
    \end{tabularx}%
\end{table}%

\begin{table}[H]
  \centering
  \caption{Causal Forest}
    \begin{tabularx}{\textwidth}{rX}
        \toprule
		1. & Apply Algorithm 1 of \citeA{Athey2017a} with the pseudo outcomes of Equation \ref{pseudo} in the so-called labeling step. \\
		[0.2em]
		2. & Calculate $\hat{\tau}_{CF}(x)$  for each observation in the validation sample according to \ref{CF} using the weights obtained in step 1. \\
	 	\bottomrule
    \end{tabularx}%
\end{table}%

\begin{table}[H]
  \centering
  \caption{Causal Forest with local centering}
    \begin{tabularx}{\textwidth}{rX}
        \toprule
		1. & Regress $D_i$ on $X_i$ to obtain a prediction model for $\hat{p}(x)$. \\
		[0.2em]
		2. & Regress $Y_i$ on $X_i$ to obtain a prediction model for $\hat{\mu}(x)$. \\
		[0.2em]
		3. & Apply Algorithm 1 of \citeA{Athey2017a} with the pseudo outcomes of Equation \ref{pseudo} but replacing $D_i$ and $Y_i$ with $D_i - p(X_i)$ and $Y_i - \mu(X_i)$. \\
		4. & Calculate $\hat{\tau}_{CF\_LC}(x)$  for each observation in the validation sample according to \ref{CF} using the weights obtained in step 3. \\
	 	\bottomrule
    \end{tabularx}%
\end{table}%

\begin{table}[H]
  \centering
  \caption{MOM with IPW}
    \begin{tabularx}{\textwidth}{rX}
        \toprule
		1. & Regress $D_i$ on $X_i$ to obtain a prediction model for $\hat{p}(x)$. \\
		[0.2em]
		2. & Create the modified outcome $Y_{i,IPW}^*$ by replacing $p(x)$ with $\hat{p}(x)$ in Equation \ref{MOMIPW}. \\
		[0.2em]
		3. & Regress $Y_{i,IPW}^*$ on $X_i$ to obtain $\hat{\tau}_{IPW}(x)$ and use it to calculate the IATE for each observation in the validation sample. \\
	 	\bottomrule
    \end{tabularx}%
\end{table}%

\begin{table}[H]
  \centering
  \caption{MOM with DR}
    \begin{tabularx}{\textwidth}{rX}
        \toprule
		1. & Regress $D_i$ on $X_i$ to obtain a prediction model for $\hat{p}(x)$. \\
		[0.2em]
		2. & Regress $Y_i$ on $X_i$ in the non-treated sample to obtain a prediction model for $\hat{\mu}(0,x)$. \\
		[0.2em]
		3. & Regress $Y_i$ on $X_i$ in the treated sample to obtain a prediction model for $\hat{\mu}(1,x)$. \\
		[0.2em]
		4. & Create the modified outcome $Y_{i,DR}^*$ by replacing the nuisance parameters by their estimates $\hat{p}(x)$, $\hat{\mu}(0,x)$, and $\hat{\mu}(1,x)$ in Equation \ref{MOMDR}. \\
		[0.2em]
		5. & Regress $Y_{i,DR}^*$ on $X_i$ to obtain $\hat{\tau}_{DR}(x)$ and use it to calculate the IATE for each observation in the validation sample. \\
	 	\bottomrule
    \end{tabularx}%
\end{table}%

The following three tables indicate two different ways to implement the estimators. Either by modifying the covariates (indicated by \textit{a}) or by modifying the outcomes (indicated by \textit{b}).

\begin{table}[H] \label{tab:mcm}
  \centering
  \caption{MCM }
    \begin{tabularx}{\textwidth}{rX}
        \toprule
		1. & Regress $D_i$ on $X_i$ to obtain a prediction model for $\hat{p}(x)$. \\
		[0.2em]
		2a. & Modify the covariates as $X_i^{MCM}=T_i/2X_i$. \\
		[0.2em]
		2b. & Modify the outcome as $Y^*_{MCM}=2 T_i Y_i$. \\
		[0.2em]
		3a. & Use $\hat{p}(x)$ and $X_i^{MCM}$ to obtain $\hat{\beta}_{MCM}$ from Equation \ref{MCMli}. \\
		[0.2em]
		3b. & Use $\hat{p}(x)$ and $Y^*_{MCM}$ to obtain $\hat{\beta}_{MCM}$ from Equation \ref{MCMgen} with $\tau(x)=x \beta$. \\
		[0.2em]
		4. & Calculate the IATE for each observation in the validation sample as $\hat{\tau}_{MCM}(x) = x\hat{\beta}_{MCM}$. \\
	 	\bottomrule
    \end{tabularx}%
\end{table}%

\begin{table}[H]
  \centering
  \caption{MCM with efficiency augmentation}
    \begin{tabularx}{\textwidth}{rX}
        \toprule
		1. & Regress $D_i$ on $X_i$ to obtain a prediction model for $\hat{p}(x)$. \\
		[0.2em]
		2. & Regress $Y_i$ on $X_i$ to obtain a prediction model for $\hat{\mu}(x)$. \\
		[0.2em]
		3a. & Modify the covariates as $X_i^{MCM}=T_i/2X_i$. \\
		[0.2em]
		3b. & Modify the outcome as $Y^*_{MCM_EA}=2T_i (Y_i - \mu(X_i))$. \\
		[0.2em]
		4a. & Use $\hat{p}(x)$ and $X_i^{MCM}$ to obtain $\hat{\beta}_{MCM\_EA}$ from Equation \ref{MCMli} with $Y_i$ being replaced by  $Y_i - \hat{\mu}(x)$. \\
		[0.2em]
		4b. & Use $\hat{p}(x)$ and $Y^*_{MCM_EA}$ to obtain $\hat{\beta}_{MCM\_EA}$ from Equation \ref{MCMgen} with $\tau(x)=x \beta$. \\
		[0.2em]
		5. & Calculate the IATE for each observation in the validation sample as $\hat{\tau}_{MCM\_EA}(x) = x\hat{\beta}_{MCM\_EA}$. \\
	 	\bottomrule
    \end{tabularx}%
\end{table}%

\begin{table}[H] \label{tab:rl}
  \centering
  \caption{R-learning}
    \begin{tabularx}{\textwidth}{rX}
        \toprule
		1. & Regress $D_i$ on $X_i$ to obtain a prediction model for $\hat{p}(x)$. \\
		[0.2em]
		2. & Regress $Y_i$ on $X_i$ to obtain a prediction model for $\hat{\mu}(x)$. \\
		[0.2em]
		3a. & Modify the covariates as $X_i^{RL}=(D_i - p(X_i))X$. \\
		[0.2em]
		3b. & Modify the outcome as $Y^*_{RL}=\dfrac{Y_i - \mu(X_i)}{D_i - p(X_i)}$. \\
		[0.2em]
		4a. & Use $\hat{p}(x)$, $\hat{\mu}(x)$ and $X_i^{RL}$ to obtain $\hat{\beta}_{RL}$ from Equation \ref{RLli}. \\
	    4b. & Use $\hat{p}(x)$, $\hat{\mu}(x)$ and $Y^*_{RL}$ to obtain $\hat{\beta}_{RL}$ from Equation \ref{RLgen} with $\tau(x)=x \beta$. \\
		[0.2em]
		5. & Calculate the IATE for each observation in the validation sample as $\hat{\tau}_{RL}(x) = x\hat{\beta}_{RL}$. \\
	 	\bottomrule
    \end{tabularx}%
\end{table}%

\FloatBarrier
\doublespacing

\section{More results}  \label{sec:final_app}

This Appendix provides the results for all settings with the full set of performance measures. Additionally to the basic measure of Section \ref{sec:performance}, we calculate and report the following performance measures for IATEs and GATEs.

To understand simulation noise, we calculate the standard error of our main measure, $\overline{MSE}$, by

\begin{equation}
	SE(\overline{MSE}) = \sqrt{\frac{1}{R} \sum_{r=1}^{R} [MSE_r - \overline{MSE}]^2},
\end{equation}

where $MSE_r = \frac{1}{N_v} \sum_{v=1}^{N_v} [\xi(x_v,y_v^0) - \hat{\tau}(x_v)_r]^2$.\footnote{This can also be regarded as the standard error of the mean Precision in Estimation of Heterogenous Effect (PEHE) introduced by \citeA{Hill2011}. Note that the mean MSE and mean PEHE result in the same number.} This measure indicates how precise the mean MSE is measured and is used to asses whether the performance differs significantly or just up to noise. Furthermore, instead of taking the mean over all validation observations $v$, we consider the median $(Median(MSE_v))$. This measure leads sometimes to different orderings of the estimators compared to $\overline{MSE}$, which indicates that outliers play a role in measuring the performance. 

Additionally to the mean absolute bias, we report the mean bias:

\begin{equation}
	\overline{Bias} = \dfrac{1}{N_v} \sum_{v=1}^{N_v} \left[ \dfrac{1}{R} \sum_{r=1}^{R}  \hat{\tau}(x_v)_r - \xi(x_v,y_v^0) \right]
\end{equation}

Additionally to the fraction of rejected JB tests, we report the mean skewness and mean kurtosis over the validation sample.

All previous measures summarize the performance over all individuals. However, the measures in the last two columns in the following tables summarize the performance on the replication level. One is the correlation between the estimated and the true ITEs (Corr.) and one is the variance ratio of the estimated and the true ITEs (Var. Ratio),

\begin{equation}
	\text{Corr.} = \dfrac{1}{R} \sum_{r=1}^{R} \rho_{\hat{\tau}_r,\xi},
\end{equation}

where $\rho_{X,Y}$ denotes the correlation between two variables and $\hat{\tau}_r$ and $\xi$ are vectors of length $N_v$ containing the estimated ITEs in replication $r$ and the true ITEs, respectively.

\begin{equation}
	\text{Var. Ratio} = \dfrac{1}{R} \sum_{r=1}^{R} \dfrac{Var(\hat{\tau}_r)}{Var(\xi)},
\end{equation}

where $Var(\cdot)$ is the variance of the respective vector.

\subsection{Results for IATE estimation} \label{sec:app-performance}

The Appendices \ref{sec:app-ite-base} to \ref{sec:app-ite-rarn} show the full results for IATE estimation in the 24 DGP-sample size combinations. The correlations as additional performance measure in columns 10 show that the estimated IATEs are mostly positively correlated with the true ITEs. This shows that the considered estimators find systematic variation. However, the size of the correlations vary for the different settings. They are larger for the DGP without noise and for larger sample sizes, as expected.

The variance ratios in columns 11 show that estimators tend to overshoot and to create IATEs that vary more than the true ITEs. This is reflected in variance ratios of above one that are particularly prevalent for ITE0 and ITE1. This suggests that estimators tend to overshoot in settings with little variation of the ITEs relative to the variation in the outcome. Especially, both MOM IPW estimators and MCM are prone to create highly variable IATEs. This is in line with the high mean SD of these methods that use only the inverse propensity score to correct for selection bias.

\clearpage

\subsubsection{ITE with selection and without random noise}  \label{sec:app-ite-base}
Table \ref{tab:app-ite0} provides additional information for the baseline ITE0. The alternative performance measures confirm the results in the main text. Furthermore, note that the substantial positive mean bias suggests that the estimators are not able to completely remove the positive selection bias which is created by the positive relation between propensity score and ITEs. 

The results for ITE1 without random noise are very similar to the baseline results for ITE1 (Table \ref{tab:app-ite1}). However, ITE2 without random noise shows some notable differences to its baseline version with random noise. Random Forest based MOM IPW seems to work quite well for ITE2 as it shows the lowest mean MSE (Table \ref{tab:app-ite2}). However, the median MSE in column three suggests that this is driven by some outliers because the outstanding performance is not observed for this measure.

\newgeometry{left=0.4in,right=0.5in,top=1in,bottom=1.2in,nohead}
\begin{landscape}
    \singlespacing

\begin{threeparttable}[t]
  \centering \footnotesize      
  \caption{Performance measure for ITE0 with selection and without random noise (baseline)}
    \begin{tabular}{lccccccccccc}
    \toprule
          & $\overline{MSE}$   & $SE(\overline{MSE})$  & Median MSE & $|\overline{Bias}|$ & $\overline{Bias}$ & $\overline{SD}$    & $JB$    & Skew. & Kurt. & Corr. & Var. ratio \\
          \midrule
          & (1)   & (2)   & (3)   & (4)   & (5)   & (6)   & (7)   & (8)   & (9)   & (10)  & (11) \\
          \midrule
          & \multicolumn{11}{c}{\textbf{1000 observations}} \\
    \midrule
    \textit{Random Forest:} &       &       &       &       &       &       &       &       &       &       &  \\
    Infeasible & \multicolumn{11}{c}{No variation in dependent variable} \\
    Conditional mean regression & 3.69  & 0.04  & 3.60  & 0.62  & 0.60  & 1.78  & 6\%   & 0.0   & 3.0   & -     & - \\
    MOM IPW & 10.52 & 0.08  & 8.52  & 2.05  & 0.71  & 2.16  & 18\%  & 0.0   & 3.1   & -     & - \\
    MOM DR & \textbf{2.00} & 0.02  & 1.94  & 0.40  & 0.40  & 1.35  & 7\%   & 0.0   & 3.0   & -     & - \\
    Causal Forest & 3.52  & 0.04  & 3.44  & 0.75  & 0.75  & 1.69  & 12\%  & 0.0   & 3.1   & -     & - \\
    Causal Forest with local centering & 3.42  & 0.03  & 3.29  & 0.34  & 0.34  & 1.81  & 10\%  & 0.0   & 3.0   & -     & - \\
    \textit{Lasso:} &       &       &       &       &       &       &       &       &       &       &  \\
    Infeasible & \multicolumn{11}{c}{No variation in dependent variable} \\
    Conditional mean regression & 11.21 & 0.11  & 10.28 & 0.69  & 0.60  & 3.19  & 91\%  & -0.1  & 3.9   & -     & - \\
    MOM IPW & 11.31 & 0.28  & 9.69  & 1.09  & 0.61  & 2.99  & 100\% & 0.7   & 18.7  & -     & - \\
    MOM DR & 45.39 & 43.11 & 37.31 & 0.60  & 0.60  & 6.31  & 100\% & 41.1  & 1794.2 & -     & - \\
    MCM   & 13.03 & 0.27  & 10.46 & 1.50  & 0.45  & 3.05  & 100\% & -0.1  & 7.5   & -     & - \\
    MCM with efficiency augmentation & 2.08  & 0.05  & 1.91  & 0.42  & 0.42  & 1.36  & 100\% & 0.0   & 8.7   & -     & - \\
    R-learning & \textbf{2.03} & 0.05  & 1.85  & 0.45  & 0.45  & 1.33  & 100\% & 0.0   & 9.4   & -     & - \\ \midrule
          & \multicolumn{11}{c}{\textbf{4000 observations}} \\
    \midrule
    \textit{Random Forest:} &       &       &       &       &       &       &       &       &       &       &  \\
    Infeasible & \multicolumn{11}{c}{No variation in dependent variable} \\
    Conditional mean regression & 2.79  & 0.04  & 2.64  & 0.61  & 0.53  & 1.49  & 4\%   & 0.0   & 3.0   & -     & - \\
    MOM IPW & 5.76  & 0.06  & 4.44  & 1.31  & 0.53  & 1.74  & 12\%  & 0.0   & 3.1   & -     & - \\
    MOM DR & 1.16  & 0.01  & 1.10  & 0.28  & 0.28  & 1.03  & 8\%   & 0.0   & 3.0   & -     & - \\
    Causal Forest & 2.31  & 0.03  & 2.24  & 0.72  & 0.70  & 1.29  & 6\%   & 0.0   & 3.0   & -     & - \\
    Causal Forest with local centering & 2.05  & 0.02  & 1.94  & 0.24  & 0.24  & 1.40  & 9\%   & 0.0   & 3.0   & -     & - \\
    \textit{Lasso:} &       &       &       &       &       &       &       &       &       &       &  \\
    Infeasible & \multicolumn{11}{c}{No variation in dependent variable} \\
    Conditional mean regression & 6.14  & 0.08  & 5.62  & 0.65  & 0.49  & 2.30  & 35\%  & 0.0   & 3.4   & -     & - \\
    MOM IPW & 5.02  & 0.17  & 4.31  & 0.82  & 0.45  & 1.93  & 100\% & 0.2   & 7.9   & -     & - \\
    MOM DR & \textbf{0.51} & 0.02  & 0.43  & 0.31  & 0.31  & 0.62  & 35\%  & 0.0   & 3.4   & -     & - \\
    MCM   & 5.79  & 0.22  & 4.28  & 1.03  & 0.34  & 1.94  & 100\% & 0.0   & 6.5   & -     & - \\
    MCM with efficiency augmentation & \textbf{0.48} & 0.02  & 0.42  & 0.26  & 0.26  & 0.62  & 97\%  & 0.0   & 8.7   & -     & - \\
    R-learning & \textbf{0.47} & 0.02  & 0.42  & 0.28  & 0.28  & 0.61  & 97\%  & 0.1   & 8.7   & -     & - \\

    \bottomrule
    \end{tabular}%
         \begin{tablenotes} \item \textit{Notes:} Table shows the performance measures defined in Sections \ref{sec:performance} and \ref{sec:app-performance} over 2000 replications for the sample size of 1000 observations and 500 replications for the sample size of 4000 observations. Bold numbers indicate the best performing estimators in terms of $\overline{MSE}$ and estimators within two standard (simulation) errors of the lowest $\overline{MSE}$. \end{tablenotes}  
  \label{tab:app-ite0}%
\end{threeparttable}

\doublespacing
    \singlespacing

\begin{threeparttable}[t]
  \centering \footnotesize
  \caption{Performance measure for ITE1 with selection and without random noise}
    \begin{tabular}{lccccccccccc}
    \toprule
          & $\overline{MSE}$   & $SE(\overline{MSE})$  & Median MSE & $|\overline{Bias}|$ & $\overline{Bias}$ & $\overline{SD}$    & $JB$    & Skew. & Kurt. & Corr. & Var. ratio \\
          \midrule
          & (1)   & (2)   & (3)   & (4)   & (5)   & (6)   & (7)   & (8)   & (9)   & (10)  & (11) \\
          \midrule
          & \multicolumn{11}{c}{\textbf{1000 observations}} \\
    \midrule
    \textit{Random Forest:} &       &       &       &       &       &       &       &       &       &       &  \\
    Infeasible & 1.34  & 0.00  & 0.92  & 0.98  & 0.00  & 0.14  & 59\%  & 0.0   & 3.1   & 0.76  & 0.48 \\
    Conditional mean regression & 7.22  & 0.05  & 4.86  & 1.54  & 1.16  & 1.81  & 7\%   & 0.0   & 3.0   & 0.14  & 0.98 \\
    MOM IPW & 10.20 & 0.08  & 7.35  & 1.80  & 1.16  & 2.24  & 17\%  & 0.0   & 3.1   & 0.60  & 4.03 \\
    MOM DR & \textbf{4.80} & 0.03  & 2.88  & 1.37  & 0.75  & 1.33  & 11\%  & 0.0   & 3.0   & 0.18  & 0.45 \\
    Causal Forest & 7.21  & 0.05  & 4.65  & 1.63  & 1.36  & 1.70  & 12\%  & 0.0   & 3.1   & 0.16  & 0.82 \\
    Causal Forest with local centering & 6.22  & 0.04  & 4.43  & 1.35  & 0.72  & 1.79  & 12\%  & 0.0   & 3.0   & 0.13  & 0.88 \\
    \textit{Lasso:} &       &       &       &       &       &       &       &       &       &       &  \\
    Infeasible & 1.17  & 0.00  & 0.68  & 0.83  & 0.01  & 0.33  & 67\%  & 0.0   & 3.5   & 0.79  & 0.60 \\
    Conditional mean regression & 14.64 & 0.12  & 13.09 & 1.41  & 1.01  & 3.30  & 90\%  & -0.1  & 4.3   & 0.15  & 3.76 \\
    MOM IPW & 13.57 & 0.30  & 11.72 & 1.21  & 0.94  & 3.24  & 100\% & 0.3   & 18.0  & 0.39  & 4.30 \\
    MOM DR & 44.64 & 39.07 & 34.97 & 1.48  & 0.87  & 6.03  & 100\% & 40.5  & 1760.1 & 0.10  & 2.04 \\
    MCM   & 13.08 & 0.22  & 11.18 & 1.11  & 0.57  & 3.19  & 100\% & -0.1  & 6.6   & 0.42  & 4.54 \\
    MCM with efficiency augmentation & 5.31  & 0.06  & 3.04  & 1.42  & 0.70  & 1.41  & 100\% & -0.1  & 9.4   & 0.11  & 0.46 \\
    R-learning & 5.35  & 0.06  & 2.83  & 1.47  & 0.79  & 1.34  & 100\% & 0.0   & 10.3  & 0.08  & 0.39 \\
    \midrule
          & \multicolumn{11}{c}{\textbf{4000 observations}} \\
    \midrule
    \textit{Random Forest:} &       &       &       &       &       &       &       &       &       &       &  \\
    Infeasible & 1.17  & 0.00  & 0.73  & 0.90  & 0.00  & 0.12  & 22\%  & 0.0   & 3.0   & 0.79  & 0.54 \\
    Conditional mean regression & 5.85  & 0.05  & 3.81  & 1.47  & 1.04  & 1.51  & 4\%   & 0.0   & 3.0   & 0.21  & 0.88 \\
    MOM IPW & 6.16  & 0.06  & 4.48  & 1.30  & 0.82  & 1.80  & 10\%  & 0.0   & 3.0   & 0.55  & 2.45 \\
    MOM DR & 3.50  & 0.02  & 2.10  & 1.24  & 0.50  & 1.02  & 11\%  & 0.0   & 3.1   & 0.29  & 0.36 \\
    Causal Forest & 5.68  & 0.05  & 3.30  & 1.55  & 1.29  & 1.31  & 7\%   & 0.0   & 3.0   & 0.25  & 0.64 \\
    Causal Forest with local centering & 4.50  & 0.03  & 3.07  & 1.25  & 0.55  & 1.40  & 10\%  & 0.0   & 3.0   & 0.21  & 0.65 \\
    \textit{Lasso:} &       &       &       &       &       &       &       &       &       &       &  \\
    Infeasible & 1.00  & 0.00  & 0.50  & 0.77  & 0.01  & 0.23  & 17\%  & 0.0   & 3.1   & 0.82  & 0.66 \\
    Conditional mean regression & 8.41  & 0.08  & 7.55  & 1.20  & 0.73  & 2.41  & 33\%  & -0.1  & 3.3   & 0.31  & 2.46 \\
    MOM IPW & 6.94  & 0.16  & 5.92  & 1.01  & 0.67  & 2.21  & 98\%  & 0.1   & 6.1   & 0.49  & 2.51 \\
    MOM DR & \textbf{3.45}  & 0.03  & 1.95  & 1.36  & 0.47  & 0.72  & 99\%  & -0.3  & 9.3   & 0.24  & 0.14 \\
    MCM   & 6.42  & 0.15  & 5.00  & 0.96  & 0.39  & 2.09  & 100\% & -0.1  & 5.6   & 0.46  & 2.33 \\
    MCM with efficiency augmentation & \textbf{3.40} & 0.02  & 2.05  & 1.35  & 0.43  & 0.72  & 99\%  & -0.1  & 7.8   & 0.24  & 0.14 \\
    R-learning & 3.56  & 0.03  & 1.87  & 1.39  & 0.53  & 0.70  & 98\%  & -0.1  & 8.2   & 0.19  & 0.12 \\

    \bottomrule
    \end{tabular}%
         \begin{tablenotes} \item \textit{Notes:} Table shows the performance measures defined in Sections \ref{sec:performance} and \ref{sec:app-performance} over 2000 replications for the sample size of 1000 observations and 500 replications for the sample size of 4000 observations. Bold numbers indicate the best performing estimators in terms of $\overline{MSE}$ and estimators within two standard (simulation) errors of the lowest $\overline{MSE}$. \end{tablenotes}  
  \label{tab:app-ite1}%
\end{threeparttable}

\doublespacing

    \singlespacing

\begin{threeparttable}[t]
  \centering \footnotesize
\caption{Performance measure for ITE2 with selection and without random noise}
    \begin{tabular}{lccccccccccc}
    \toprule
          & $\overline{MSE}$   & $SE(\overline{MSE})$  & Median MSE & $|\overline{Bias}|$ & $\overline{Bias}$ & $\overline{SD}$    & $JB$    & Skew. & Kurt. & Corr. & Var. ratio \\
              \midrule
          & (1)   & (2)   & (3)   & (4)   & (5)   & (6)   & (7)   & (8)   & (9)   & (10)  & (11) \\
          \midrule
          & \multicolumn{11}{c}{\textbf{1000 observations}} \\
    \midrule
    Infeasible & 18.84 & 0.01  & 12.79 & 3.66  & -0.04 & 0.54  & 39\%  & 0.0   & 3.1   & 0.74  & 0.47 \\
    Conditional mean regression & 37.71 & 0.11  & 17.19 & 4.55  & 2.24  & 1.89  & 20\%  & -0.1  & 3.0   & 0.50  & 0.15 \\
    MOM IPW & \textbf{30.30} & 0.09  & 19.54 & 4.10  & 2.25  & 2.37  & 17\%  & 0.0   & 3.1   & 0.66  & 0.59 \\
    MOM DR & 37.78 & 0.07  & 12.38 & 4.61  & 1.64  & 1.36  & 29\%  & 0.0   & 3.1   & 0.47  & 0.06 \\
    Causal Forest & 43.79 & 0.12  & 16.98 & 4.94  & 2.85  & 1.77  & 35\%  & -0.1  & 3.1   & 0.41  & 0.10 \\
    Causal Forest with local centering & 37.05 & 0.09  & 14.05 & 4.46  & 1.59  & 1.90  & 36\%  & 0.0   & 3.1   & 0.45  & 0.12 \\
    \textit{Lasso:} &       &       &       &       &       &       &       &       &       &       &  \\
    Infeasible & 16.81 & 0.01  & 10.78 & 3.22  & -0.03 & 1.23  & 71\%  & 0.0   & 3.4   & 0.77  & 0.57 \\
    Conditional mean regression & 37.10 & 0.12  & 26.65 & 3.93  & 1.46  & 3.68  & 84\%  & -0.1  & 3.8   & 0.52  & 0.64 \\
    MOM IPW & 38.53 & 0.29  & 28.54 & 3.86  & 1.77  & 3.95  & 100\% & -0.1  & 10.4  & 0.55  & 0.71 \\
    MOM DR & 78.97 & 39.21 & 47.70 & 4.54  & 1.32  & 6.25  & 100\% & 37.1  & 1574.6 & 0.42  & 0.27 \\
    MCM   & 36.49 & 0.15  & 25.60 & 3.91  & 0.97  & 3.60  & 100\% & -0.1  & 5.7   & 0.50  & 0.53 \\
    MCM with efficiency augmentation & 38.48 & 0.09  & 14.41 & 4.55  & 1.30  & 1.91  & 100\% & -0.1  & 6.7   & 0.41  & 0.11 \\
    R-learning & 40.18 & 0.10  & 14.21 & 4.68  & 1.53  & 1.80  & 100\% & 0.0   & 7.7   & 0.38  & 0.09 \\
    \midrule
          & \multicolumn{11}{c}{\textbf{4000 observations}} \\
    \midrule
    \textit{Random Forest:} &       &       &       &       &       &       &       &       &       &       &  \\
    Infeasible & 16.68 & 0.01  & 11.34 & 3.43  & -0.04 & 0.45  & 15\%  & 0.0   & 3.0   & 0.78  & 0.52 \\
    Conditional mean regression & 27.75 & 0.09  & 16.09 & 4.08  & 1.62  & 1.58  & 6\%   & 0.0   & 3.0   & 0.64  & 0.31 \\
    MOM IPW & \textbf{24.08} & 0.07  & 14.71 & 3.69  & 1.47  & 1.92  & 10\%  & 0.0   & 3.0   & 0.70  & 0.50 \\
    MOM DR & 27.19 & 0.07  & 11.84 & 4.14  & 0.97  & 1.15  & 12\%  & 0.0   & 3.0   & 0.66  & 0.18 \\
    Causal Forest & 34.53 & 0.15  & 16.01 & 4.46  & 2.47  & 1.51  & 17\%  & -0.1  & 3.0   & 0.60  & 0.17 \\
    Causal Forest with local centering & 25.17 & 0.08  & 14.39 & 3.96  & 0.97  & 1.61  & 10\%  & 0.0   & 3.0   & 0.67  & 0.29 \\
    \textit{Lasso:} &       &       &       &       &       &       &       &       &       &       &  \\
    Infeasible & 14.49 & 0.01  & 8.35  & 3.00  & -0.03 & 0.87  & 16\%  & 0.0   & 3.1   & 0.81  & 0.63 \\
    Conditional mean regression & 24.47 & 0.07  & 17.85 & 3.48  & 0.84  & 2.58  & 19\%  & 0.0   & 3.1   & 0.68  & 0.69 \\
    MOM IPW & 25.33 & 0.12  & 18.22 & 3.47  & 1.18  & 2.71  & 76\%  & -0.1  & 4.2   & 0.68  & 0.68 \\
    MOM DR & 25.62 & 0.08  & 12.79 & 4.02  & 0.62  & 1.30  & 85\%  & -0.1  & 4.9   & 0.66  & 0.23 \\
    MCM   & 27.86 & 0.12  & 17.80 & 3.78  & 0.62  & 2.50  & 95\%  & -0.1  & 4.4   & 0.61  & 0.38 \\
    MCM with efficiency augmentation & 26.79 & 0.08  & 12.27 & 4.09  & 0.71  & 1.30  & 88\%  & 0.0   & 4.4   & 0.65  & 0.20 \\
    R-learning & 27.77 & 0.09  & 12.04 & 4.13  & 0.87  & 1.32  & 92\%  & 0.0   & 4.9   & 0.64  & 0.18 \\

    \bottomrule
    \end{tabular}%
         \begin{tablenotes} \item \textit{Notes:} Table shows the performance measures defined in Sections \ref{sec:performance} and \ref{sec:app-performance} over 2000 replications for the sample size of 1000 observations and 500 replications for the sample size of 4000 observations. Bold numbers indicate the best performing estimators in terms of $\overline{MSE}$ and estimators within two standard (simulation) errors of the lowest $\overline{MSE}$. \end{tablenotes}  
  \label{tab:app-ite2}%
\end{threeparttable}

\doublespacing

\end{landscape}
\restoregeometry 

\subsubsection{ITE with selection and random noise}   \label{sec:app-ite-rn}
\doublespacing
ITE0 with random noise (Table \ref{tab:app-ite0-rn}) shows the same pattern as the baseline ITE0 without noise, only that the levels of mean MSE and mean bias are higher. This is expected because this ITE0 consists mainly of irreducible noise.

Tables \ref{tab:app-ite1-rn} and  \ref{tab:app-ite2-rn} provide additional information for the baseline ITE1 and ITE2, respectively. The alternative performance measures confirm the results in the main text of ITE1. However, Random Forest conditional mean regression and Causal Forest show highly competitive median MSE for both sample sizes, which is in contrast to their mean MSE. 

\newgeometry{left=0.4in,right=0.5in,top=1in,bottom=1.2in,nohead}
\begin{landscape}
    \singlespacing

\begin{threeparttable}[t]
  \centering \footnotesize      
  \caption{Performance measure for ITE0 with selection and random noise}
    \begin{tabular}{lccccccccccc}
    \toprule
          & $\overline{MSE}$   & $SE(\overline{MSE})$  & Median MSE & $|\overline{Bias}|$ & $\overline{Bias}$ & $\overline{SD}$    & $JB$    & Skew. & Kurt. & Corr. & Var. ratio \\
          \midrule
          & (1)   & (2)   & (3)   & (4)   & (5)   & (6)   & (7)   & (8)   & (9)   & (10)  & (11) \\
          \midrule
          & \multicolumn{11}{c}{\textbf{1000 observations}} \\
    \midrule
    \textit{Random Forest:} &       &       &       &       &       &       &       &       &       &       &  \\
    Infeasible & 0.79  & 0.00  & 0.82  & 0.62  & 0.00  & 0.05  & 83\%  & -0.2  & 3.2   & 0.04  & 0.00 \\
    Conditional mean regression & 4.50  & 0.04  & 3.75  & 0.85  & 0.60  & 1.78  & 6\%   & 0.0   & 3.0   & -0.01 & 3.47 \\
    MOM IPW & 11.62 & 0.08  & 8.84  & 2.17  & 0.71  & 2.18  & 17\%  & 0.0   & 3.1   & -0.03 & 12.11 \\
    MOM DR & \textbf{2.74} & 0.02  & 2.11  & 0.75  & 0.39  & 1.33  & 8\%   & 0.0   & 3.0   & 0.00  & 1.65 \\
    Causal Forest & 4.30  & 0.04  & 3.47  & 0.90  & 0.74  & 1.69  & 13\%  & 0.0   & 3.1   & -0.01 & 2.96 \\
    Causal Forest with local centering & 4.16  & 0.03  & 3.59  & 0.73  & 0.33  & 1.80  & 11\%  & 0.0   & 3.0   & 0.00  & 3.34 \\
    \textit{Lasso:} &       &       &       &       &       &       &       &       &       &       &  \\
    Infeasible & 0.79  & 0.00  & 0.86  & 0.62  & 0.00  & 0.05  & 100\% & -1.3  & 21.3  & 0.02  & 0.00 \\
    Conditional mean regression & 11.85 & 0.11  & 10.88 & 0.90  & 0.60  & 3.17  & 90\%  & -0.1  & 4.0   & -0.01 & 12.78 \\
    MOM IPW & 12.20 & 0.30  & 10.45 & 1.25  & 0.60  & 2.99  & 100\% & 0.7   & 19.8  & 0.00  & 13.12 \\
    MOM DR & 1346.38 & 1343.32 & 24.83 & 0.87  & 0.63  & 7.92  & 100\% & 39.5  & 1703.1 & 0.00  & 1636.00 \\
    MCM   & 13.79 & 0.26  & 11.26 & 1.57  & 0.42  & 3.05  & 100\% & -0.1  & 7.4   & 0.00  & 15.41 \\
    MCM with efficiency augmentation & 2.83  & 0.06  & 2.17  & 0.76  & 0.42  & 1.34  & 100\% & 0.0   & 9.0   & 0.00  & 1.50 \\
    R-learning & 2.80  & 0.05  & 2.11  & 0.77  & 0.45  & 1.32  & 100\% & 0.0   & 9.7   & 0.00  & 1.45 \\
    \midrule
          & \multicolumn{11}{c}{\textbf{4000 observations}} \\
    \midrule
    \textit{Random Forest:} &       &       &       &       &       &       &       &       &       &       &  \\
    Infeasible & 0.78  & 0.00  & 0.78  & 0.62  & 0.00  & 0.04  & 30\%  & -0.2  & 3.1   & 0.06  & 0.00 \\
    Conditional mean regression & 3.58  & 0.03  & 2.86  & 0.86  & 0.53  & 1.49  & 4\%   & 0.0   & 3.0   & -0.01 & 2.97 \\
    MOM IPW & 6.71  & 0.06  & 4.89  & 1.50  & 0.53  & 1.76  & 12\%  & 0.0   & 3.1   & -0.02 & 6.83 \\
    MOM DR & 1.92  & 0.01  & 1.39  & 0.71  & 0.27  & 1.02  & 7\%   & 0.0   & 3.0   & 0.00  & 1.19 \\
    Causal Forest & 3.09  & 0.03  & 2.26  & 0.89  & 0.70  & 1.29  & 6\%   & 0.0   & 3.0   & -0.01 & 2.10 \\
    Causal Forest with local centering & 2.81  & 0.02  & 2.31  & 0.70  & 0.23  & 1.39  & 9\%   & 0.0   & 3.0   & 0.00  & 2.32 \\
    \textit{Lasso:} &       &       &       &       &       &       &       &       &       &       &  \\
    Infeasible & 0.78  & 0.00  & 0.82  & 0.62  & 0.00  & 0.04  & 100\% & -0.1  & 7.1   & 0.04  & 0.00 \\
    Conditional mean regression & 6.84  & 0.08  & 6.12  & 0.88  & 0.49  & 2.28  & 35\%  & 0.0   & 3.4   & -0.01 & 7.17 \\
    MOM IPW & 5.81  & 0.17  & 4.81  & 1.03  & 0.45  & 1.92  & 100\% & 0.2   & 7.8   & 0.00  & 5.91 \\
    MOM DR & 1.31  & 0.02  & 0.69  & 0.72  & 0.31  & 0.63  & 94\%  & -0.1  & 11.1  & 0.00  & 0.33 \\
    MCM   & 6.34  & 0.20  & 4.64  & 1.16  & 0.31  & 1.90  & 100\% & -0.1  & 6.4   & 0.00  & 6.75 \\
    MCM with efficiency augmentation & \textbf{1.26} & 0.02  & 0.73  & 0.71  & 0.26  & 0.62  & 96\%  & 0.0   & 8.3   & 0.00  & 0.31 \\
    R-learning & \textbf{1.26} & 0.02  & 0.72  & 0.71  & 0.28  & 0.62  & 96\%  & 0.1   & 8.1   & 0.00  & 0.30 \\

    \bottomrule
    \end{tabular}%
         \begin{tablenotes} \item \textit{Notes:} Table shows the performance measures defined in Sections \ref{sec:performance} and \ref{sec:app-performance} over 2000 replications for the sample size of 1000 observations and 500 replications for the sample size of 4000 observations. Bold numbers indicate the best performing estimators in terms of $\overline{MSE}$ and estimators within two standard (simulation) errors of the lowest $\overline{MSE}$. \end{tablenotes}  
  \label{tab:app-ite0-rn}%
\end{threeparttable}

\doublespacing
    \singlespacing

\begin{threeparttable}[t]
  \centering \footnotesize
  \caption{Performance measure for ITE1 with selection and random noise  (baseline)}
    \begin{tabular}{lccccccccccc}
    \toprule
          & $\overline{MSE}$   & $SE(\overline{MSE})$  & Median MSE & $|\overline{Bias}|$ & $\overline{Bias}$ & $\overline{SD}$    & $JB$    & Skew. & Kurt. & Corr. & Var. ratio \\
          \midrule
          & (1)   & (2)   & (3)   & (4)   & (5)   & (6)   & (7)   & (8)   & (9)   & (10)  & (11) \\
          \midrule
          & \multicolumn{11}{c}{\textbf{1000 observations}} \\
    \midrule
    \textit{Random Forest:} &       &       &       &       &       &       &       &       &       &       &  \\
    Infeasible & 2.98  & 0.00  & 1.03  & 1.29  & 0.01  & 0.15  & 71\%  & -0.2  & 3.2   & 0.23  & 0.03 \\
    Conditional mean regression & 7.04  & 0.04  & 4.58  & 1.45  & 0.80  & 1.78  & 8\%   & 0.0   & 3.0   & 0.02  & 0.88 \\
    MOM IPW & 12.92 & 0.08  & 8.65  & 2.26  & 0.87  & 2.20  & 16\%  & 0.0   & 3.1   & 0.18  & 3.34 \\
    MOM DR & \textbf{5.08} & 0.02  & 2.92  & 1.36  & 0.52  & 1.33  & 8\%   & 0.0   & 3.0   & 0.03  & 0.42 \\
    Causal Forest & 6.86  & 0.04  & 4.23  & 1.49  & 0.96  & 1.68  & 12\%  & 0.0   & 3.1   & 0.03  & 0.75 \\
    Causal Forest with local centering & 6.50  & 0.03  & 4.48  & 1.35  & 0.48  & 1.79  & 12\%  & 0.0   & 3.1   & 0.02  & 0.83 \\
    \textit{Lasso:} &       &       &       &       &       &       &       &       &       &       &  \\
    Infeasible & 3.00  & 0.00  & 1.08  & 1.28  & 0.01  & 0.21  & 100\% & -0.5  & 7.7   & 0.21  & 0.04 \\
    Conditional mean regression & 14.26 & 0.11  & 11.96 & 1.46  & 0.76  & 3.16  & 90\%  & -0.1  & 4.1   & 0.02  & 3.23 \\
    MOM IPW & 15.69 & 1.46  & 11.75 & 1.56  & 0.73  & 3.12  & 100\% & 0.1   & 30.3  & 0.12  & 3.99 \\
    MOM DR & 48.76 & 43.14 & 38.41 & 1.40  & 0.70  & 6.32  & 100\% & 40.9  & 1783.4 & 0.01  & 2.17 \\
    MCM   & 15.31 & 0.26  & 12.32 & 1.72  & 0.46  & 3.10  & 100\% & -0.1  & 7.1   & 0.14  & 4.10 \\
    MCM with efficiency augmentation & 5.27  & 0.06  & 3.14  & 1.37  & 0.52  & 1.36  & 100\% & -0.1  & 9.2   & 0.02  & 0.40 \\
    R-learning & 5.16  & 0.05  & 2.89  & 1.38  & 0.58  & 1.29  & 100\% & 0.0   & 9.5   & 0.01  & 0.34 \\
    \midrule
          & \multicolumn{11}{c}{\textbf{4000 observations}} \\
    \midrule
    \textit{Random Forest:} &       &       &       &       &       &       &       &       &       &       &  \\
    Infeasible & 2.93  & 0.00  & 1.27  & 1.30  & 0.01  & 0.11  & 26\%  & -0.1  & 3.1   & 0.25  & 0.06 \\
    Conditional mean regression & 6.05  & 0.04  & 3.63  & 1.44  & 0.73  & 1.49  & 4\%   & 0.0   & 3.0   & 0.02  & 0.76 \\
    MOM IPW & 8.42  & 0.06  & 5.38  & 1.76  & 0.63  & 1.77  & 11\%  & 0.0   & 3.0   & 0.15  & 1.92 \\
    MOM DR & 4.17  & 0.02  & 2.02  & 1.32  & 0.35  & 1.01  & 9\%   & 0.0   & 3.0   & 0.05  & 0.30 \\
    Causal Forest & 5.61  & 0.04  & 3.01  & 1.48  & 0.91  & 1.29  & 6\%   & 0.0   & 3.0   & 0.05  & 0.54 \\
    Causal Forest with local centering & 5.10  & 0.02  & 3.08  & 1.32  & 0.35  & 1.39  & 10\%  & 0.0   & 3.0   & 0.03  & 0.59 \\
    \textit{Lasso:} &       &       &       &       &       &       &       &       &       &       &  \\
    Infeasible & 2.93  & 0.00  & 1.08  & 1.28  & 0.01  & 0.16  & 83\%  & -0.1  & 4.0   & 0.25  & 0.06 \\
    Conditional mean regression & 9.20  & 0.08  & 7.16  & 1.43  & 0.60  & 2.30  & 36\%  & 0.0   & 3.4   & 0.04  & 1.87 \\
    MOM IPW & 8.03  & 0.17  & 5.81  & 1.46  & 0.53  & 2.01  & 100\% & 0.1   & 7.4   & 0.13  & 1.75 \\
    MOM DR & \textbf{3.66} & 0.02  & 1.31  & 1.34  & 0.37  & 0.64  & 96\%  & -0.3  & 10.1  & 0.03  & 0.09 \\
    MCM   & 8.14  & 0.19  & 5.44  & 1.51  & 0.32  & 1.96  & 100\% & -0.1  & 6.2   & 0.15  & 1.90 \\
    MCM with efficiency augmentation & \textbf{3.62} & 0.02  & 1.22  & 1.33  & 0.33  & 0.63  & 98\%  & 0.0   & 8.7   & 0.03  & 0.08 \\
    R-learning & \textbf{3.65} & 0.02  & 1.07  & 1.34  & 0.38  & 0.63  & 98\%  & 0.1   & 8.6   & 0.02  & 0.08 \\

    \bottomrule
    \end{tabular}%
         \begin{tablenotes} \item \textit{Notes:} Table shows the performance measures defined in Sections \ref{sec:performance} and \ref{sec:app-performance} over 2000 replications for the sample size of 1000 observations and 500 replications for the sample size of 4000 observations. Bold numbers indicate the best performing estimators in terms of $\overline{MSE}$ and estimators within two standard (simulation) errors of the lowest $\overline{MSE}$. \end{tablenotes}  
  \label{tab:app-ite1-rn}%
\end{threeparttable}

\doublespacing

    \singlespacing

\begin{threeparttable}[t]
  \centering \footnotesize
\caption{Performance measure for ITE2 with selection and random noise (baseline)}
    \begin{tabular}{lccccccccccc}
    \toprule
          & $\overline{MSE}$   & $SE(\overline{MSE})$  & Median MSE & $|\overline{Bias}|$ & $\overline{Bias}$ & $\overline{SD}$    & $JB$    & Skew. & Kurt. & Corr. & Var. ratio \\
              \midrule
          & (1)   & (2)   & (3)   & (4)   & (5)   & (6)   & (7)   & (8)   & (9)   & (10)  & (11) \\
          \midrule
          & \multicolumn{11}{c}{\textbf{1000 observations}} \\
    \midrule
    Infeasible & 38.46 & 0.01  & 8.96  & 4.43  & 0.09  & 0.52  & 66\%  & -0.2  & 3.2   & 0.19  & 0.02 \\
    Conditional mean regression & 43.74 & 0.05  & 12.15 & 4.58  & 1.24  & 1.74  & 8\%   & 0.0   & 3.0   & 0.04  & 0.07 \\
    MOM IPW & 46.83 & 0.08  & 19.41 & 4.83  & 1.23  & 2.23  & 16\%  & 0.0   & 3.1   & 0.17  & 0.31 \\
    MOM DR & \textbf{41.45} & 0.03  & 12.25 & 4.50  & 0.82  & 1.32  & 10\%  & 0.0   & 3.0   & 0.05  & 0.03 \\
    Causal Forest & 43.87 & 0.06  & 11.76 & 4.61  & 1.43  & 1.66  & 12\%  & 0.0   & 3.1   & 0.04  & 0.06 \\
    Causal Forest with local centering & 42.84 & 0.04  & 13.72 & 4.50  & 0.81  & 1.78  & 12\%  & 0.0   & 3.0   & 0.04  & 0.07 \\
    \textit{Lasso:} &       &       &       &       &       &       &       &       &       &       &  \\
    Infeasible & 38.66 & 0.01  & 8.89  & 4.42  & 0.08  & 0.71  & 100\% & -0.6  & 8.3   & 0.18  & 0.03 \\
    Conditional mean regression & 50.11 & 0.11  & 22.97 & 4.52  & 1.12  & 3.15  & 92\%  & -0.1  & 4.1   & 0.04  & 0.26 \\
    MOM IPW & 49.82 & 0.25  & 24.14 & 4.50  & 1.02  & 3.20  & 100\% & 0.5   & 13.9  & 0.11  & 0.33 \\
    MOM DR & 537.16 & 494.76 & 21.20 & 4.55  & 0.92  & 5.04  & 100\% & 31.9  & 1310.9 & 0.03  & 11.94 \\
    MCM   & 49.25 & 0.22  & 23.98 & 4.47  & 0.52  & 3.18  & 100\% & -0.1  & 6.8   & 0.13  & 0.35 \\
    MCM with efficiency augmentation & 41.99 & 0.06  & 12.63 & 4.51  & 0.79  & 1.41  & 100\% & 0.0   & 9.2   & 0.04  & 0.04 \\
    R-learning & 42.13 & 0.06  & 11.73 & 4.54  & 0.89  & 1.35  & 100\% & 0.1   & 10.1  & 0.03  & 0.03 \\
    \midrule
          & \multicolumn{11}{c}{\textbf{4000 observations}} \\
    \midrule
    \textit{Random Forest:} &       &       &       &       &       &       &       &       &       &       &  \\
    Infeasible & 37.86 & 0.00  & 6.30  & 4.43  & 0.07  & 0.41  & 31\%  & -0.2  & 3.1   & 0.22  & 0.04 \\
    Conditional mean regression & 42.26 & 0.05  & 11.94 & 4.54  & 1.12  & 1.46  & 5\%   & 0.0   & 3.0   & 0.06  & 0.06 \\
    MOM IPW & 42.69 & 0.06  & 13.91 & 4.54  & 0.88  & 1.80  & 11\%  & 0.0   & 3.0   & 0.16  & 0.18 \\
    MOM DR & \textbf{40.03} & 0.02  & 12.21 & 4.45  & 0.56  & 1.03  & 11\%  & 0.0   & 3.1   & 0.10  & 0.03 \\
    Causal Forest & 42.34 & 0.05  & 11.14 & 4.58  & 1.36  & 1.29  & 7\%   & 0.0   & 3.0   & 0.07  & 0.05 \\
    Causal Forest with local centering & 41.05 & 0.03  & 12.80 & 4.46  & 0.63  & 1.40  & 9\%   & 0.0   & 3.0   & 0.07  & 0.05 \\
    \textit{Lasso:} &       &       &       &       &       &       &       &       &       &       &  \\
    Infeasible & 37.84 & 0.00  & 6.68  & 4.40  & 0.07  & 0.53  & 84\%  & -0.1  & 4.2   & 0.22  & 0.04 \\
    Conditional mean regression & 44.31 & 0.07  & 16.52 & 4.46  & 0.82  & 2.33  & 34\%  & -0.1  & 3.3   & 0.10  & 0.17 \\
    MOM IPW & 43.21 & 0.15  & 15.02 & 4.43  & 0.74  & 2.17  & 97\%  & 0.1   & 5.8   & 0.14  & 0.19 \\
    MOM DR & 40.11 & 0.03  & 12.55 & 4.48  & 0.54  & 0.76  & 97\%  & -0.2  & 9.9   & 0.08  & 0.01 \\
    MCM   & 42.63 & 0.16  & 14.91 & 4.41  & 0.34  & 2.07  & 100\% & -0.1  & 5.6   & 0.14  & 0.18 \\
    MCM with efficiency augmentation & \textbf{40.04} & 0.03  & 12.45 & 4.47  & 0.51  & 0.75  & 99\%  & 0.0   & 7.5   & 0.08  & 0.01 \\
    R-learning & 40.25 & 0.03  & 11.78 & 4.49  & 0.63  & 0.74  & 98\%  & -0.1  & 7.9   & 0.07  & 0.01 \\

    \bottomrule
    \end{tabular}%
         \begin{tablenotes} \item \textit{Notes:} Table shows the performance measures defined in Sections \ref{sec:performance} and \ref{sec:app-performance} over 2000 replications for the sample size of 1000 observations and 500 replications for the sample size of 4000 observations. Bold numbers indicate the best performing estimators in terms of $\overline{MSE}$ and estimators within two standard (simulation) errors of the lowest $\overline{MSE}$. \end{tablenotes}  
  \label{tab:app-ite2-rn}%
\end{threeparttable}

\doublespacing

\end{landscape}
\restoregeometry 

\subsubsection{ITE with random assignment and without random noise}  \label{sec:app-ite-ra}
\doublespacing
Table \ref{tab:app-ite0-ra} provides the results for the baseline ITE0 with random assignment. The relative performance order remains unchanged. However, the striking difference is that the mean absolute bias and the mean bias are both close to zero for all estimators. This shows that the interpretation of the results for DGPs with selectivity in \ref{sec:app-ite-base} is correct and that the observed biases are driven by remaining selection bias.

The comparison of ITE1 with its selective equivalents shows that the remaining mean absolute bias is lower because of the absence of selection bias, which is also reflected in the close to zero mean bias. However, the mean absolute bias remains substantial, which suggest that approximation errors and the irreducible noise play a significant role. The relative performance of the estimators is similar for ITE1. This is not true for ITE2 where conditional mean regression based on Random Forests show the lowest mean MSE. This suggests that settings with large and informative ITEs favor conditional mean regression because the systematic part in the outcomes can be exploited. However, this is the least realistic setting and this locally good performance of conditional mean regression is not confirmed for ITE2 with random noise in the next section.

\newgeometry{left=0.4in,right=0.5in,top=1in,bottom=1.2in,nohead}
\begin{landscape}
    \singlespacing

\begin{threeparttable}[t]
  \centering \footnotesize      
  \caption{Performance measure for ITE0 with random assignment and without random noise}
    \begin{tabular}{lccccccccccc}
    \toprule
          & $\overline{MSE}$   & $SE(\overline{MSE})$  & Median MSE & $|\overline{Bias}|$ & $\overline{Bias}$ & $\overline{SD}$    & $JB$    & Skew. & Kurt. & Corr. & Var. ratio \\
          \midrule
          & (1)   & (2)   & (3)   & (4)   & (5)   & (6)   & (7)   & (8)   & (9)   & (10)  & (11) \\
          \midrule
          & \multicolumn{11}{c}{\textbf{1000 observations}} \\
    \midrule
     Infeasible & \multicolumn{11}{c}{No variation in dependent variable} \\
    Conditional mean regression & 3.25  & 0.03  & 3.15  & 0.03  & 0.01  & 1.80  & 6\%   & 0.0   & 3.0   & -     & - \\
    MOM IPW & 4.04  & 0.03  & 3.84  & 0.03  & 0.01  & 1.99  & 10\%  & 0.0   & 3.1   & -     & - \\
    MOM DR & 1.88  & 0.02  & 1.81  & 0.02  & 0.01  & 1.37  & 6\%   & 0.0   & 3.0   & -     & - \\
    Causal Forest & 2.92  & 0.03  & 2.80  & 0.02  & 0.01  & 1.70  & 10\%  & 0.0   & 3.1   & -     & - \\
    Causal Forest with local centering & 2.97  & 0.02  & 2.83  & 0.03  & 0.02  & 1.71  & 11\%  & 0.0   & 3.1   & -     & - \\
    \textit{Lasso:} &       &       &       &       &       &       &       &       &       &       &  \\
    Infeasible & \multicolumn{11}{c}{No variation in dependent variable} \\
    Conditional mean regression & 10.40 & 0.11  & 9.48  & 0.06  & 0.01  & 3.16  & 87\%  & 0.0   & 4.0   & -     & - \\
    MOM IPW & 3.83  & 0.14  & 3.29  & 0.03  & 0.01  & 1.90  & 100\% & -0.1  & 19.4  & -     & - \\
    MOM DR & 1.65  & 0.05  & 1.52  & 0.02  & 0.01  & 1.27  & 100\% & 0.0   & 9.4   & -     & - \\
    MCM   & 3.62  & 0.11  & 3.09  & 0.03  & 0.01  & 1.84  & 100\% & 0.1   & 15.4  & -     & - \\
    MCM with efficiency augmentation & \textbf{1.58}  & 0.04  & 1.43  & 0.02  & 0.02  & 1.24  & 100\% & 0.0   & 8.4   & -     & - \\
    R-learning & \textbf{1.55} & 0.04  & 1.42  & 0.02  & 0.02  & 1.23  & 100\% & 0.0   & 8.1   & -     & - \\\midrule
          & \multicolumn{11}{c}{\textbf{4000 observations}} \\
    \midrule
    \textit{Random Forest:} &       &       &       &       &       &       &       &       &       &       &  \\
    Infeasible & \multicolumn{11}{c}{No variation in dependent variable} \\
    Conditional mean regression & 2.29  & 0.03  & 2.20  & 0.06  & 0.02  & 1.50  & 6\%   & 0.0   & 3.0   & -     & - \\
    MOM IPW & 2.47  & 0.02  & 2.28  & 0.06  & 0.03  & 1.55  & 7\%   & 0.0   & 3.0   & -     & - \\
    MOM DR & 0.99  & 0.01  & 0.94  & 0.04  & 0.02  & 0.99  & 6\%   & 0.0   & 3.0   & -     & - \\
    Causal Forest & 1.68  & 0.01  & 1.59  & 0.05  & 0.03  & 1.29  & 7\%   & 0.0   & 3.0   & -     & - \\
    Causal Forest with local centering & 1.73  & 0.01  & 1.62  & 0.05  & 0.02  & 1.31  & 8\%   & 0.0   & 3.0   & -     & - \\
    \textit{Lasso:} &       &       &       &       &       &       &       &       &       &       &  \\
    Infeasible & \multicolumn{11}{c}{No variation in dependent variable} \\
    Conditional mean regression & 5.64  & 0.07  & 5.16  & 0.09  & 0.02  & 2.31  & 32\%  & 0.0   & 3.3   & -     & - \\
    MOM IPW & 0.80  & 0.05  & 0.67  & 0.04  & 0.03  & 0.87  & 100\% & 0.0   & 12.6  & -     & - \\
    MOM DR & \textbf{0.35}  & 0.02  & 0.32  & 0.03  & 0.02  & 0.59  & 32\%  & 0.0   & 3.3   & -     & - \\
    MCM   & 0.79  & 0.05  & 0.63  & 0.04  & 0.03  & 0.85  & 100\% & 0.0   & 13.2  & -     & - \\
    MCM with efficiency augmentation & \textbf{0.36} & 0.02  & 0.31  & 0.02  & 0.02  & 0.59  & 98\%  & 0.0   & 8.8   & -     & - \\
    R-learning & \textbf{0.36} & 0.02  & 0.31  & 0.02  & 0.02  & 0.59  & 99\%  & -0.1  & 9.1   & -     & - \\

    \bottomrule
    \end{tabular}%
         \begin{tablenotes} \item \textit{Notes:} Table shows the performance measures defined in Sections \ref{sec:performance} and \ref{sec:app-performance} over 2000 replications for the sample size of 1000 observations and 500 replications for the sample size of 4000 observations. Bold numbers indicate the best performing estimators in terms of $\overline{MSE}$ and estimators within two standard (simulation) errors of the lowest $\overline{MSE}$. \end{tablenotes}  
  \label{tab:app-ite0-ra}%
\end{threeparttable}

\doublespacing
    \singlespacing

\begin{threeparttable}[t]
  \centering \footnotesize
  \caption{Performance measure for ITE1 with random assignment and without random noise}
    \begin{tabular}{lccccccccccc}
    \toprule
          & $\overline{MSE}$   & $SE(\overline{MSE})$  & Median MSE & $|\overline{Bias}|$ & $\overline{Bias}$ & $\overline{SD}$    & $JB$    & Skew. & Kurt. & Corr. & Var. ratio \\
          \midrule
          & (1)   & (2)   & (3)   & (4)   & (5)   & (6)   & (7)   & (8)   & (9)   & (10)  & (11) \\
          \midrule
          & \multicolumn{11}{c}{\textbf{1000 observations}} \\
    \midrule
     \textit{Random Forest:} &       &       &       &       &       &       &       &       &       &       &  \\
    Infeasible & 1.34  & 0.00  & 0.92  & 0.98  & 0.00  & 0.14  & 59\%  & 0.0   & 3.1   & 0.76  & 0.48 \\
    Conditional mean regression & 5.54  & 0.03  & 4.64  & 1.21  & 0.05  & 1.81  & 6\%   & 0.0   & 3.0   & 0.18  & 0.93 \\
    MOM IPW & 6.56  & 0.03  & 5.88  & 1.25  & 0.05  & 2.02  & 10\%  & 0.0   & 3.1   & 0.12  & 1.16 \\
    MOM DR & \textbf{4.27}  & 0.02  & 3.08  & 1.25  & 0.05  & 1.35  & 6\%   & 0.0   & 3.0   & 0.18  & 0.45 \\
    Causal Forest & 5.24  & 0.03  & 4.41  & 1.22  & 0.05  & 1.71  & 11\%  & 0.0   & 3.1   & 0.17  & 0.79 \\
    Causal Forest with local centering & 5.23  & 0.02  & 4.40  & 1.22  & 0.06  & 1.71  & 12\%  & 0.0   & 3.1   & 0.17  & 0.81 \\
    \textit{Lasso:} &       &       &       &       &       &       &       &       &       &       &  \\
    Infeasible & 1.17  & 0.00  & 0.68  & 0.83  & 0.01  & 0.33  & 66\%  & 0.0   & 3.5   & 0.79  & 0.60 \\
    Conditional mean regression & 12.87 & 0.11  & 12.01 & 1.12  & 0.02  & 3.26  & 86\%  & 0.0   & 4.1   & 0.15  & 3.53 \\
    MOM IPW & 6.72  & 0.12  & 5.61  & 1.31  & 0.03  & 1.93  & 100\% & 0.0   & 14.8  & 0.06  & 1.06 \\
    MOM DR & 4.57  & 0.04  & 3.25  & 1.31  & 0.03  & 1.32  & 100\% & 0.0   & 9.0   & 0.11  & 0.38 \\
    MCM   & 6.64  & 0.12  & 5.52  & 1.34  & -0.02 & 1.85  & 100\% & 0.0   & 16.1  & 0.03  & 0.98 \\
    MCM with efficiency augmentation & 4.49  & 0.04  & 3.22  & 1.32  & 0.04  & 1.28  & 100\% & 0.0   & 8.0   & 0.11  & 0.36 \\
    R-learning & 4.52 & 0.04  & 3.29  & 1.32  & 0.04  & 1.29  & 100\% & -0.1  & 8.2   & 0.11  & 0.37 \\
    \midrule
          & \multicolumn{11}{c}{\textbf{4000 observations}} \\
    \midrule
    \textit{Random Forest:} &       &       &       &       &       &       &       &       &       &       &  \\
    Infeasible & 1.17  & 0.00  & 0.73  & 0.90  & 0.00  & 0.12  & 22\%  & 0.0   & 3.0   & 0.79  & 0.54 \\
    Conditional mean regression & 4.24  & 0.03  & 3.51  & 1.13  & 0.06  & 1.52  & 5\%   & 0.0   & 3.0   & 0.29  & 0.86 \\
    MOM IPW & 4.81  & 0.02  & 4.07  & 1.21  & 0.07  & 1.58  & 7\%   & 0.0   & 3.0   & 0.18  & 0.83 \\
    MOM DR & 3.20  & 0.01  & 2.18  & 1.21  & 0.05  & 0.99  & 7\%   & 0.0   & 3.0   & 0.29  & 0.34 \\
    Causal Forest & 3.78  & 0.02  & 2.96  & 1.17  & 0.06  & 1.31  & 7\%   & 0.0   & 3.0   & 0.27  & 0.60 \\
    Causal Forest with local centering & 3.78  & 0.02  & 2.99  & 1.16  & 0.06  & 1.32  & 9\%   & 0.0   & 3.0   & 0.27  & 0.61 \\
    \textit{Lasso:} &       &       &       &       &       &       &       &       &       &       &  \\
    Infeasible & 1.00  & 0.00  & 0.49  & 0.77  & 0.01  & 0.23  & 15\%  & 0.0   & 3.1   & 0.82  & 0.66 \\
    Conditional mean regression & 7.39  & 0.07  & 6.73  & 0.96  & 0.03  & 2.40  & 26\%  & 0.0   & 3.3   & 0.32  & 2.31 \\
    MOM IPW & 3.71  & 0.05  & 2.18  & 1.32  & 0.05  & 0.92  & 100\% & -0.2  & 11.9  & 0.12  & 0.25 \\
    MOM DR & \textbf{3.15}  & 0.02  & 1.79  & 1.29  & 0.04  & 0.70  & 100\% & -0.2  & 7.1   & 0.26  & 0.14 \\
    MCM   & 3.80  & 0.06  & 2.15  & 1.34  & -0.01 & 0.88  & 100\% & -0.1  & 14.1  & 0.06  & 0.23 \\
    MCM with efficiency augmentation & 3.20 & 0.02  & 1.81  & 1.30  & 0.04  & 0.69  & 100\% & -0.2  & 8.0   & 0.23  & 0.13 \\
    R-learning & \textbf{3.19}  & 0.02  & 1.81  & 1.30  & 0.04  & 0.69  & 99\%  & -0.2  & 7.9   & 0.24  & 0.13 \\

    \bottomrule
    \end{tabular}%
         \begin{tablenotes} \item \textit{Notes:} Table shows the performance measures defined in Sections \ref{sec:performance} and \ref{sec:app-performance} over 2000 replications for the sample size of 1000 observations and 500 replications for the sample size of 4000 observations. Bold numbers indicate the best performing estimators in terms of $\overline{MSE}$ and estimators within two standard (simulation) errors of the lowest $\overline{MSE}$. \end{tablenotes}  
  \label{tab:app-ite1-ra}%
\end{threeparttable}

\doublespacing

    \singlespacing

\begin{threeparttable}[t]
  \centering \footnotesize
\caption{Performance measure for ITE2 with random assignment and without random noise}
    \begin{tabular}{lccccccccccc}
    \toprule
          & $\overline{MSE}$   & $SE(\overline{MSE})$  & Median MSE & $|\overline{Bias}|$ & $\overline{Bias}$ & $\overline{SD}$    & $JB$    & Skew. & Kurt. & Corr. & Var. ratio \\
              \midrule
          & (1)   & (2)   & (3)   & (4)   & (5)   & (6)   & (7)   & (8)   & (9)   & (10)  & (11) \\
          \midrule
          & \multicolumn{11}{c}{\textbf{1000 observations}} \\
    \midrule
    \textit{Random Forest:} &       &       &       &       &       &       &       &       &       &       &  \\
    Infeasible & 18.84 & 0.01  & 12.79 & 3.66  & -0.04 & 0.54  & 39\%  & 0.0   & 3.1   & 0.74  & 0.47 \\
    Conditional mean regression & \textbf{26.71} & 0.05  & 15.28 & 3.93  & 0.04  & 1.92  & 14\%  & 0.0   & 3.0   & 0.62  & 0.29 \\
    MOM IPW & 36.35 & 0.06  & 17.07 & 4.42  & 0.09  & 2.06  & 15\%  & 0.0   & 3.1   & 0.38  & 0.13 \\
    MOM DR & 31.96 & 0.06  & 13.75 & 4.32  & 0.08  & 1.45  & 19\%  & 0.0   & 3.0   & 0.56  & 0.10 \\
    Causal Forest & 29.83 & 0.07  & 15.28 & 4.13  & 0.06  & 1.92  & 32\%  & 0.0   & 3.1   & 0.57  & 0.19 \\
    Causal Forest with local centering & 29.16 & 0.07  & 15.57 & 4.09  & 0.06  & 1.93  & 38\%  & 0.0   & 3.1   & 0.58  & 0.21 \\
    \textit{Lasso:} &       &       &       &       &       &       &       &       &       &       &  \\
    Infeasible & 16.81 & 0.01  & 10.77 & 3.22  & -0.03 & 1.23  & 72\%  & 0.0   & 3.4   & 0.77  & 0.57 \\
    Conditional mean regression & 31.81 & 0.10  & 25.66 & 3.64  & -0.02 & 3.54  & 78\%  & 0.0   & 3.6   & 0.57  & 0.66 \\
    MOM IPW & 41.74 & 0.12  & 20.93 & 4.61  & 0.03  & 2.37  & 100\% & -0.2  & 12.2  & 0.25  & 0.14 \\
    MOM DR & 33.63 & 0.18  & 16.60 & 4.30  & 0.02  & 2.00  & 100\% & -0.2  & 17.0  & 0.51  & 0.15 \\
    MCM   & 43.79 & 0.12  & 20.62 & 4.76  & -0.03 & 2.11  & 100\% & -0.1  & 11.6  & 0.13  & 0.10 \\
    MCM with efficiency augmentation & 34.08 & 0.08  & 16.63 & 4.34  & 0.02  & 1.92  & 100\% & 0.0   & 5.6   & 0.49  & 0.14 \\
    R-learning & 34.08 & 0.09  & 16.66 & 4.34  & 0.03  & 1.93  & 100\% & 0.0   & 7.0   & 0.49  & 0.14 \\
    \midrule
          & \multicolumn{11}{c}{\textbf{4000 observations}} \\
    \midrule
    \textit{Random Forest:} &       &       &       &       &       &       &       &       &       &       &  \\
    Infeasible & 16.68 & 0.01  & 11.34 & 3.43  & -0.04 & 0.45  & 15\%  & 0.0   & 3.0   & 0.78  & 0.52 \\
    Conditional mean regression & \textbf{21.70} & 0.03  & 13.64 & 3.65  & 0.01  & 1.57  & 8\%   & 0.0   & 3.0   & 0.70  & 0.51 \\
    MOM IPW & 29.57 & 0.09  & 13.96 & 4.15  & 0.08  & 1.69  & 8\%   & 0.0   & 3.0   & 0.57  & 0.17 \\
    MOM DR & 23.51 & 0.05  & 13.50 & 3.91  & 0.03  & 1.16  & 8\%   & 0.0   & 3.0   & 0.69  & 0.25 \\
    Causal Forest & 22.07 & 0.04  & 14.11 & 3.75  & 0.01  & 1.53  & 8\%   & 0.0   & 3.0   & 0.70  & 0.41 \\
    Causal Forest with local centering & \textbf{21.75} & 0.04  & 14.24 & 3.72  & 0.00  & 1.54  & 10\%  & 0.0   & 3.0   & 0.70  & 0.43 \\
    \textit{Lasso:} &       &       &       &       &       &       &       &       &       &       &  \\
    Infeasible & 14.49 & 0.01  & 8.33  & 3.00  & -0.03 & 0.86  & 15\%  & 0.0   & 3.1   & 0.81  & 0.63 \\
    Conditional mean regression & 22.63 & 0.06  & 16.83 & 3.30  & -0.02 & 2.56  & 23\%  & 0.0   & 3.2   & 0.70  & 0.70 \\
    MOM IPW & 30.93 & 0.14  & 15.01 & 4.22  & 0.02  & 1.70  & 99\%  & 0.0   & 5.4   & 0.56  & 0.15 \\
    MOM DR & 23.61 & 0.06  & 13.54 & 3.86  & -0.01 & 1.38  & 85\%  & 0.0   & 4.1   & 0.69  & 0.27 \\
    MCM   & 36.54 & 0.14  & 15.71 & 4.56  & -0.03 & 1.36  & 100\% & 0.0   & 7.7   & 0.43  & 0.06 \\
    MCM with efficiency augmentation & 24.21 & 0.06  & 13.76 & 3.91  & 0.00  & 1.36  & 89\%  & -0.1  & 4.4   & 0.68  & 0.26 \\
    R-learning & 24.21 & 0.06  & 13.76 & 3.91  & 0.00  & 1.36  & 89\%  & 0.0   & 4.4   & 0.68  & 0.26 \\

    \bottomrule
    \end{tabular}%
         \begin{tablenotes} \item \textit{Notes:} Table shows the performance measures defined in Sections \ref{sec:performance} and \ref{sec:app-performance} over 2000 replications for the sample size of 1000 observations and 500 replications for the sample size of 4000 observations. Bold numbers indicate the best performing estimators in terms of $\overline{MSE}$ and estimators within two standard (simulation) errors of the lowest $\overline{MSE}$. \end{tablenotes}  
  \label{tab:app-ite2-ra}%
\end{threeparttable}

\doublespacing

\end{landscape}
\restoregeometry 

\subsubsection{ITE with random assignment and random noise}   \label{sec:app-ite-rarn}
\doublespacing

The relative performances for the ITEs with noise but randomized treatment assignment are very close to their selective equivalents. The mean SDs in the selective and randomized settings are very similar. Also within the randomized setting, the mean absolute biases are nearly identical for all estimators. The differences between selective and randomized settings are thus only driven by different capabilities of the estimators to correct for selection bias.

\newgeometry{left=0.4in,right=0.5in,top=1in,bottom=1.2in,nohead}
\begin{landscape}
    \singlespacing

\begin{threeparttable}[t]
  \centering \footnotesize      
  \caption{Performance measure for ITE0 with random assignment and random noise}
    \begin{tabular}{lccccccccccc}
    \toprule
          & $\overline{MSE}$   & $SE(\overline{MSE})$  & Median MSE & $|\overline{Bias}|$ & $\overline{Bias}$ & $\overline{SD}$    & $JB$    & Skew. & Kurt. & Corr. & Var. ratio \\
          \midrule
          & (1)   & (2)   & (3)   & (4)   & (5)   & (6)   & (7)   & (8)   & (9)   & (10)  & (11) \\
          \midrule
          & \multicolumn{11}{c}{\textbf{1000 observations}} \\
    \midrule
    \textit{Random Forest:} &       &       &       &       &       &       &       &       &       &       &  \\
    Infeasible & 0.79  & 0.00  & 0.82  & 0.62  & 0.00  & 0.05  & 83\%  & -0.2  & 3.2   & 0.04  & 0.00 \\
    Conditional mean regression & 4.04  & 0.03  & 3.75  & 0.62  & 0.02  & 1.80  & 6\%   & 0.0   & 3.0   & 0.00  & 3.36 \\
    MOM IPW & 4.89  & 0.03  & 4.51  & 0.62  & 0.01  & 2.01  & 10\%  & 0.0   & 3.1   & 0.00  & 4.40 \\
    MOM DR & 2.64  & 0.02  & 2.43  & 0.62  & 0.01  & 1.36  & 6\%   & 0.0   & 3.0   & 0.00  & 1.67 \\
    Causal Forest & 3.71  & 0.03  & 3.43  & 0.62  & 0.01  & 1.70  & 9\%   & 0.0   & 3.1   & 0.00  & 2.90 \\
    Causal Forest with local centering & 3.71  & 0.02  & 3.41  & 0.62  & 0.01  & 1.70  & 10\%  & 0.0   & 3.1   & 0.00  & 2.97 \\
    \textit{Lasso:} &       &       &       &       &       &       &       &       &       &       &  \\
    Infeasible & 0.79  & 0.00  & 0.86  & 0.62  & 0.00  & 0.05  & 100\% & -1.3  & 21.4  & 0.02  & 0.00 \\
    Conditional mean regression & 11.07 & 0.10  & 10.13 & 0.63  & 0.01  & 3.14  & 87\%  & 0.0   & 4.0   & 0.00  & 12.32 \\
    MOM IPW & 4.54  & 0.12  & 3.92  & 0.62  & 0.01  & 1.89  & 100\% & 0.1   & 15.4  & 0.00  & 3.94 \\
    MOM DR & 2.45  & 0.05  & 2.17  & 0.62  & 0.02  & 1.27  & 100\% & 0.1   & 9.7   & 0.00  & 1.33 \\
    MCM   & 4.50  & 0.11  & 3.86  & 0.61  & -0.02 & 1.86  & 100\% & 0.1   & 14.3  & 0.00  & 3.88 \\
    MCM with efficiency augmentation & \textbf{2.35} & 0.04  & 2.07  & 0.62  & 0.02  & 1.23  & 100\% & 0.0   & 8.3   & 0.00  & 1.22 \\
    R-learning & \textbf{2.32} & 0.04  & 2.03  & 0.62  & 0.02  & 1.22  & 100\% & 0.1   & 8.1   & 0.00  & 1.18 \\
    \midrule
          & \multicolumn{11}{c}{\textbf{4000 observations}} \\
    \midrule
    \textit{Random Forest:} &       &       &       &       &       &       &       &       &       &       &  \\
    Infeasible & 0.78  & 0.00  & 0.78  & 0.62  & 0.00  & 0.04  & 30\%  & -0.2  & 3.1   & 0.06  & 0.00 \\
    Conditional mean regression & 3.08  & 0.03  & 2.76  & 0.63  & 0.03  & 1.50  & 5\%   & 0.0   & 3.0   & 0.00  & 2.76 \\
    MOM IPW & 3.30  & 0.02  & 2.91  & 0.63  & 0.03  & 1.57  & 7\%   & 0.0   & 3.0   & 0.00  & 3.03 \\
    MOM DR & 1.75  & 0.01  & 1.54  & 0.63  & 0.02  & 0.98  & 6\%   & 0.0   & 3.0   & 0.00  & 1.09 \\
    Causal Forest & 2.47  & 0.01  & 2.19  & 0.63  & 0.02  & 1.29  & 7\%   & 0.0   & 3.0   & 0.00  & 1.97 \\
    Causal Forest with local centering & 2.49  & 0.01  & 2.22  & 0.63  & 0.02  & 1.30  & 7\%   & 0.0   & 3.0   & 0.00  & 2.01 \\
    \textit{Lasso:} &       &       &       &       &       &       &       &       &       &       &  \\
    Infeasible & 0.78  & 0.00  & 0.82  & 0.62  & 0.00  & 0.04  & 100\% & -0.1  & 7.1   & 0.04  & 0.00 \\
    Conditional mean regression & 6.33  & 0.07  & 5.78  & 0.64  & 0.02  & 2.29  & 33\%  & 0.0   & 3.4   & 0.00  & 6.90 \\
    MOM IPW & 1.57  & 0.05  & 1.30  & 0.63  & 0.03  & 0.86  & 100\% & -0.1  & 13.2  & 0.00  & 0.81 \\
    MOM DR & \textbf{1.15} & 0.02  & 1.05  & 0.62  & 0.02  & 0.59  & 98\%  & 0.1   & 8.2   & 0.00  & 0.29 \\
    MCM   & 1.55  & 0.05  & 1.30  & 0.62  & 0.00  & 0.84  & 100\% & 0.0   & 14.2  & 0.00  & 0.79 \\
    MCM with efficiency augmentation & \textbf{1.13} & 0.02  & 1.03  & 0.62  & 0.02  & 0.58  & 97\%  & 0.0   & 9.0   & 0.00  & 0.27 \\
    R-learning & \textbf{1.13} & 0.02  & 1.03  & 0.62  & 0.02  & 0.57  & 96\%  & -0.1  & 8.7   & 0.00  & 0.27 \\

    \bottomrule
    \end{tabular}%
         \begin{tablenotes} \item \textit{Notes:} Table shows the performance measures defined in Sections \ref{sec:performance} and \ref{sec:app-performance} over 2000 replications for the sample size of 1000 observations and 500 replications for the sample size of 4000 observations. Bold numbers indicate the best performing estimators in terms of $\overline{MSE}$ and estimators within two standard (simulation) errors of the lowest $\overline{MSE}$. \end{tablenotes}  
  \label{tab:app-ite0-ra-rn}%
\end{threeparttable}

\doublespacing
    \singlespacing

\begin{threeparttable}[t]
  \centering \footnotesize
  \caption{Performance measure for ITE1 with random assignment and random noise}
    \begin{tabular}{lccccccccccc}
    \toprule
          & $\overline{MSE}$   & $SE(\overline{MSE})$  & Median MSE & $|\overline{Bias}|$ & $\overline{Bias}$ & $\overline{SD}$    & $JB$    & Skew. & Kurt. & Corr. & Var. ratio \\
          \midrule
          & (1)   & (2)   & (3)   & (4)   & (5)   & (6)   & (7)   & (8)   & (9)   & (10)  & (11) \\
          \midrule
          & \multicolumn{11}{c}{\textbf{1000 observations}} \\
    \midrule
    \textit{Random Forest:} &       &       &       &       &       &       &       &       &       &       &  \\
    Infeasible & 2.98  & 0.00  & 1.03  & 1.29  & 0.01  & 0.15  & 71\%  & -0.2  & 3.2   & 0.23  & 0.03 \\
    Conditional mean regression & 6.28  & 0.03  & 4.22  & 1.28  & 0.04  & 1.79  & 7\%   & 0.0   & 3.0   & 0.02  & 0.84 \\
    MOM IPW & 7.16  & 0.03  & 5.50  & 1.27  & 0.03  & 2.01  & 10\%  & 0.0   & 3.0   & 0.01  & 1.11 \\
    MOM DR & 4.88  & 0.02  & 2.76  & 1.27  & 0.03  & 1.35  & 6\%   & 0.0   & 3.0   & 0.02  & 0.42 \\
    Causal Forest & 5.94  & 0.03  & 3.91  & 1.27  & 0.03  & 1.70  & 10\%  & 0.0   & 3.1   & 0.02  & 0.73 \\
    Causal Forest with local centering & 5.93  & 0.02  & 3.92  & 1.27  & 0.03  & 1.70  & 12\%  & 0.0   & 3.1   & 0.02  & 0.74 \\
    \textit{Lasso:} &       &       &       &       &       &       &       &       &       &       &  \\
    Infeasible & 3.00  & 0.00  & 1.08  & 1.28  & 0.01  & 0.21  & 100\% & -0.5  & 7.7   & 0.21  & 0.04 \\
    Conditional mean regression & 13.26 & 0.10  & 11.73 & 1.28  & 0.03  & 3.14  & 87\%  & 0.0   & 4.0   & 0.02  & 3.10 \\
    MOM IPW & 6.99  & 0.13  & 5.34  & 1.28  & 0.02  & 1.92  & 100\% & -0.2  & 17.2  & 0.01  & 1.04 \\
    MOM DR & 4.78  & 0.05  & 2.74  & 1.28  & 0.03  & 1.28  & 100\% & 0.1   & 9.8   & 0.01  & 0.34 \\
    MCM   & 6.78  & 0.11  & 5.27  & 1.27  & -0.04 & 1.85  & 100\% & 0.1   & 15.1  & 0.00  & 0.96 \\
    MCM with efficiency augmentation & \textbf{4.69} & 0.04  & 2.67  & 1.28  & 0.03  & 1.24  & 100\% & 0.0   & 8.5   & 0.01  & 0.32 \\
    R-learning & \textbf{4.65} & 0.04  & 2.64  & 1.28  & 0.03  & 1.23  & 100\% & 0.0   & 8.4   & 0.01  & 0.31 \\
    \midrule
          & \multicolumn{11}{c}{\textbf{4000 observations}} \\
    \midrule
    \textit{Random Forest:} &       &       &       &       &       &       &       &       &       &       &  \\
    Infeasible & 2.93  & 0.00  & 1.27  & 1.30  & 0.01  & 0.11  & 26\%  & -0.1  & 3.1   & 0.25  & 0.06 \\
    Conditional mean regression & 5.31  & 0.03  & 3.33  & 1.28  & 0.05  & 1.50  & 5\%   & 0.0   & 3.0   & 0.03  & 0.70 \\
    MOM IPW & 5.57  & 0.02  & 3.74  & 1.28  & 0.04  & 1.57  & 6\%   & 0.0   & 3.0   & 0.02  & 0.77 \\
    MOM DR & 4.00  & 0.01  & 1.90  & 1.28  & 0.04  & 0.98  & 6\%   & 0.0   & 3.0   & 0.04  & 0.28 \\
    Causal Forest & 4.70  & 0.01  & 2.67  & 1.28  & 0.04  & 1.29  & 7\%   & 0.0   & 3.0   & 0.03  & 0.50 \\
    Causal Forest with local centering & 4.71  & 0.01  & 2.69  & 1.28  & 0.04  & 1.29  & 8\%   & 0.0   & 3.0   & 0.03  & 0.51 \\
    \textit{Lasso:} &       &       &       &       &       &       &       &       &       &       &  \\
    Infeasible & 2.93  & 0.00  & 1.08  & 1.28  & 0.01  & 0.16  & 82\%  & -0.1  & 4.0   & 0.25  & 0.06 \\
    Conditional mean regression & 8.67  & 0.08  & 7.30  & 1.28  & 0.03  & 2.33  & 32\%  & 0.0   & 3.3   & 0.04  & 1.83 \\
    MOM IPW & 3.89  & 0.05  & 1.86  & 1.28  & 0.04  & 0.86  & 100\% & 0.0   & 11.7  & 0.01  & 0.21 \\
    MOM DR & \textbf{3.46} & 0.02  & 1.41  & 1.28  & 0.04  & 0.59  & 98\%  & -0.1  & 8.8   & 0.02  & 0.07 \\
    MCM   & 3.92  & 0.05  & 1.86  & 1.27  & -0.02 & 0.86  & 100\% & 0.0   & 13.8  & 0.01  & 0.21 \\
    MCM with efficiency augmentation & \textbf{3.45} & 0.02  & 1.44  & 1.28  & 0.03  & 0.58  & 99\%  & 0.0   & 9.0   & 0.02  & 0.07 \\
    R-learning & \textbf{3.46} & 0.02  & 1.43  & 1.28  & 0.03  & 0.59  & 98\%  & -0.1  & 8.7   & 0.02  & 0.07 \\

    \bottomrule
    \end{tabular}%
         \begin{tablenotes} \item \textit{Notes:} Table shows the performance measures defined in Sections \ref{sec:performance} and \ref{sec:app-performance} over 2000 replications for the sample size of 1000 observations and 500 replications for the sample size of 4000 observations. Bold numbers indicate the best performing estimators in terms of $\overline{MSE}$ and estimators within two standard (simulation) errors of the lowest $\overline{MSE}$. \end{tablenotes}  
  \label{tab:app-ite1-ra-rn}%
\end{threeparttable}

\doublespacing

    \singlespacing

\begin{threeparttable}[t]
  \centering \footnotesize
\caption{Performance measure for ITE2 with random assignment and random noise}
    \begin{tabular}{lccccccccccc}
    \toprule
          & $\overline{MSE}$   & $SE(\overline{MSE})$  & Median MSE & $|\overline{Bias}|$ & $\overline{Bias}$ & $\overline{SD}$    & $JB$    & Skew. & Kurt. & Corr. & Var. ratio \\
              \midrule
          & (1)   & (2)   & (3)   & (4)   & (5)   & (6)   & (7)   & (8)   & (9)   & (10)  & (11) \\
          \midrule
          & \multicolumn{11}{c}{\textbf{1000 observations}} \\
    \midrule
    \textit{Random Forest:} &       &       &       &       &       &       &       &       &       &       &  \\
    Infeasible & 38.46 & 0.01  & 8.96  & 4.43  & 0.09  & 0.52  & 66\%  & -0.2  & 3.2   & 0.19  & 0.02 \\
    Conditional mean regression & 41.88 & 0.03  & 13.82 & 4.44  & 0.12  & 1.76  & 6\%   & 0.0   & 3.0   & 0.06  & 0.07 \\
    MOM IPW & 43.04 & 0.03  & 14.92 & 4.44  & 0.12  & 2.01  & 11\%  & 0.0   & 3.0   & 0.04  & 0.09 \\
    MOM DR & \textbf{40.76} & 0.02  & 12.86 & 4.44  & 0.10  & 1.35  & 5\%   & 0.0   & 3.0   & 0.06  & 0.04 \\
    Causal Forest & 41.67 & 0.03  & 13.85 & 4.44  & 0.11  & 1.69  & 9\%   & 0.0   & 3.0   & 0.06  & 0.06 \\
    Causal Forest with local centering & 41.68 & 0.02  & 13.84 & 4.44  & 0.12  & 1.70  & 10\%  & 0.0   & 3.0   & 0.06  & 0.06 \\
    \textit{Lasso:} &       &       &       &       &       &       &       &       &       &       &  \\
    Infeasible & 38.66 & 0.01  & 8.88  & 4.42  & 0.08  & 0.71  & 100\% & -0.6  & 8.3   & 0.18  & 0.03 \\
    Conditional mean regression & 48.39 & 0.10  & 21.71 & 4.43  & 0.11  & 3.13  & 89\%  & 0.0   & 3.9   & 0.05  & 0.26 \\
    MOM IPW & 43.34 & 0.13  & 16.17 & 4.46  & 0.10  & 1.94  & 100\% & -0.2  & 16.9  & 0.02  & 0.08 \\
    MOM DR & 41.11 & 0.04  & 13.50 & 4.46  & 0.11  & 1.32  & 100\% & 0.0   & 7.9   & 0.04  & 0.03 \\
    MCM   & 43.26 & 0.12  & 16.45 & 4.47  & -0.08 & 1.87  & 100\% & 0.0   & 14.3  & 0.01  & 0.08 \\
    MCM with efficiency augmentation & 41.09 & 0.04  & 13.51 & 4.46  & 0.11  & 1.30  & 100\% & 0.0   & 8.2   & 0.04  & 0.03 \\
    R-learning & 41.09 & 0.04  & 13.50 & 4.46  & 0.11  & 1.30  & 100\% & 0.0   & 8.2   & 0.04  & 0.03 \\
    \midrule
          & \multicolumn{11}{c}{\textbf{4000 observations}} \\
    \midrule
    \textit{Random Forest:} &       &       &       &       &       &       &       &       &       &       &  \\
    Infeasible & 37.86 & 0.00  & 6.30  & 4.43  & 0.07  & 0.41  & 31\%  & -0.2  & 3.1   & 0.22  & 0.04 \\
    Conditional mean regression & 40.51 & 0.03  & 11.74 & 4.43  & 0.13  & 1.47  & 5\%   & 0.0   & 3.0   & 0.10  & 0.07 \\
    MOM IPW & 41.23 & 0.02  & 13.48 & 4.43  & 0.12  & 1.57  & 6\%   & 0.0   & 3.0   & 0.06  & 0.06 \\
    MOM DR & \textbf{39.63} & 0.02  & 11.70 & 4.43  & 0.10  & 0.99  & 6\%   & 0.0   & 3.0   & 0.10  & 0.03 \\
    Causal Forest & 40.15 & 0.02  & 11.72 & 4.43  & 0.12  & 1.30  & 7\%   & 0.0   & 3.0   & 0.09  & 0.05 \\
    Causal Forest with local centering & 40.17 & 0.02  & 11.68 & 4.43  & 0.11  & 1.32  & 8\%   & 0.0   & 3.0   & 0.09  & 0.05 \\
    \textit{Lasso:} &       &       &       &       &       &       &       &       &       &       &  \\
    Infeasible & 37.84 & 0.00  & 6.69  & 4.40  & 0.07  & 0.53  & 84\%  & -0.1  & 4.2   & 0.22  & 0.04 \\
    Conditional mean regression & 43.27 & 0.07  & 14.64 & 4.42  & 0.09  & 2.33  & 27\%  & 0.0   & 3.3   & 0.11  & 0.18 \\
    MOM IPW & 40.27 & 0.05  & 12.33 & 4.46  & 0.12  & 0.94  & 100\% & -0.3  & 11.6  & 0.05  & 0.02 \\
    MOM DR & \textbf{39.67} & 0.02  & 11.83 & 4.45  & 0.11  & 0.73  & 99\%  & -0.1  & 6.8   & 0.09  & 0.01 \\
    MCM   & 40.38 & 0.05  & 11.46 & 4.47  & -0.06 & 0.87  & 100\% & -0.3  & 13.4  & 0.02  & 0.02 \\
    MCM with efficiency augmentation & 39.74 & 0.02  & 11.81 & 4.45  & 0.11  & 0.73  & 100\% & 0.0   & 7.8   & 0.08  & 0.01 \\
    R-learning & 39.73 & 0.02  & 11.86 & 4.45  & 0.11  & 0.73  & 100\% & 0.0   & 7.7   & 0.08  & 0.01 \\

    \bottomrule
    \end{tabular}%
         \begin{tablenotes} \item \textit{Notes:} Table shows the performance measures defined in Sections \ref{sec:performance} and \ref{sec:app-performance} over 2000 replications for the sample size of 1000 observations and 500 replications for the sample size of 4000 observations. Bold numbers indicate the best performing estimators in terms of $\overline{MSE}$ and estimators within two standard (simulation) errors of the lowest $\overline{MSE}$. \end{tablenotes}  
  \label{tab:app-ite2-ra-rn}%
\end{threeparttable}

\doublespacing

\end{landscape}
\restoregeometry 

\subsubsection{ITE without censoring}  \label{sec:app-ite-nc}
\doublespacing
This appendix shows the results for an alternative DGP that ignores the natural bounds of our outcome variable. It takes the following form similar to Equations \ref{small_omega} to \ref{xi}:

\begin{equation} \label{small_omega}
	\omega(x) = sin \left(1.25 \pi \dfrac{p^{HLM}(x)}{max(p^{HLM}(x))} \right),
\end{equation}

\begin{equation} \label{omega}
	\Omega(x) = \alpha \dfrac{\omega(x) - \bar{\omega}}{SD(\omega(x))} + \varepsilon_i,
\end{equation}

\begin{equation}
\xi_{nc}(x) =  \left\{ \begin{array}{l l}
    \lfloor \Omega(x)  & \text{if $  \Omega(x) - \lfloor \Omega(x) < u_i$}\\
     \Omega(x) \rceil & \text{if $  \Omega(x) - \lfloor\Omega(x) \geq u_i$}
  \end{array} \right.
\end{equation}

where $u_i$ is uniformly distributed between zero and one. It is similar to the baseline ITE1 without the censoring.

We run this robustness check for two purposes. First, to investigate whether our results are sensitive to such modifications of the DGP. Second, we know the true IATE in this formulation because $\xi_{nc}(x)$ does not depend on the non-treated outcome $Y_i^0$ via censoring.

Table \ref{tab:app-ite1_nc} shows the performance measure as in all tables above comparing the estimated IATEs to the \textit{true ITEs} in the validation sample. The only striking difference to the other results is the very high mean MSE of Lasso MOM IPW. This is driven by extreme outliers as indicated by the median MSE that is comparable to other methods.

Table \ref{tab:app-iate1_nc} shows the performance measures when comparing the estimated IATEs to the \textit{true IATEs}. The only differences compared to Table \ref{tab:app-ite1_nc} are the lower MSE measures that are driven by lower mean absolute biases. Although the level of these measures is changed, the ordering remains the same. This is expected because we get rid of the irreducible noise component that enters as bias if the true ITEs is considered as benchmark. This exercise illustrates that the performance measures based on the true ITEs as benchmark lead to the same conclusions as if we would know the true IATE.

\newgeometry{left=0.4in,right=0.5in,top=1in,bottom=1.2in,nohead}
\begin{landscape}
    \singlespacing

\begin{threeparttable}[t]
  \centering \footnotesize
  \caption{Performance measure for ITE1 without censoring}
    \begin{tabular}{lccccccccccc}
    \toprule
          & $\overline{MSE}$   & $SE(\overline{MSE})$  & Median MSE & $|\overline{Bias}|$ & $\overline{Bias}$ & $\overline{SD}$    & $JB$    & Skew. & Kurt. & Corr. & Var. ratio \\
          \midrule
          & (1)   & (2)   & (3)   & (4)   & (5)   & (6)   & (7)   & (8)   & (9)   & (10)  & (11) \\
          \midrule
          & \multicolumn{11}{c}{\textbf{1000 observations}} \\
    \midrule
    \textit{Random Forest:} &       &       &       &       &       &       &       &       &       &       &  \\
    Infeasible & 2.17  & 0.00  & 0.92  & 1.15  & 0.00  & 0.19  & 79\%  & -0.1  & 3.2   & 0.77  & 0.49 \\
    Conditional mean regression & 9.50  & 0.06  & 5.98  & 1.92  & 1.34  & 1.86  & 7\%   & 0.0   & 3.0   & 0.18  & 0.66 \\
    MOM IPW & 10.83 & 0.08  & 7.73  & 1.85  & 1.31  & 2.29  & 17\%  & 0.0   & 3.1   & 0.62  & 2.60 \\
    MOM DR & \textbf{6.75} & 0.03  & 4.28  & 1.78  & 0.88  & 1.35  & 12\%  & 0.0   & 3.0   & 0.22  & 0.29 \\
    Causal Forest & 9.57  & 0.06  & 5.79  & 2.00  & 1.55  & 1.75  & 14\%  & 0.0   & 3.1   & 0.19  & 0.54 \\
    Causal Forest with local centering & 8.21  & 0.04  & 5.89  & 1.76  & 0.84  & 1.84  & 15\%  & 0.0   & 3.1   & 0.16  & 0.56 \\
    \textit{Lasso:} &       &       &       &       &       &       &       &       &       &       &  \\
    Infeasible & 1.86  & 0.00  & 0.90  & 1.01  & 0.00  & 0.42  & 75\%  & -0.1  & 3.5   & 0.80  & 0.61 \\
    Conditional mean regression & 16.87 & 0.12  & 14.74 & 1.68  & 1.12  & 3.43  & 90\%  & -0.1  & 4.0   & 0.20  & 2.50 \\
    MOM IPW & 1048.99 & 1034.37 & 12.59 & 1.41  & 1.10  & 5.09  & 100\% & 1.0   & 41.5  & 0.41  & 200.92 \\
    MOM DR & 8.83  & 0.80  & 6.12  & 1.86  & 0.80  & 1.80  & 100\% & -8.4  & 223.4 & 0.13  & 0.38 \\
    MCM   & 13.67 & 0.21  & 11.87 & 1.27  & 0.68  & 3.17  & 100\% & -0.1  & 6.9   & 0.42  & 2.68 \\
    MCM with efficiency augmentation & 7.35  & 0.05  & 5.17  & 1.85  & 0.79  & 1.43  & 100\% & -0.2  & 7.8   & 0.15  & 0.28 \\
    R-learning & 7.59  & 0.06  & 5.12  & 1.89  & 0.90  & 1.39  & 100\% & -0.1  & 10.6  & 0.11  & 0.26 \\
    \midrule
          & \multicolumn{11}{c}{\textbf{4000 observations}} \\
    \midrule
    \textit{Random Forest:} &       &       &       &       &       &       &       &       &       &       &  \\
    Infeasible & 1.91  & 0.00  & 0.86  & 1.07  & 0.00  & 0.16  & 29\%  & -0.1  & 3.1   & 0.80  & 0.55 \\
    Conditional mean regression & 7.80  & 0.06  & 4.77  & 1.83  & 1.20  & 1.56  & 5\%   & 0.0   & 3.0   & 0.26  & 0.61 \\
    MOM IPW & 6.89  & 0.05  & 4.87  & 1.42  & 0.91  & 1.84  & 9\%   & 0.0   & 3.0   & 0.59  & 1.66 \\
    MOM DR & \textbf{5.12} & 0.03  & 3.28  & 1.63  & 0.58  & 1.05  & 15\%  & 0.0   & 3.1   & 0.35  & 0.25 \\
    Causal Forest & 7.77  & 0.06  & 4.26  & 1.91  & 1.47  & 1.36  & 8\%   & 0.0   & 3.0   & 0.29  & 0.44 \\
    Causal Forest with local centering & 6.14  & 0.03  & 4.24  & 1.62  & 0.64  & 1.44  & 11\%  & 0.0   & 3.1   & 0.28  & 0.44 \\
    \textit{Lasso:} &       &       &       &       &       &       &       &       &       &       &  \\
    Infeasible & 1.59  & 0.00  & 0.73  & 0.95  & 0.00  & 0.28  & 16\%  & 0.0   & 3.1   & 0.83  & 0.67 \\
    Conditional mean regression & 9.93  & 0.09  & 8.44  & 1.38  & 0.78  & 2.53  & 35\%  & -0.1  & 3.3   & 0.38  & 1.79 \\
    MOM IPW & 7.63  & 0.14  & 6.27  & 1.19  & 0.74  & 2.19  & 97\%  & 0.1   & 5.4   & 0.53  & 1.61 \\
    MOM DR & 5.34  & 0.04  & 3.08  & 1.76  & 0.51  & 0.85  & 99\%  & -0.4  & 9.3   & 0.31  & 0.14 \\
    MCM   & 7.34  & 0.14  & 5.80  & 1.20  & 0.49  & 2.09  & 100\% & -0.1  & 5.4   & 0.49  & 1.44 \\
    MCM with efficiency augmentation & 5.28  & 0.03  & 2.95  & 1.77  & 0.47  & 0.82  & 99\%  & -0.2  & 7.6   & 0.30  & 0.13 \\
    R-learning & 5.49  & 0.03  & 3.22  & 1.81  & 0.60  & 0.79  & 99\%  & -0.2  & 7.5   & 0.27  & 0.11 \\

    \bottomrule
    \end{tabular}%
         \begin{tablenotes} \item \textit{Notes:} Table shows the performance measures defined in Sections \ref{sec:performance} and \ref{sec:app-performance} over 2000 replications for the sample size of 1000 observations and 500 replications for the sample size of 4000 observations. Bold numbers indicate the best performing estimators in terms of $\overline{MSE}$ and estimators within two standard (simulation) errors of the lowest $\overline{MSE}$. \end{tablenotes}  
  \label{tab:app-ite1_nc}%
\end{threeparttable}

\doublespacing

    \singlespacing

\begin{threeparttable}[t]
  \centering \footnotesize
  \caption{Performance measure for ITE1 without censoring with true IATE as benchmark}
    \begin{tabular}{lccccccccccc}
    \toprule
          & $\overline{MSE}$   & $SE(\overline{MSE})$  & Median MSE & $|\overline{Bias}|$ & $\overline{Bias}$ & $\overline{SD}$    & $JB$    & Skew. & Kurt. & Corr. & Var. ratio \\
          \midrule
          & (1)   & (2)   & (3)   & (4)   & (5)   & (6)   & (7)   & (8)   & (9)   & (10)  & (11) \\
          \midrule
          & \multicolumn{11}{c}{\textbf{1000 observations}} \\
    \midrule
    \textit{Random Forest:} &       &       &       &       &       &       &       &       &       &       &  \\
    Infeasible & 0.98  & 0.00  & 0.48  & 0.77  & 0.00  & 0.19  & 79\%  & -0.1  & 3.2   & 0.77  & 0.49 \\
    Conditional mean regression & 8.31  & 0.06  & 5.26  & 1.69  & 1.34  & 1.86  & 7\%   & 0.0   & 3.0   & 0.18  & 0.66 \\
    MOM IPW & 9.64  & 0.08  & 7.19  & 1.66  & 1.31  & 2.29  & 17\%  & 0.0   & 3.1   & 0.62  & 2.60 \\
    MOM DR & \textbf{5.56} & 0.03  & 3.42  & 1.56  & 0.88  & 1.35  & 12\%  & 0.0   & 3.0   & 0.22  & 0.29 \\
    Causal Forest & 8.37  & 0.06  & 4.96  & 1.77  & 1.55  & 1.75  & 14\%  & 0.0   & 3.1   & 0.19  & 0.54 \\
    Causal Forest with local centering & 7.01  & 0.04  & 5.09  & 1.54  & 0.84  & 1.84  & 15\%  & 0.0   & 3.1   & 0.16  & 0.56 \\
    \textit{Lasso:} &       &       &       &       &       &       &       &       &       &       &  \\
    Infeasible & 0.68  & 0.00  & 0.41  & 0.53  & 0.00  & 0.42  & 75\%  & -0.1  & 3.5   & 0.80  & 0.61 \\
    Conditional mean regression & 15.68 & 0.12  & 14.08 & 1.44  & 1.12  & 3.43  & 90\%  & -0.1  & 4.0   & 0.20  & 2.50 \\
    MOM IPW & 1047.83 & 1034.38 & 11.76 & 1.19  & 1.10  & 5.09  & 100\% & 1.0   & 41.5  & 0.41  & 200.92 \\
    MOM DR & 7.64  & 0.80  & 4.77  & 1.68  & 0.80  & 1.80  & 100\% & -8.4  & 223.4 & 0.13  & 0.38 \\
    MCM   & 12.49 & 0.21  & 10.94 & 0.95  & 0.68  & 3.17  & 100\% & -0.1  & 6.9   & 0.42  & 2.68 \\
    MCM with efficiency augmentation & 6.16  & 0.05  & 3.91  & 1.66  & 0.79  & 1.43  & 100\% & -0.2  & 7.8   & 0.15  & 0.28 \\
    R-learning & 6.40  & 0.06  & 3.72  & 1.71  & 0.90  & 1.39  & 100\% & -0.1  & 10.6  & 0.11  & 0.26 \\
    \midrule
          & \multicolumn{11}{c}{\textbf{4000 observations}} \\
    \midrule
    \textit{Random Forest:} &       &       &       &       &       &       &       &       &       &       &  \\
    Infeasible & 0.72  & 0.00  & 0.33  & 0.64  & 0.00  & 0.16  & 29\%  & -0.1  & 3.1   & 0.80  & 0.55 \\
    Conditional mean regression & 6.60  & 0.06  & 4.18  & 1.59  & 1.20  & 1.56  & 5\%   & 0.0   & 3.0   & 0.27  & 0.61 \\
    MOM IPW & 5.72  & 0.05  & 4.34  & 1.15  & 0.91  & 1.84  & 9\%   & 0.0   & 3.0   & 0.59  & 1.66 \\
    MOM DR & \textbf{3.92} & 0.03  & 2.76  & 1.41  & 0.58  & 1.05  & 15\%  & 0.0   & 3.1   & 0.35  & 0.25 \\
    Causal Forest & 6.57  & 0.06  & 3.57  & 1.67  & 1.47  & 1.36  & 8\%   & 0.0   & 3.0   & 0.29  & 0.44 \\
    Causal Forest with local centering & 4.94  & 0.03  & 3.68  & 1.39  & 0.64  & 1.44  & 11\%  & 0.0   & 3.1   & 0.28  & 0.44 \\
    \textit{Lasso:} &       &       &       &       &       &       &       &       &       &       &  \\
    Infeasible & 0.42  & 0.00  & 0.20  & 0.39  & 0.00  & 0.28  & 16\%  & 0.0   & 3.1   & 0.83  & 0.67 \\
    Conditional mean regression & 8.75  & 0.09  & 7.84  & 1.10  & 0.78  & 2.53  & 35\%  & -0.1  & 3.3   & 0.38  & 1.79 \\
    MOM IPW & 6.47  & 0.14  & 5.63  & 0.88  & 0.74  & 2.19  & 97\%  & 0.1   & 5.4   & 0.52  & 1.61 \\
    MOM DR & 4.15  & 0.04  & 2.94  & 1.56  & 0.50  & 0.85  & 99\%  & -0.4  & 9.3   & 0.31  & 0.14 \\
    MCM   & 6.17  & 0.14  & 4.92  & 0.87  & 0.49  & 2.09  & 100\% & -0.1  & 5.4   & 0.49  & 1.44 \\
    MCM with efficiency augmentation & 4.09  & 0.03  & 3.00  & 1.57  & 0.47  & 0.82  & 99\%  & -0.2  & 7.6   & 0.30  & 0.13 \\
    R-learning & 4.31  & 0.03  & 2.87  & 1.61  & 0.60  & 0.79  & 99\%  & -0.2  & 7.5   & 0.27  & 0.11 \\

    \bottomrule
    \end{tabular}%
         \begin{tablenotes} \item \textit{Notes:} Table shows the performance measures defined in Sections \ref{sec:performance} and \ref{sec:app-performance} over 2000 replications for the sample size of 1000 observations and 500 replications for the sample size of 4000 observations. Bold numbers indicate the best performing estimators in terms of $\overline{MSE}$ and estimators within two standard (simulation) errors of the lowest $\overline{MSE}$. \end{tablenotes}  
  \label{tab:app-iate1_nc}%
\end{threeparttable}

\doublespacing

\end{landscape}
\restoregeometry 

\FloatBarrier

\subsection{Results for GATE estimation} \label{sec:app-gate}
\doublespacing
The Appendices \ref{sec:app-gate-base} to \ref{sec:app-gate-rarn} show the full results for GATE estimation in the 24 DGP-sample size combinations. The performance measures are the same as for IATE estimation. However, we omit the infeasible benchmark because it is not clear how it should be constructed for GATEs.

\subsubsection{GATEs from ITE with selection and without random noise} \label{sec:app-gate-base}

The additional performance measures for the baseline ITE0 do not change the conclusions in the main text. The mean bias shows that the substantial positive biases remain due to selection bias. This is in line with the IATE results and shows that averaging the IATEs does not remove the selection bias.

Also the GATE estimation of ITE1 and ITE2 without random noise shows patterns that are already observed for their baseline equivalents with noise. We observe that some estimators that show low mean absolute bias but high mean SD for the IATEs become competitive by averaging out the noise. In particular, Random Forest MOM IPW shows very low mean MSE for ITE1 with 4,000 observations and ITE2 with 1,000 observations. Similarly, Lasso mean regression is locally very successful for ITE2 in the 4,000 observations sample. While such competitive performances are usually not consistent for mean MSE and median MSE in the case of IATE estimation, they are confirmed for GATE estimation. However, we observe no systematic pattern that explains under which circumstances which noisy IATE estimator provides a good GATE estimator. Thus, the four consistently well performing IATE estimators seem to be also the dominant choice for the GATE estimation.

\newgeometry{left=0.4in,right=0.5in,top=1in,bottom=1.2in,nohead}
\begin{landscape}
    \singlespacing

\begin{threeparttable}[t]
  \centering \footnotesize      
  \caption{Performance measures for GATE of ITE0 with selection and without random noise  (baseline)}
    \begin{tabular}{lccccccccccc}
    \toprule
          & $\overline{MSE}$   & $SE(\overline{MSE})$  & Median MSE & $|\overline{Bias}|$ & $\overline{Bias}$ & $\overline{SD}$    & $JB$    & Skew. & Kurt. & Corr. & Var. ratio \\
          \midrule
          & (1)   & (2)   & (3)   & (4)   & (5)   & (6)   & (7)   & (8)   & (9)   & (10)  & (11) \\
          \midrule
          & \multicolumn{11}{c}{\textbf{1000 observations}} \\
    \midrule
    \textit{Random Forest:} &       &       &       &       &       &       &       &       &       &       &  \\
    Conditional mean regression & 1.77  & 0.03  & 1.32  & 0.55  & 0.55  & 1.19  & 22\%  & 0.0   & 3.1   & -     & - \\
    MOM IPW & 4.44  & 0.05  & 2.12  & 1.59  & 0.05  & 1.16  & 47\%  & -0.1  & 3.2   & -     & - \\
    MOM DR & \textbf{0.87} & 0.02  & 0.93  & 0.38  & 0.38  & 0.85  & 20\%  & 0.0   & 3.1   & -     & - \\
    Causal Forest & 1.44  & 0.03  & 1.22  & 0.70  & 0.70  & 0.96  & 17\%  & 0.0   & 3.1   & -     & - \\
    Causal Forest with local centering & 1.08  & 0.02  & 1.03  & 0.33  & 0.33  & 0.99  & 8\%   & 0.0   & 3.0   & -     & - \\
    \textit{Lasso:} &       &       &       &       &       &       &       &       &       &       &  \\
    Conditional mean regression & 3.35  & 0.04  & 1.77  & 0.55  & 0.54  & 1.70  & 34\%  & 0.0   & 3.1   & -     & - \\
    MOM IPW & 3.15  & 0.07  & 1.72  & 0.78  & 0.32  & 1.50  & 100\% & -0.1  & 4.6   & -     & - \\
    MOM DR & 38.85 & 37.73 & 6.13  & 0.59  & 0.59  & 6.20  & 100\% & 43.1  & 1901.5 & -     & - \\
    MCM   & 4.56  & 0.10  & 1.93  & 1.20  & 0.03  & 1.62  & 100\% & -0.3  & 4.0   & -     & - \\
    MCM with efficiency augmentation & 1.04  & 0.03  & 1.02  & 0.41  & 0.41  & 0.93  & 66\%  & -0.1  & 3.5   & -     & - \\
    R-learning & 1.04  & 0.03  & 1.01  & 0.44  & 0.44  & 0.92  & 73\%  & -0.1  & 3.7   & -     & - \\
    \midrule
          & \multicolumn{11}{c}{\textbf{4000 observations}} \\
    \midrule
    \textit{Random Forest:} &       &       &       &       &       &       &       &       &       &       &  \\
    Conditional mean regression & 1.07  & 0.03  & 1.01  & 0.49  & 0.47  & 0.86  & 8\%   & 0.0   & 3.1   & -     & - \\
    MOM IPW & 1.12  & 0.02  & 0.92  & 0.67  & 0.21  & 0.66  & 6\%   & 0.0   & 3.0   & -     & - \\
    MOM DR & 0.30  & 0.01  & 0.55  & 0.25  & 0.25  & 0.48  & 14\%  & 0.0   & 3.1   & -     & - \\
    Causal Forest & 0.74  & 0.03  & 0.86  & 0.64  & 0.64  & 0.53  & 0\%   & 0.0   & 2.9   & -     & - \\
    Causal Forest with local centering & 0.35  & 0.01  & 0.59  & 0.22  & 0.22  & 0.54  & 3\%   & 0.0   & 3.0   & -     & - \\
    \textit{Lasso:} &       &       &       &       &       &       &       &       &       &       &  \\
    Conditional mean regression & 1.45  & 0.03  & 1.16  & 0.47  & 0.42  & 1.06  & 6\%   & 0.0   & 3.1   & -     & - \\
    MOM IPW & 1.19  & 0.04  & 1.06  & 0.53  & 0.26  & 0.87  & 86\%  & -0.1  & 3.7   & -     & - \\
    MOM DR & 0.30  & 0.02  & 0.54  & 0.31  & 0.31  & 0.45  & 38\%  & 0.0   & 3.4   & -     & - \\
    MCM   & 1.65  & 0.06  & 1.14  & 0.75  & 0.07  & 0.92  & 94\%  & -0.1  & 3.8   & -     & - \\
    MCM with efficiency augmentation & \textbf{0.27} & 0.01  & 0.51  & 0.26  & 0.26  & 0.45  & 55\%  & -0.1  & 3.7   & -     & - \\
    R-learning & \textbf{0.27} & 0.02  & 0.52  & 0.28  & 0.28  & 0.44  & 25\%  & 0.1   & 3.4   & -     & - \\
    \bottomrule
    \end{tabular}%
         \begin{tablenotes} \item \textit{Notes:} Table shows the performance measures defined in Sections \ref{sec:performance} and \ref{sec:app-performance} over 2000 replications for the sample size of 1000 observations and 500 replications for the sample size of 4000 observations. Bold numbers indicate the best performing estimators in terms of $\overline{MSE}$ and estimators within two standard (simulation) errors of the lowest $\overline{MSE}$. \end{tablenotes}  
  \label{tab:app-gate0-base}%
\end{threeparttable}

\doublespacing
    \singlespacing

\begin{threeparttable}[t]
  \centering \footnotesize
  \caption{Performance measures for GATE of ITE1 with selection and without random noise}
    \begin{tabular}{lccccccccccc}
    \toprule
          & $\overline{MSE}$   & $SE(\overline{MSE})$  & Median MSE & $|\overline{Bias}|$ & $\overline{Bias}$ & $\overline{SD}$    & $JB$    & Skew. & Kurt. & Corr. & Var. ratio \\
          \midrule
          & (1)   & (2)   & (3)   & (4)   & (5)   & (6)   & (7)   & (8)   & (9)   & (10)  & (11) \\
          \midrule
          & \multicolumn{11}{c}{\textbf{1000 observations}} \\
    \midrule
    \textit{Random Forest:} &       &       &       &       &       &       &       &       &       &       &  \\
    Conditional mean regression & 4.39  & 0.06  & 1.84  & 1.41  & 1.40  & 1.20  & 22\%  & 0.0   & 3.1   & -0.11 & 0.33 \\
    MOM IPW & 3.19  & 0.04  & 1.55  & 1.10  & 0.72  & 1.18  & 44\%  & -0.1  & 3.2   & 0.56  & 1.85 \\
    MOM DR & \textbf{2.66} & 0.04  & 1.30  & 1.09  & 1.04  & 0.83  & 22\%  & -0.1  & 3.1   & -0.06 & 0.09 \\
    Causal Forest & 4.40  & 0.06  & 1.85  & 1.62  & 1.62  & 0.96  & 20\%  & 0.0   & 3.1   & -0.22 & 0.14 \\
    Causal Forest with local centering & 2.87  & 0.04  & 1.36  & 1.08  & 1.01  & 0.98  & 14\%  & 0.0   & 3.0   & -0.03 & 0.12 \\
    \textit{Lasso:} &       &       &       &       &       &       &       &       &       &       &  \\
    Conditional mean regression & 5.11  & 0.06  & 2.24  & 1.17  & 1.15  & 1.76  & 42\%  & -0.1  & 3.1   & 0.07  & 0.95 \\
    MOM IPW & 3.50  & 0.06  & 1.83  & 0.89  & 0.89  & 1.60  & 100\% & -0.1  & 4.2   & 0.32  & 1.15 \\
    MOM DR & 37.54 & 34.31 & 6.07  & 1.26  & 1.21  & 5.91  & 100\% & 42.8  & 1887.4 & -1.16 & 0.09 \\
    MCM   & 3.43  & 0.07  & 1.71  & 0.58  & 0.45  & 1.68  & 100\% & -0.3  & 3.9   & 0.43  & 1.52 \\
    MCM with efficiency augmentation & 3.06  & 0.04  & 1.42  & 1.16  & 1.03  & 0.95  & 78\%  & -0.1  & 3.8   & -1.42 & 0.08 \\
    R-learning & 3.34  & 0.05  & 1.47  & 1.23  & 1.14  & 0.93  & 78\%  & -0.1  & 3.8   & -1.62 & 0.06 \\
    \midrule
          & \multicolumn{11}{c}{\textbf{4000 observations}} \\
    \midrule
    \textit{Random Forest:} &       &       &       &       &       &       &       &       &       &       &  \\
    Conditional mean regression & 3.15  & 0.05  & 1.46  & 1.26  & 1.23  & 0.88  & 6\%   & 0.0   & 3.1   & 0.03  & 0.31 \\
    MOM IPW & \textbf{1.09}  & 0.03  & 0.93  & 0.69  & 0.69  & 0.68  & 16\%  & 0.0   & 3.1   & 0.52  & 0.65 \\
    MOM DR & 1.47  & 0.03  & 0.90  & 0.87  & 0.75  & 0.48  & 25\%  & 0.0   & 3.2   & 0.26  & 0.08 \\
    Causal Forest & 3.27  & 0.06  & 1.62  & 1.51  & 1.51  & 0.54  & 3\%   & 0.0   & 3.0   & -0.05 & 0.12 \\
    Causal Forest with local centering & 1.67  & 0.03  & 0.96  & 0.91  & 0.80  & 0.55  & 8\%   & 0.0   & 3.1   & 0.19  & 0.09 \\
    \textit{Lasso:} &       &       &       &       &       &       &       &       &       &       &  \\
    Conditional mean regression & 2.08  & 0.04  & 1.38  & 0.77  & 0.76  & 1.11  & 8\%   & 0.0   & 3.1   & 0.34  & 0.67 \\
    MOM IPW & 1.43  & 0.04  & 1.20  & 0.65  & 0.65  & 0.96  & 64\%  & 0.0   & 3.5   & 0.46  & 0.74 \\
    MOM DR & 1.90  & 0.04  & 0.89  & 1.03  & 0.79  & 0.50  & 69\%  & -0.1  & 3.5   & -1.41 & 0.04 \\
    MCM   & 1.28 & 0.04  & 1.00  & 0.42  & 0.40  & 0.97  & 67\%  & -0.1  & 3.4   & 0.47  & 0.72 \\
    MCM with efficiency augmentation & 1.81  & 0.04  & 0.89  & 1.01  & 0.75  & 0.49  & 69\%  & -0.1  & 3.4   & -0.77 & 0.04 \\
    R-learning & 2.09  & 0.04  & 0.94  & 1.09  & 0.87  & 0.48  & 66\%  & -0.1  & 3.8   & -1.29 & 0.03 \\
    \bottomrule
    \end{tabular}%
         \begin{tablenotes} \item \textit{Notes:} Table shows the performance measures defined in Sections \ref{sec:performance} and \ref{sec:app-performance} over 2000 replications for the sample size of 1000 observations and 500 replications for the sample size of 4000 observations. Bold numbers indicate the best performing estimators in terms of $\overline{MSE}$ and estimators within two standard (simulation) errors of the lowest $\overline{MSE}$. \end{tablenotes}  
  \label{tab:app-gate1-base}%
\end{threeparttable}

\doublespacing

    \singlespacing

\begin{threeparttable}[t]
  \centering \footnotesize
\caption{Performance measures for GATE of ITE2 with selection and without random noise}
    \begin{tabular}{lccccccccccc}
    \toprule
          & $\overline{MSE}$   & $SE(\overline{MSE})$  & Median MSE & $|\overline{Bias}|$ & $\overline{Bias}$ & $\overline{SD}$    & $JB$    & Skew. & Kurt. & Corr. & Var. ratio \\
          \midrule
          & (1)   & (2)   & (3)   & (4)   & (5)   & (6)   & (7)   & (8)   & (9)   & (10)  & (11) \\
          \midrule
          & \multicolumn{11}{c}{\textbf{1000 observations}} \\
    \midrule
    \textit{Random Forest:} &       &       &       &       &       &       &       &       &       &       &  \\
    Conditional mean regression & 19.52 & 0.14  & 3.50  & 3.30  & 3.05  & 1.29  & 28\%  & 0.0   & 3.1   & 0.28  & 0.08 \\
    MOM IPW & \textbf{8.38} & 0.09  & 2.54  & 2.30  & 2.30  & 1.24  & 45\%  & -0.1  & 3.1   & 0.56  & 0.34 \\
    MOM DR & 19.64 & 0.09  & 2.71  & 3.44  & 2.68  & 0.85  & 41\%  & -0.1  & 3.1   & 0.24  & 0.03 \\
    Causal Forest & 27.12 & 0.15  & 3.87  & 3.97  & 3.84  & 1.02  & 44\%  & -0.1  & 3.2   & 0.00  & 0.04 \\
    Causal Forest with local centering & 17.59 & 0.12  & 2.75  & 3.20  & 2.54  & 1.06  & 47\%  & -0.1  & 3.2   & 0.32  & 0.05 \\
    \textit{Lasso:} &       &       &       &       &       &       &       &       &       &       &  \\
    Conditional mean regression & 9.28  & 0.10  & 2.51  & 1.88  & 1.77  & 1.88  & 30\%  & 0.0   & 3.1   & 0.52  & 0.28 \\
    MOM IPW & 9.81  & 0.12  & 2.77  & 2.14  & 2.11  & 1.84  & 97\%  & -0.1  & 3.7   & 0.51  & 0.29 \\
    MOM DR & 51.94 & 34.59 & 6.64  & 3.25  & 2.35  & 5.97  & 100\% & 41.5  & 1810.7 & 0.11  & 0.05 \\
    MCM   & 9.29  & 0.12  & 2.35  & 1.95  & 1.53  & 1.80  & 94\%  & -0.2  & 3.6   & 0.52  & 0.21 \\
    MCM with efficiency augmentation & 17.87 & 0.13  & 2.93  & 3.27  & 2.34  & 1.18  & 94\%  & -0.2  & 3.6   & 0.12  & 0.04 \\
    R-learning & 20.45 & 0.14  & 2.86  & 3.50  & 2.63  & 1.14  & 84\%  & -0.2  & 3.6   & -0.20 & 0.03 \\
    \midrule
          & \multicolumn{11}{c}{\textbf{4000 observations}} \\
    \midrule
    \textit{Random Forest:} &       &       &       &       &       &       &       &       &       &       &  \\
    Conditional mean regression & 8.44  & 0.12  & 2.91  & 2.26  & 2.10  & 0.94  & 19\%  & 0.0   & 3.1   & 0.55  & 0.20 \\
    MOM IPW & 5.23  & 0.08  & 1.62  & 1.78  & 1.68  & 0.73  & 17\%  & -0.1  & 3.1   & 0.60  & 0.25 \\
    MOM DR & 7.77  & 0.09  & 1.92  & 2.18  & 1.60  & 0.58  & 2\%   & 0.0   & 3.0   & 0.59  & 0.13 \\
    Causal Forest & 16.66 & 0.19  & 3.38  & 3.21  & 3.16  & 0.72  & 16\%  & 0.0   & 2.9   & 0.40  & 0.10 \\
    Causal Forest with local centering & 5.36  & 0.09  & 1.55  & 1.76  & 1.41  & 0.71  & 3\%   & 0.1   & 3.0   & 0.62  & 0.20 \\
    \textit{Lasso:} &       &       &       &       &       &       &       &       &       &       &  \\
    Conditional mean regression & \textbf{2.64} & 0.05  & 1.48  & 0.96  & 0.89  & 1.11  & 5\%   & 0.0   & 3.0   & 0.65  & 0.40 \\
    MOM IPW & 3.51  & 0.07  & 1.72  & 1.35  & 1.33  & 1.05  & 25\%  & 0.0   & 3.2   & 0.63  & 0.36 \\
    MOM DR & 5.50  & 0.09  & 1.81  & 1.85  & 1.19  & 0.71  & 27\%  & 0.0   & 3.2   & 0.64  & 0.16 \\
    MCM   & 6.41  & 0.13  & 1.88  & 1.85  & 1.21  & 1.08  & 48\%  & -0.1  & 3.4   & 0.61  & 0.17 \\
    MCM with efficiency augmentation & 6.60  & 0.09  & 2.02  & 2.05  & 1.35  & 0.69  & 36\%  & 0.0   & 3.2   & 0.62  & 0.14 \\
    R-learning & 7.86  & 0.11  & 2.17  & 2.23  & 1.55  & 0.72  & 41\%  & 0.1   & 3.2   & 0.60  & 0.12 \\
    \bottomrule
    \end{tabular}%
         \begin{tablenotes} \item \textit{Notes:} Table shows the performance measures defined in Sections \ref{sec:performance} and \ref{sec:app-performance} over 2000 replications for the sample size of 1000 observations and 500 replications for the sample size of 4000 observations. Bold numbers indicate the best performing estimators in terms of $\overline{MSE}$ and estimators within two standard (simulation) errors of the lowest $\overline{MSE}$. \end{tablenotes}  
  \label{tab:app-gate2-base}%
\end{threeparttable}

\doublespacing

\end{landscape}
\restoregeometry 
\doublespacing

\subsubsection{GATEs from ITE with selection and random noise}  \label{sec:app-gate-rn}

The GATE results for ITE0 with random noise are very similar to the baseline without noise. Besides Lasso MOM DR that performs even worse, all estimators show very similar performance with differences only at the second digit of most performance measures. This shows how the additional noise is averaged out on the group level.

The additional performance measures for the baseline ITE1 and ITE2 confirm the results that are discussed in the main text.

\newgeometry{left=0.4in,right=0.5in,top=1in,bottom=1.2in,nohead}
\begin{landscape}
    \singlespacing

\begin{threeparttable}[t]
  \centering \footnotesize      
  \caption{Performance measures for GATE of ITE0 with selection and random noise}
    \begin{tabular}{lccccccccccc}
    \toprule
          & $\overline{MSE}$   & $SE(\overline{MSE})$  & Median MSE & $|\overline{Bias}|$ & $\overline{Bias}$ & $\overline{SD}$    & $JB$    & Skew. & Kurt. & Corr. & Var. ratio \\
          \midrule
          & (1)   & (2)   & (3)   & (4)   & (5)   & (6)   & (7)   & (8)   & (9)   & (10)  & (11) \\
          \midrule
          & \multicolumn{11}{c}{\textbf{1000 observations}} \\
    \midrule
    \textit{Random Forest:} &       &       &       &       &       &       &       &       &       &       &  \\
    Conditional mean regression & 1.75  & 0.03  & 1.30  & 0.53  & 0.53  & 1.18  & 25\%  & 0.0   & 3.1   & 0.01  & 1.09 \\
    MOM IPW & 4.57  & 0.05  & 2.09  & 1.61  & 0.02  & 1.16  & 41\%  & -0.1  & 3.2   & -0.01 & 4.95 \\
    MOM DR & \textbf{0.84} & 0.02  & 0.92  & 0.35  & 0.35  & 0.84  & 20\%  & 0.0   & 3.1   & 0.02  & 0.26 \\
    Causal Forest & 1.41  & 0.03  & 1.18  & 0.68  & 0.68  & 0.95  & 16\%  & 0.0   & 3.1   & 0.03  & 0.40 \\
    Causal Forest with local centering & 1.06  & 0.02  & 1.02  & 0.30  & 0.30  & 0.98  & 20\%  & 0.0   & 3.1   & 0.02  & 0.41 \\
    \textit{Lasso:} &       &       &       &       &       &       &       &       &       &       &  \\
    Conditional mean regression & 3.28  & 0.04  & 1.74  & 0.54  & 0.52  & 1.69  & 33\%  & 0.0   & 3.1   & 0.01  & 3.00 \\
    MOM IPW & 3.14  & 0.07  & 1.71  & 0.80  & 0.29  & 1.49  & 100\% & -0.1  & 4.5   & 0.02  & 2.90 \\
    MOM DR & 72.02 & 70.92 & 6.06  & 0.60  & 0.60  & 7.64  & 100\% & 43.0  & 1900.7 & 0.13  & 17.68 \\
    MCM   & 4.55  & 0.09  & 1.91  & 1.20  & -0.03 & 1.61  & 98\%  & -0.3  & 3.9   & 0.01  & 4.74 \\
    MCM with efficiency augmentation & 1.03  & 0.03  & 1.01  & 0.39  & 0.39  & 0.93  & 75\%  & 0.0   & 3.6   & 0.09  & 0.22 \\
    R-learning & 1.03  & 0.03  & 1.01  & 0.42  & 0.42  & 0.92  & 73\%  & 0.0   & 3.7   & 0.11  & 0.21 \\
    \midrule
          & \multicolumn{11}{c}{\textbf{4000 observations}} \\
    \midrule
    \textit{Random Forest:} &       &       &       &       &       &       &       &       &       &       &  \\
    Conditional mean regression & 1.06  & 0.03  & 1.01  & 0.48  & 0.45  & 0.86  & 6\%   & 0.0   & 3.0   & 0.01  & 0.88 \\
    MOM IPW & 1.16  & 0.02  & 0.96  & 0.69  & 0.19  & 0.66  & 6\%   & 0.0   & 3.0   & -0.01 & 1.17 \\
    MOM DR & 0.30 & 0.01  & 0.54  & 0.24  & 0.23  & 0.48  & 13\%  & 0.0   & 3.0   & 0.03  & 0.12 \\
    Causal Forest & 0.72  & 0.03  & 0.85  & 0.62  & 0.62  & 0.53  & 2\%   & 0.1   & 2.9   & 0.03  & 0.21 \\
    Causal Forest with local centering & 0.35  & 0.01  & 0.59  & 0.21  & 0.20  & 0.54  & 5\%   & 0.1   & 3.0   & 0.02  & 0.17 \\
    \textit{Lasso:} &       &       &       &       &       &       &       &       &       &       &  \\
    Conditional mean regression & 1.43  & 0.03  & 1.15  & 0.45  & 0.41  & 1.05  & 14\%  & 0.0   & 3.1   & 0.01  & 1.39 \\
    MOM IPW & 1.20  & 0.04  & 1.06  & 0.54  & 0.24  & 0.87  & 91\%  & 0.0   & 3.9   & 0.02  & 1.23 \\
    MOM DR & 0.30  & 0.02  & 0.54  & 0.29  & 0.29  & 0.45  & 42\%  & 0.1   & 3.5   & 0.20  & 0.05 \\
    MCM   & 1.60  & 0.06  & 1.10  & 0.75  & 0.03  & 0.90  & 89\%  & -0.2  & 3.6   & 0.02  & 1.78 \\
    MCM with efficiency augmentation & \textbf{0.27} & 0.01  & 0.51  & 0.24  & 0.24  & 0.45  & 28\%  & 0.0   & 3.4   & 0.14  & 0.05 \\
    R-learning & \textbf{0.28} & 0.01  & 0.52  & 0.26  & 0.26  & 0.44  & 14\%  & 0.1   & 3.2   & 0.19  & 0.04 \\
    \bottomrule

    \end{tabular}%
         \begin{tablenotes} \item \textit{Notes:} Table shows the performance measures defined in Sections \ref{sec:performance} and \ref{sec:app-performance} over 2000 replications for the sample size of 1000 observations and 500 replications for the sample size of 4000 observations. Bold numbers indicate the best performing estimators in terms of $\overline{MSE}$ and estimators within two standard (simulation) errors of the lowest $\overline{MSE}$. \end{tablenotes}  
  \label{tab:app-gate0-rn}%
\end{threeparttable}

\doublespacing
    \singlespacing

\begin{threeparttable}[t]
  \centering \footnotesize
  \caption{Performance measures for GATE of ITE1 with selection and random noise (baseline)}
    \begin{tabular}{lccccccccccc}
    \toprule
          & $\overline{MSE}$   & $SE(\overline{MSE})$  & Median MSE & $|\overline{Bias}|$ & $\overline{Bias}$ & $\overline{SD}$    & $JB$    & Skew. & Kurt. & Corr. & Var. ratio \\
          \midrule
          & (1)   & (2)   & (3)   & (4)   & (5)   & (6)   & (7)   & (8)   & (9)   & (10)  & (11) \\
          \midrule
          & \multicolumn{11}{c}{\textbf{1000 observations}} \\
    \midrule
    \textit{Random Forest:} &       &       &       &       &       &       &       &       &       &       &  \\
    Conditional mean regression & 2.30  & 0.04  & 1.41  & 0.85  & 0.84  & 1.18  & 20\%  & 0.0   & 3.1   & -0.04 & 0.28 \\
    MOM IPW & 3.84  & 0.04  & 1.89  & 1.41  & 0.26  & 1.17  & 41\%  & -0.1  & 3.1   & 0.20  & 1.43 \\
    MOM DR & \textbf{1.16} & 0.02  & 1.01  & 0.59  & 0.59  & 0.83  & 20\%  & 0.0   & 3.1   & -0.03 & 0.07 \\
    Causal Forest & 2.04  & 0.04  & 1.32  & 1.01  & 1.01  & 0.95  & 19\%  & 0.0   & 3.1   & -0.06 & 0.11 \\
    Causal Forest with local centering & 1.38  & 0.03  & 1.11  & 0.56  & 0.56  & 0.97  & 17\%  & 0.0   & 3.1   & -0.02 & 0.10 \\
    \textit{Lasso:} &       &       &       &       &       &       &       &       &       &       &  \\
    Conditional mean regression & 3.68  & 0.05  & 1.92  & 0.78  & 0.76  & 1.69  & 41\%  & -0.1  & 3.1   & -0.01 & 0.77 \\
    MOM IPW & 3.03  & 0.06  & 1.70  & 0.65  & 0.51  & 1.53  & 100\% & -0.2  & 5.3   & 0.09  & 0.84 \\
    MOM DR & 39.33 & 37.80 & 6.18  & 0.79  & 0.79  & 6.20  & 100\% & 43.0  & 1899.5 & -0.30 & 0.08 \\
    MCM   & 3.98  & 0.09  & 1.79  & 0.93  & 0.11  & 1.65  & 97\%  & -0.2  & 4.1   & 0.15  & 1.32 \\
    MCM with efficiency augmentation & 1.40  & 0.03  & 1.11  & 0.62  & 0.61  & 0.93  & 80\%  & -0.1  & 3.7   & -0.32 & 0.06 \\
    R-learning & 1.45  & 0.03  & 1.13  & 0.68  & 0.67  & 0.91  & 75\%  & 0.0   & 3.6   & -0.31 & 0.05 \\
    \midrule
          & \multicolumn{11}{c}{\textbf{4000 observations}} \\
    \midrule
    \textit{Random Forest:} &       &       &       &       &       &       &       &       &       &       &  \\
    Conditional mean regression & 1.53  & 0.04  & 1.11  & 0.76  & 0.74  & 0.86  & 6\%   & 0.0   & 3.1   & -0.02 & 0.23 \\
    MOM IPW & 0.99  & 0.02  & 0.81  & 0.54  & 0.37  & 0.67  & 11\%  & 0.0   & 3.0   & 0.17  & 0.39 \\
    MOM DR & \textbf{0.49} & 0.02  & 0.61  & 0.42  & 0.41  & 0.48  & 17\%  & 0.0   & 3.1   & 0.02  & 0.04 \\
    Causal Forest & 1.26  & 0.04  & 1.00  & 0.93  & 0.93  & 0.53  & 2\%   & 0.0   & 3.0   & -0.03 & 0.06 \\
    Causal Forest with local centering & 0.58  & 0.02  & 0.67  & 0.44  & 0.42  & 0.54  & 5\%   & 0.1   & 3.0   & 0.01  & 0.05 \\
    \textit{Lasso:} &       &       &       &       &       &       &       &       &       &       &  \\
    Conditional mean regression & 1.66  & 0.03  & 1.27  & 0.60  & 0.57  & 1.06  & 14\%  & 0.0   & 3.1   & 0.03  & 0.39 \\
    MOM IPW & 1.17  & 0.04  & 1.04  & 0.47  & 0.38  & 0.89  & 77\%  & 0.0   & 3.7   & 0.13  & 0.41 \\
    MOM DR & 0.59  & 0.02  & 0.61  & 0.50  & 0.47  & 0.46  & 42\%  & 0.0   & 3.5   & -0.37 & 0.02 \\
    MCM   & 1.31  & 0.05  & 1.03  & 0.52  & 0.13  & 0.93  & 91\%  & -0.1  & 3.6   & 0.16  & 0.53 \\
    MCM with efficiency augmentation & 0.55  & 0.02  & 0.58  & 0.47  & 0.42  & 0.45  & 39\%  & 0.0   & 3.3   & -0.39 & 0.01 \\
    R-learning & 0.61  & 0.02  & 0.61  & 0.51  & 0.48  & 0.45  & 48\%  & 0.0   & 3.4   & -0.60 & 0.01 \\
    \bottomrule
    \end{tabular}%
         \begin{tablenotes} \item \textit{Notes:} Table shows the performance measures defined in Sections \ref{sec:performance} and \ref{sec:app-performance} over 2000 replications for the sample size of 1000 observations and 500 replications for the sample size of 4000 observations. Bold numbers indicate the best performing estimators in terms of $\overline{MSE}$ and estimators within two standard (simulation) errors of the lowest $\overline{MSE}$. \end{tablenotes}  
  \label{tab:app-gate1-rn}%
\end{threeparttable}

\doublespacing

    \singlespacing

\begin{threeparttable}[t]
  \centering \footnotesize
\caption{Performance measures for GATE of ITE2 with selection and random noise (baseline)}
    \begin{tabular}{lccccccccccc}
    \toprule
          & $\overline{MSE}$   & $SE(\overline{MSE})$  & Median MSE & $|\overline{Bias}|$ & $\overline{Bias}$ & $\overline{SD}$    & $JB$    & Skew. & Kurt. & Corr. & Var. ratio \\
          \midrule
          & (1)   & (2)   & (3)   & (4)   & (5)   & (6)   & (7)   & (8)   & (9)   & (10)  & (11) \\
          \midrule
          & \multicolumn{11}{c}{\textbf{1000 observations}} \\
    \midrule
    \textit{Random Forest:} &       &       &       &       &       &       &       &       &       &       &  \\
    Conditional mean regression & 5.28  & 0.06  & 1.70  & 1.57  & 1.51  & 1.15  & 20\%  & 0.0   & 3.1   & -0.05 & 0.02 \\
    MOM IPW & 3.75  & 0.04  & 1.61  & 1.25  & 0.77  & 1.18  & 41\%  & -0.1  & 3.1   & 0.16  & 0.15 \\
    MOM DR & \textbf{3.50} & 0.04  & 1.32  & 1.29  & 1.10  & 0.84  & 23\%  & -0.1  & 3.1   & -0.02 & 0.01 \\
    Causal Forest & 5.33  & 0.06  & 1.65  & 1.71  & 1.69  & 0.94  & 25\%  & 0.0   & 3.1   & -0.07 & 0.01 \\
    Causal Forest with local centering & 3.74  & 0.05  & 1.40  & 1.28  & 1.10  & 0.98  & 11\%  & 0.0   & 3.0   & -0.01 & 0.01 \\
    \textit{Lasso:} &       &       &       &       &       &       &       &       &       &       &  \\
    Conditional mean regression & 5.73  & 0.06  & 2.15  & 1.34  & 1.29  & 1.71  & 42\%  & -0.1  & 3.1   & 0.02  & 0.07 \\
    MOM IPW & 4.00  & 0.06  & 1.94  & 1.01  & 0.95  & 1.59  & 100\% & -0.1  & 4.1   & 0.09  & 0.09 \\
    MOM DR & 30.36 & 26.20 & 4.28  & 1.43  & 1.24  & 4.71  & 100\% & 40.2  & 1737.8 & -0.47 & 0.13 \\
    MCM   & 3.65  & 0.06  & 1.84  & 0.66  & 0.37  & 1.67  & 100\% & -0.3  & 4.0   & 0.13  & 0.12 \\
    MCM with efficiency augmentation & 3.94  & 0.05  & 1.42  & 1.35  & 1.10  & 0.95  & 72\%  & -0.1  & 3.7   & -0.39 & 0.01 \\
    R-learning & 4.27  & 0.05  & 1.46  & 1.43  & 1.22  & 0.93  & 72\%  & -0.1  & 3.7   & -0.45 & 0.01 \\
    \midrule
          & \multicolumn{11}{c}{\textbf{4000 observations}} \\
    \midrule
    \textit{Random Forest:} &       &       &       &       &       &       &       &       &       &       &  \\
    Conditional mean regression & 3.97  & 0.06  & 1.45  & 1.44  & 1.36  & 0.85  & 22\%  & 0.0   & 3.2   & 0.01  & 0.02 \\
    MOM IPW & 1.72  & 0.03  & 1.14  & 0.95  & 0.76  & 0.68  & 14\%  & -0.1  & 3.0   & 0.15  & 0.05 \\
    MOM DR & 2.19  & 0.03  & 0.95  & 1.08  & 0.81  & 0.49  & 23\%  & 0.0   & 3.1   & 0.08  & 0.01 \\
    Causal Forest & 4.17  & 0.06  & 1.47  & 1.60  & 1.59  & 0.53  & 8\%   & 0.0   & 3.0   & -0.02 & 0.01 \\
    Causal Forest with local centering & 2.43  & 0.04  & 1.00  & 1.12  & 0.90  & 0.55  & 9\%   & 0.0   & 3.0   & 0.06  & 0.01 \\
    \textit{Lasso:} &       &       &       &       &       &       &       &       &       &       &  \\
    Conditional mean regression & 2.59  & 0.05  & 1.39  & 0.95  & 0.89  & 1.09  & 11\%  & 0.0   & 3.0   & 0.11  & 0.05 \\
    MOM IPW & 1.92  & 0.04  & 1.33  & 0.82  & 0.71  & 0.96  & 45\%  & 0.1   & 3.4   & 0.14  & 0.06 \\
    MOM DR & 2.71  & 0.04  & 1.02  & 1.23  & 0.85  & 0.52  & 69\%  & -0.1  & 3.6   & -0.33 & 0.00 \\
    MCM   & \textbf{1.58} & 0.04  & 1.15  & 0.61  & 0.33  & 0.97  & 81\%  & -0.1  & 3.6   & 0.16  & 0.06 \\
    MCM with efficiency augmentation & 2.63  & 0.04  & 1.04  & 1.22  & 0.81  & 0.51  & 67\%  & -0.1  & 3.4   & -0.24 & 0.00 \\
    R-learning & 2.95  & 0.04  & 1.07  & 1.29  & 0.94  & 0.50  & 50\%  & -0.1  & 3.5   & -0.33 & 0.00 \\
    \bottomrule
    \end{tabular}%
         \begin{tablenotes} \item \textit{Notes:} Table shows the performance measures defined in Sections \ref{sec:performance} and \ref{sec:app-performance} over 2000 replications for the sample size of 1000 observations and 500 replications for the sample size of 4000 observations. Bold numbers indicate the best performing estimators in terms of $\overline{MSE}$ and estimators within two standard (simulation) errors of the lowest $\overline{MSE}$. \end{tablenotes}  
  \label{tab:app-gate2-rn}%
\end{threeparttable}

\doublespacing

\end{landscape}
\restoregeometry 
\doublespacing

\subsubsection{GATEs from ITE with random assignment and without random noise} 

The GATE estimation with random assignment and without noise shows similar patterns as for the IATE estimation discussed in \ref{sec:app-ite-ra}. However, the outstanding performance of Lasso conditional mean regression for the large ITE2 is even more pronounced with mean absolute biases of less than halve of the next best estimator. Furthermore, both versions of Causal Forest perform best for ITE1 with 4,000 observations.

\label{sec:app-gate-ra}
\newgeometry{left=0.4in,right=0.5in,top=1in,bottom=1.2in,nohead}
\begin{landscape}
    \singlespacing

\begin{threeparttable}[t]
  \centering \footnotesize      
  \caption{Performance measures for GATE of ITE0 with random assignment and without random noise}
    \begin{tabular}{lccccccccccc}
    \toprule
          & $\overline{MSE}$   & $SE(\overline{MSE})$  & Median MSE & $|\overline{Bias}|$ & $\overline{Bias}$ & $\overline{SD}$    & $JB$    & Skew. & Kurt. & Corr. & Var. ratio \\
          \midrule
          & (1)   & (2)   & (3)   & (4)   & (5)   & (6)   & (7)   & (8)   & (9)   & (10)  & (11) \\
          \midrule
          & \multicolumn{11}{c}{\textbf{1000 observations}} \\
    \midrule
    Conditional mean regression & 1.43  & 0.03  & 1.20  & 0.02  & 0.01  & 1.19  & 23\%  & 0.0   & 3.1   & -     & - \\
    MOM IPW & 1.13  & 0.02  & 1.05  & 0.01  & 0.01  & 1.06  & 28\%  & 0.0   & 3.1   & -     & - \\
    MOM DR & \textbf{0.75} & 0.02  & 0.87  & 0.01  & 0.01  & 0.87  & 0\%   & 0.0   & 3.0   & -     & - \\
    Causal Forest & 0.94  & 0.02  & 0.97  & 0.01  & 0.01  & 0.97  & 11\%  & 0.0   & 3.1   & -     & - \\
    Causal Forest with local centering & 0.90  & 0.02  & 0.94  & 0.02  & 0.01  & 0.94  & 2\%   & 0.0   & 3.0   & -     & - \\
    \textit{Lasso:} &       &       &       &       &       &       &       &       &       &       &  \\
    Conditional mean regression & 2.90  & 0.04  & 1.64  & 0.02  & 0.01  & 1.68  & 20\%  & 0.0   & 3.1   & -     & - \\
    MOM IPW & 1.09  & 0.03  & 1.00  & 0.01  & 0.01  & 1.04  & 100\% & 0.0   & 4.3   & -     & - \\
    MOM DR & 0.78  & 0.02  & 0.88  & 0.01  & 0.01  & 0.88  & 61\%  & 0.0   & 3.5   & -     & - \\
    MCM   & 1.04  & 0.03  & 0.96  & 0.01  & 0.01  & 1.01  & 100\% & 0.0   & 4.3   & -     & - \\
    MCM with efficiency augmentation & \textbf{0.74} & 0.02  & 0.85  & 0.02  & 0.02  & 0.86  & 64\%  & 0.0   & 3.4   & -     & - \\
    R-learning & \textbf{0.73} & 0.02  & 0.85  & 0.02  & 0.02  & 0.86  & 56\%  & 0.0   & 3.4   & -     & - \\
    \midrule
          & \multicolumn{11}{c}{\textbf{4000 observations}} \\
    \midrule
    \textit{Random Forest:} &       &       &       &       &       &       &       &       &       &       &  \\
    Conditional mean regression & 0.76  & 0.02  & 0.84  & 0.04  & 0.02  & 0.86  & 6\%   & 0.0   & 3.1   & -     & - \\
    MOM IPW & 0.33  & 0.01  & 0.57  & 0.03  & 0.03  & 0.57  & 11\%  & 0.0   & 3.2   & -     & - \\
    MOM DR & 0.20  & 0.01  & 0.45  & 0.02  & 0.02  & 0.45  & 2\%   & 0.0   & 3.0   & -     & - \\
    Causal Forest & 0.26  & 0.01  & 0.52  & 0.03  & 0.03  & 0.51  & 0\%   & 0.1   & 3.0   & -     & - \\
    Causal Forest with local centering & 0.25  & 0.01  & 0.50  & 0.02  & 0.02  & 0.50  & 3\%   & 0.1   & 3.0   & -     & - \\
    \textit{Lasso:} &       &       &       &       &       &       &       &       &       &       &  \\
    Conditional mean regression & 1.13  & 0.02  & 1.03  & 0.05  & 0.01  & 1.05  & 6\%   & 0.0   & 3.0   & -     & - \\
    MOM IPW & 0.24  & 0.01  & 0.46  & 0.03  & 0.03  & 0.48  & 59\%  & 0.1   & 4.7   & -     & - \\
    MOM DR & \textbf{0.17} & 0.01  & 0.41  & 0.02  & 0.02  & 0.41  & 6\%   & 0.1   & 3.0   & -     & - \\
    MCM   & 0.23  & 0.01  & 0.45  & 0.03  & 0.03  & 0.47  & 56\%  & 0.2   & 4.4   & -     & - \\
    MCM with efficiency augmentation & \textbf{0.16} & 0.01  & 0.40  & 0.02  & 0.02  & 0.40  & 14\%  & 0.1   & 3.1   & -     & - \\
    R-learning & \textbf{0.16} & 0.01  & 0.40  & 0.02  & 0.02  & 0.40  & 17\%  & 0.1   & 3.1   & -     & - \\
    \bottomrule

    \end{tabular}%
         \begin{tablenotes} \item \textit{Notes:} Table shows the performance measures defined in Sections \ref{sec:performance} and \ref{sec:app-performance} over 2000 replications for the sample size of 1000 observations and 500 replications for the sample size of 4000 observations. Bold numbers indicate the best performing estimators in terms of $\overline{MSE}$ and estimators within two standard (simulation) errors of the lowest $\overline{MSE}$. \end{tablenotes}  
  \label{tab:app-gate0-ra}%
\end{threeparttable}

\doublespacing
    \singlespacing

\begin{threeparttable}[t]
  \centering \footnotesize
  \caption{Performance measures for GATE of ITE1 with random assignment and without random noise}
    \begin{tabular}{lccccccccccc}
    \toprule
          & $\overline{MSE}$   & $SE(\overline{MSE})$  & Median MSE & $|\overline{Bias}|$ & $\overline{Bias}$ & $\overline{SD}$    & $JB$    & Skew. & Kurt. & Corr. & Var. ratio \\
          \midrule
          & (1)   & (2)   & (3)   & (4)   & (5)   & (6)   & (7)   & (8)   & (9)   & (10)  & (11) \\
          \midrule
          & \multicolumn{11}{c}{\textbf{1000 observations}} \\
    \midrule
    \textit{Random Forest:} &       &       &       &       &       &       &       &       &       &       &  \\
    Conditional mean regression & 2.27  & 0.03  & 1.42  & 0.79  & 0.31  & 1.19  & 17\%  & 0.0   & 3.1   & 0.24  & 0.30 \\
    MOM IPW & 2.12  & 0.03  & 1.33  & 0.85  & 0.34  & 1.05  & 22\%  & 0.0   & 3.1   & 0.19  & 0.18 \\
    MOM DR & \textbf{1.74} & 0.02  & 1.19  & 0.85  & 0.34  & 0.86  & 2\%   & 0.0   & 3.0   & 0.28  & 0.09 \\
    Causal Forest & 1.84  & 0.03  & 1.24  & 0.80  & 0.32  & 0.97  & 9\%   & 0.0   & 3.1   & 0.30  & 0.13 \\
    Causal Forest with local centering & \textbf{1.78} & 0.02  & 1.23  & 0.80  & 0.33  & 0.94  & 6\%   & 0.0   & 3.0   & 0.29  & 0.13 \\
    \textit{Lasso:} &       &       &       &       &       &       &       &       &       &       &  \\
    Conditional mean regression & 3.47  & 0.04  & 1.84  & 0.57  & 0.20  & 1.71  & 16\%  & 0.0   & 3.1   & 0.22  & 0.86 \\
    MOM IPW & 2.39  & 0.03  & 1.38  & 0.97  & 0.38  & 1.04  & 100\% & -0.1  & 4.3   & 0.02  & 0.15 \\
    MOM DR & 2.09  & 0.03  & 1.27  & 0.97  & 0.38  & 0.90  & 77\%  & -0.1  & 3.5   & 0.06  & 0.07 \\
    MCM   & 2.43  & 0.03  & 1.41  & 1.02  & 0.34  & 1.02  & 100\% & 0.0   & 4.4   & 0.02  & 0.13 \\
    MCM with efficiency augmentation & 2.05  & 0.03  & 1.25  & 0.97  & 0.39  & 0.87  & 67\%  & 0.0   & 3.5   & 0.06  & 0.06 \\
    R-learning & 2.05  & 0.02  & 1.25  & 0.97  & 0.39  & 0.87  & 75\%  & -0.1  & 3.5   & 0.06  & 0.06 \\
    \midrule
          & \multicolumn{11}{c}{\textbf{4000 observations}} \\
    \midrule
    \textit{Random Forest:} &       &       &       &       &       &       &       &       &       &       &  \\
    Conditional mean regression & 1.34  & 0.03  & 1.10  & 0.65  & 0.26  & 0.87  & 8\%   & 0.0   & 3.1   & 0.37  & 0.30 \\
    MOM IPW & 1.22  & 0.02  & 0.94  & 0.79  & 0.33  & 0.58  & 11\%  & 0.0   & 3.2   & 0.34  & 0.10 \\
    MOM DR & 1.04  & 0.02  & 0.87  & 0.78  & 0.32  & 0.45  & 0\%   & 0.1   & 3.0   & 0.47  & 0.07 \\
    Causal Forest & \textbf{0.97} & 0.02  & 0.84  & 0.70  & 0.30  & 0.52  & 8\%   & 0.0   & 3.1   & 0.48  & 0.10 \\
    Causal Forest with local centering & \textbf{0.95} & 0.02  & 0.83  & 0.70  & 0.29  & 0.52  & 5\%   & 0.0   & 3.1   & 0.48  & 0.10 \\
    \textit{Lasso:} &       &       &       &       &       &       &       &       &       &       &  \\
    Conditional mean regression & 1.28  & 0.02  & 1.13  & 0.25  & 0.09  & 1.07  & 5\%   & 0.0   & 3.0   & 0.47  & 0.61 \\
    MOM IPW & 1.58  & 0.02  & 1.07  & 0.98  & 0.40  & 0.51  & 80\%  & -0.2  & 4.9   & -0.01 & 0.04 \\
    MOM DR & 1.36  & 0.02  & 0.97  & 0.92  & 0.37  & 0.46  & 48\%  & -0.1  & 3.3   & 0.12  & 0.03 \\
    MCM   & 1.63  & 0.02  & 1.09  & 1.02  & 0.36  & 0.48  & 69\%  & 0.0   & 4.7   & 0.05  & 0.03 \\
    MCM with efficiency augmentation & 1.39  & 0.02  & 0.98  & 0.93  & 0.38  & 0.45  & 53\%  & -0.1  & 3.3   & 0.21  & 0.03 \\
    R-learning & 1.38  & 0.02  & 0.98  & 0.93  & 0.38  & 0.45  & 44\%  & -0.1  & 3.3   & 0.34  & 0.03 \\
    \bottomrule
    \end{tabular}%
         \begin{tablenotes} \item \textit{Notes:} Table shows the performance measures defined in Sections \ref{sec:performance} and \ref{sec:app-performance} over 2000 replications for the sample size of 1000 observations and 500 replications for the sample size of 4000 observations. Bold numbers indicate the best performing estimators in terms of $\overline{MSE}$ and estimators within two standard (simulation) errors of the lowest $\overline{MSE}$. \end{tablenotes}  
  \label{tab:app-gate1-ra}%
\end{threeparttable}

\doublespacing

    \singlespacing

\begin{threeparttable}[t]
  \centering \footnotesize
\caption{Performance measures for GATE of ITE2 with random assignment and without random noise}
    \begin{tabular}{lccccccccccc}
    \toprule
          & $\overline{MSE}$   & $SE(\overline{MSE})$  & Median MSE & $|\overline{Bias}|$ & $\overline{Bias}$ & $\overline{SD}$    & $JB$    & Skew. & Kurt. & Corr. & Var. ratio \\
          \midrule
          & (1)   & (2)   & (3)   & (4)   & (5)   & (6)   & (7)   & (8)   & (9)   & (10)  & (11) \\
          \midrule
          & \multicolumn{11}{c}{\textbf{1000 observations}} \\
    \midrule
    \textit{Random Forest:} &       &       &       &       &       &       &       &       &       &       &  \\
    Conditional mean regression & 5.26  & 0.05  & 2.03  & 1.66  & 0.53  & 1.26  & 8\%   & 0.0   & 3.0   & 0.66  & 0.19 \\
    MOM IPW & 13.35 & 0.06  & 2.98  & 3.01  & 1.09  & 1.06  & 27\%  & 0.0   & 3.1   & 0.56  & 0.04 \\
    MOM DR & 10.96 & 0.06  & 2.59  & 2.74  & 0.99  & 0.92  & 39\%  & 0.0   & 3.1   & 0.64  & 0.05 \\
    Causal Forest & 8.07  & 0.07  & 2.22  & 2.21  & 0.77  & 1.15  & 59\%  & -0.1  & 3.1   & 0.66  & 0.11 \\
    Causal Forest with local centering & 7.41  & 0.07  & 2.12  & 2.10  & 0.73  & 1.13  & 55\%  & -0.1  & 3.1   & 0.66  & 0.12 \\
    \textit{Lasso:} &       &       &       &       &       &       &       &       &       &       &  \\
    Conditional mean regression & \textbf{4.52} & 0.04  & 1.97  & 0.95  & 0.21  & 1.74  & 19\%  & 0.0   & 3.1   & 0.65  & 0.33 \\
    MOM IPW & 16.87 & 0.10  & 3.37  & 3.39  & 1.20  & 1.25  & 100\% & -0.1  & 4.4   & 0.43  & 0.03 \\
    MOM DR & 10.78 & 0.09  & 2.72  & 2.65  & 0.93  & 1.20  & 100\% & -0.1  & 3.6   & 0.64  & 0.07 \\
    MCM   & 19.02 & 0.09  & 3.64  & 3.67  & 1.24  & 1.13  & 100\% & -0.1  & 4.4   & 0.16  & 0.02 \\
    MCM with efficiency augmentation & 11.17 & 0.09  & 2.79  & 2.71  & 0.96  & 1.18  & 100\% & -0.1  & 3.4   & 0.60  & 0.06 \\
    R-learning & 11.14 & 0.09  & 2.79  & 2.71  & 0.96  & 1.18  & 100\% & -0.1  & 3.4   & 0.62  & 0.06 \\
    \midrule
          & \multicolumn{11}{c}{\textbf{4000 observations}} \\
    \midrule
    \textit{Random Forest:} &       &       &       &       &       &       &       &       &       &       &  \\
    Conditional mean regression & 1.70  & 0.03  & 1.01  & 0.73  & 0.13  & 0.88  & 17\%  & 0.0   & 3.1   & 0.69  & 0.37 \\
    MOM IPW & 8.08  & 0.09  & 2.14  & 2.36  & 0.84  & 0.67  & 11\%  & 0.0   & 3.2   & 0.67  & 0.08 \\
    MOM DR & 3.63  & 0.06  & 1.46  & 1.49  & 0.48  & 0.59  & 3\%   & 0.1   & 3.0   & 0.69  & 0.19 \\
    Causal Forest & 1.97  & 0.03  & 1.19  & 1.02  & 0.23  & 0.67  & 5\%   & 0.1   & 3.0   & 0.70  & 0.30 \\
    Causal Forest with local centering & 1.74  & 0.03  & 1.18  & 0.96  & 0.18  & 0.65  & 3\%   & 0.1   & 3.0   & 0.70  & 0.32 \\
    \textit{Lasso:} &       &       &       &       &       &       &       &       &       &       &  \\
    Conditional mean regression & \textbf{1.45} & 0.02  & 1.17  & 0.44  & 0.02  & 1.05  & 11\%  & 0.0   & 3.0   & 0.70  & 0.41 \\
    MOM IPW & 8.95  & 0.15  & 2.47  & 2.47  & 0.88  & 0.86  & 47\%  & -0.1  & 3.3   & 0.66  & 0.08 \\
    MOM DR & 3.49  & 0.05  & 1.51  & 1.48  & 0.49  & 0.69  & 25\%  & 0.0   & 3.2   & 0.69  & 0.19 \\
    MCM   & 14.26 & 0.14  & 3.18  & 3.23  & 1.10  & 0.71  & 83\%  & 0.0   & 3.8   & 0.58  & 0.02 \\
    MCM with efficiency augmentation & 3.73  & 0.06  & 1.56  & 1.54  & 0.52  & 0.68  & 25\%  & 0.0   & 3.2   & 0.69  & 0.18 \\
    R-learning & 3.74  & 0.06  & 1.56  & 1.54  & 0.53  & 0.69  & 22\%  & 0.0   & 3.2   & 0.69  & 0.18 \\
    \bottomrule
    \end{tabular}%
         \begin{tablenotes} \item \textit{Notes:} Table shows the performance measures defined in Sections \ref{sec:performance} and \ref{sec:app-performance} over 2000 replications for the sample size of 1000 observations and 500 replications for the sample size of 4000 observations. Bold numbers indicate the best performing estimators in terms of $\overline{MSE}$ and estimators within two standard (simulation) errors of the lowest $\overline{MSE}$. \end{tablenotes}  
  \label{tab:app-gate2-ra}%
\end{threeparttable}

\doublespacing

\end{landscape}
\restoregeometry 
\doublespacing

\subsubsection{GATEs from ITE with random assignment and random noise}  \label{sec:app-gate-rarn}

The relative performances for the GATE estimators for ITEs with noise but randomized treatment assignment are very close to their selective equivalents. The only exception is ITE2 with 4,000 observations where the two Causal Forest versions perform best instead of MCM in the selective case. The latter is in the randomized setting the worst estimator. This emphasizes again that the averaging of noisy estimators is not always successful.

\newgeometry{left=0.4in,right=0.5in,top=1in,bottom=1.2in,nohead}
\begin{landscape}
    \singlespacing

\begin{threeparttable}[t]
  \centering \footnotesize      
  \caption{Performance measures for GATE of ITE0 with random assignment and random noise}
    \begin{tabular}{lccccccccccc}
    \toprule
          & $\overline{MSE}$   & $SE(\overline{MSE})$  & Median MSE & $|\overline{Bias}|$ & $\overline{Bias}$ & $\overline{SD}$    & $JB$    & Skew. & Kurt. & Corr. & Var. ratio \\
          \midrule
          & (1)   & (2)   & (3)   & (4)   & (5)   & (6)   & (7)   & (8)   & (9)   & (10)  & (11) \\
          \midrule
          & \multicolumn{11}{c}{\textbf{1000 observations}} \\
    \midrule
    \textit{Random Forest:} &       &       &       &       &       &       &       &       &       &       &  \\
    Conditional mean regression & 1.43  & 0.02  & 1.20  & 0.08  & 0.00  & 1.19  & 20\%  & 0.0   & 3.1   & 0.00  & 1.02 \\
    MOM IPW & 1.15  & 0.02  & 1.06  & 0.08  & -0.01 & 1.06  & 30\%  & 0.0   & 3.1   & 0.00  & 0.63 \\
    MOM DR & \textbf{0.76} & 0.02  & 0.87  & 0.08  & -0.01 & 0.86  & 8\%   & 0.0   & 3.1   & 0.01  & 0.26 \\
    Causal Forest & 0.95  & 0.02  & 0.97  & 0.08  & -0.01 & 0.97  & 13\%  & 0.0   & 3.1   & 0.00  & 0.37 \\
    Causal Forest with local centering & 0.89  & 0.02  & 0.94  & 0.08  & 0.00  & 0.94  & 6\%   & 0.0   & 3.0   & 0.01  & 0.37 \\
    \textit{Lasso:} &       &       &       &       &       &       &       &       &       &       &  \\
    Conditional mean regression & 2.87  & 0.03  & 1.63  & 0.08  & 0.00  & 1.67  & 20\%  & 0.0   & 3.1   & 0.00  & 2.86 \\
    MOM IPW & 1.09  & 0.03  & 1.01  & 0.08  & -0.01 & 1.03  & 100\% & 0.1   & 4.4   & 0.01  & 0.54 \\
    MOM DR & 0.79  & 0.02  & 0.88  & 0.09  & 0.00  & 0.88  & 63\%  & 0.0   & 3.6   & 0.04  & 0.21 \\
    MCM   & 1.06  & 0.03  & 0.97  & 0.09  & -0.04 & 1.02  & 100\% & 0.0   & 4.2   & 0.01  & 0.50 \\
    MCM with efficiency augmentation & \textbf{0.75} & 0.02  & 0.86  & 0.09  & 0.00  & 0.86  & 61\%  & 0.0   & 3.5   & 0.02  & 0.17 \\
    R-learning & \textbf{0.74} & 0.02  & 0.86  & 0.09  & 0.00  & 0.85  & 58\%  & 0.0   & 3.5   & 0.02  & 0.17 \\
    \midrule
          & \multicolumn{11}{c}{\textbf{4000 observations}} \\
    \midrule
    \textit{Random Forest:} &       &       &       &       &       &       &       &       &       &       &  \\
    Conditional mean regression & 0.77  & 0.02  & 0.84  & 0.09  & 0.01  & 0.86  & 11\%  & 0.0   & 3.1   & 0.00  & 0.80 \\
    MOM IPW & 0.35  & 0.01  & 0.59  & 0.08  & 0.01  & 0.58  & 11\%  & 0.0   & 3.2   & 0.01  & 0.25 \\
    MOM DR & 0.21  & 0.01  & 0.45  & 0.08  & 0.00  & 0.44  & 0\%   & 0.1   & 3.0   & 0.01  & 0.10 \\
    Causal Forest & 0.27  & 0.01  & 0.51  & 0.08  & 0.01  & 0.50  & 3\%   & 0.1   & 3.0   & 0.01  & 0.15 \\
    Causal Forest with local centering & 0.25  & 0.01  & 0.50  & 0.08  & 0.01  & 0.49  & 5\%   & 0.1   & 3.0   & 0.01  & 0.15 \\
    \textit{Lasso:} &       &       &       &       &       &       &       &       &       &       &  \\
    Conditional mean regression & 1.12  & 0.02  & 1.03  & 0.09  & 0.01  & 1.04  & 9\%   & 0.0   & 3.1   & 0.01  & 1.24 \\
    MOM IPW & 0.24  & 0.01  & 0.47  & 0.08  & 0.01  & 0.48  & 63\%  & 0.1   & 4.4   & 0.03  & 0.11 \\
    MOM DR & \textbf{0.18} & 0.01  & 0.42  & 0.09  & 0.00  & 0.41  & 6\%   & 0.1   & 3.0   & 0.03  & 0.04 \\
    MCM   & 0.23  & 0.01  & 0.46  & 0.09  & -0.02 & 0.47  & 59\%  & 0.1   & 4.4   & 0.00  & 0.10 \\
    MCM with efficiency augmentation & \textbf{0.17} & 0.01  & 0.41  & 0.09  & 0.00  & 0.40  & 22\%  & 0.1   & 3.1   & 0.03  & 0.03 \\
    R-learning & \textbf{0.17} & 0.01  & 0.41  & 0.08  & 0.00  & 0.40  & 13\%  & 0.1   & 3.0   & 0.06  & 0.03 \\
    \bottomrule
    \end{tabular}%
         \begin{tablenotes} \item \textit{Notes:} Table shows the performance measures defined in Sections \ref{sec:performance} and \ref{sec:app-performance} over 2000 replications for the sample size of 1000 observations and 500 replications for the sample size of 4000 observations. Bold numbers indicate the best performing estimators in terms of $\overline{MSE}$ and estimators within two standard (simulation) errors of the lowest $\overline{MSE}$. \end{tablenotes}  
  \label{tab:app-gate0-rarn}%
\end{threeparttable}

\doublespacing
    \singlespacing

\begin{threeparttable}[t]
  \centering \footnotesize
  \caption{Performance measures for GATE of ITE1 with random assignment and random noise}
    \begin{tabular}{lccccccccccc}
    \toprule
          & $\overline{MSE}$   & $SE(\overline{MSE})$  & Median MSE & $|\overline{Bias}|$ & $\overline{Bias}$ & $\overline{SD}$    & $JB$    & Skew. & Kurt. & Corr. & Var. ratio \\
          \midrule
          & (1)   & (2)   & (3)   & (4)   & (5)   & (6)   & (7)   & (8)   & (9)   & (10)  & (11) \\
          \midrule
          & \multicolumn{11}{c}{\textbf{1000 observations}} \\
    \midrule
    \textit{Random Forest:} &       &       &       &       &       &       &       &       &       &       &  \\
    Conditional mean regression & 1.55  & 0.02  & 1.23  & 0.32  & 0.12  & 1.18  & 17\%  & 0.1   & 3.1   & 0.02  & 0.26 \\
    MOM IPW & 1.27  & 0.02  & 1.09  & 0.32  & 0.11  & 1.06  & 23\%  & 0.0   & 3.1   & 0.02  & 0.16 \\
    MOM DR & \textbf{0.89} & 0.02  & 0.91  & 0.32  & 0.11  & 0.86  & 5\%   & 0.0   & 3.0   & 0.03  & 0.07 \\
    Causal Forest & 1.07  & 0.02  & 1.01  & 0.31  & 0.11  & 0.96  & 6\%   & 0.0   & 3.1   & 0.03  & 0.09 \\
    Causal Forest with local centering & 1.02  & 0.02  & 0.99  & 0.31  & 0.11  & 0.94  & 6\%   & 0.0   & 3.0   & 0.03  & 0.10 \\
    \textit{Lasso:} &       &       &       &       &       &       &       &       &       &       &  \\
    Conditional mean regression & 2.93  & 0.03  & 1.65  & 0.26  & 0.09  & 1.66  & 19\%  & 0.0   & 3.1   & 0.02  & 0.72 \\
    MOM IPW & 1.25  & 0.03  & 1.07  & 0.35  & 0.12  & 1.03  & 100\% & 0.0   & 4.4   & 0.01  & 0.14 \\
    MOM DR & 0.98  & 0.02  & 0.95  & 0.36  & 0.13  & 0.88  & 64\%  & 0.0   & 3.6   & -0.06 & 0.05 \\
    MCM   & 1.24  & 0.03  & 1.04  & 0.37  & 0.07  & 1.01  & 100\% & 0.0   & 4.4   & 0.00  & 0.13 \\
    MCM with efficiency augmentation & 0.94  & 0.02  & 0.93  & 0.37  & 0.13  & 0.86  & 59\%  & 0.0   & 3.5   & -0.04 & 0.05 \\
    R-learning & \textbf{0.93} & 0.02  & 0.92  & 0.36  & 0.13  & 0.86  & 64\%  & 0.0   & 3.5   & -0.09 & 0.04 \\
    \midrule
          & \multicolumn{11}{c}{\textbf{4000 observations}} \\
    \midrule
    \textit{Random Forest:} &       &       &       &       &       &       &       &       &       &       &  \\
    Conditional mean regression & 0.89  & 0.02  & 0.92  & 0.29  & 0.12  & 0.86  & 11\%  & 0.0   & 3.1   & 0.03  & 0.21 \\
    MOM IPW & 0.48  & 0.01  & 0.65  & 0.31  & 0.12  & 0.58  & 13\%  & 0.0   & 3.1   & 0.03  & 0.07 \\
    MOM DR & \textbf{0.34} & 0.01  & 0.52  & 0.31  & 0.11  & 0.45  & 0\%   & 0.1   & 3.0   & 0.05  & 0.03 \\
    Causal Forest & 0.39  & 0.01  & 0.58  & 0.29  & 0.11  & 0.51  & 3\%   & 0.1   & 3.0   & 0.06  & 0.04 \\
    Causal Forest with local centering & 0.38  & 0.01  & 0.57  & 0.29  & 0.11  & 0.50  & 3\%   & 0.1   & 3.0   & 0.06  & 0.04 \\
    \textit{Lasso:} &       &       &       &       &       &       &       &       &       &       &  \\
    Conditional mean regression & 1.19  & 0.02  & 1.07  & 0.19  & 0.04  & 1.05  & 6\%   & 0.0   & 3.0   & 0.06  & 0.34 \\
    MOM IPW & 0.43  & 0.01  & 0.60  & 0.37  & 0.14  & 0.48  & 70\%  & 0.1   & 4.3   & -0.08 & 0.03 \\
    MOM DR & \textbf{0.36} & 0.01  & 0.52  & 0.36  & 0.14  & 0.41  & 14\%  & 0.0   & 3.0   & -0.10 & 0.01 \\
    MCM   & 0.42  & 0.01  & 0.60  & 0.37  & 0.09  & 0.47  & 59\%  & 0.1   & 4.3   & 0.00  & 0.03 \\
    MCM with efficiency augmentation & \textbf{0.36} & 0.01  & 0.52  & 0.36  & 0.13  & 0.40  & 14\%  & 0.1   & 3.1   & -0.10 & 0.01 \\
    R-learning & \textbf{0.36} & 0.01  & 0.51  & 0.37  & 0.14  & 0.40  & 17\%  & 0.1   & 3.1   & -0.04 & 0.01 \\
    \bottomrule
    \end{tabular}%
         \begin{tablenotes} \item \textit{Notes:} Table shows the performance measures defined in Sections \ref{sec:performance} and \ref{sec:app-performance} over 2000 replications for the sample size of 1000 observations and 500 replications for the sample size of 4000 observations. Bold numbers indicate the best performing estimators in terms of $\overline{MSE}$ and estimators within two standard (simulation) errors of the lowest $\overline{MSE}$. \end{tablenotes}  
  \label{tab:app-gate1-rarn}%
\end{threeparttable}

\doublespacing

    \singlespacing

\begin{threeparttable}[t]
  \centering \footnotesize
\caption{Performance measures for GATE of ITE2 with random assignment and random noise}
    \begin{tabular}{lccccccccccc}
    \toprule
          & $\overline{MSE}$   & $SE(\overline{MSE})$  & Median MSE & $|\overline{Bias}|$ & $\overline{Bias}$ & $\overline{SD}$    & $JB$    & Skew. & Kurt. & Corr. & Var. ratio \\
          \midrule
          & (1)   & (2)   & (3)   & (4)   & (5)   & (6)   & (7)   & (8)   & (9)   & (10)  & (11) \\
          \midrule
          & \multicolumn{11}{c}{\textbf{1000 observations}} \\
    \midrule
    \textit{Random Forest:} &       &       &       &       &       &       &       &       &       &       &  \\
    Conditional mean regression & 2.89  & 0.03  & 1.47  & 1.04  & 0.41  & 1.16  & 14\%  & 0.0   & 3.1   & 0.06  & 0.02 \\
    MOM IPW & 2.82  & 0.03  & 1.38  & 1.07  & 0.40  & 1.05  & 17\%  & 0.0   & 3.1   & 0.05  & 0.01 \\
    MOM DR & \textbf{2.44} & 0.02  & 1.24  & 1.07  & 0.39  & 0.86  & 6\%   & 0.0   & 3.0   & 0.07  & 0.01 \\
    Causal Forest & 2.51  & 0.03  & 1.29  & 1.03  & 0.38  & 0.96  & 11\%  & 0.0   & 3.1   & 0.08  & 0.01 \\
    Causal Forest with local centering & \textbf{2.46} & 0.03  & 1.28  & 1.02  & 0.39  & 0.94  & 8\%   & 0.0   & 3.0   & 0.08  & 0.01 \\
    \textit{Lasso:} &       &       &       &       &       &       &       &       &       &       &  \\
    Conditional mean regression & 3.83  & 0.04  & 1.85  & 0.81  & 0.31  & 1.67  & 13\%  & 0.0   & 3.0   & 0.08  & 0.07 \\
    MOM IPW & 3.18  & 0.03  & 1.47  & 1.19  & 0.43  & 1.04  & 100\% & -0.1  & 4.6   & -0.07 & 0.01 \\
    MOM DR & 2.85  & 0.03  & 1.38  & 1.17  & 0.43  & 0.90  & 77\%  & 0.0   & 3.5   & -0.01 & 0.01 \\
    MCM   & 3.15  & 0.03  & 1.51  & 1.22  & 0.27  & 1.01  & 100\% & 0.0   & 4.2   & 0.01  & 0.01 \\
    MCM with efficiency augmentation & 2.83  & 0.03  & 1.37  & 1.18  & 0.44  & 0.88  & 72\%  & -0.1  & 3.5   & -0.07 & 0.01 \\
    R-learning & 2.81  & 0.03  & 1.37  & 1.17  & 0.43  & 0.88  & 77\%  & 0.0   & 3.5   & -0.06 & 0.01 \\
    \midrule
          & \multicolumn{11}{c}{\textbf{4000 observations}} \\
    \midrule
    \textit{Random Forest:} &       &       &       &       &       &       &       &       &       &       &  \\
    Conditional mean regression & 1.86  & 0.03  & 1.13  & 0.89  & 0.36  & 0.84  & 9\%   & 0.0   & 3.1   & 0.12  & 0.03 \\
    MOM IPW & 1.87  & 0.02  & 1.00  & 1.01  & 0.39  & 0.58  & 8\%   & 0.0   & 3.1   & 0.10  & 0.01 \\
    MOM DR & 1.67  & 0.02  & 0.91  & 0.99  & 0.36  & 0.45  & 0\%   & 0.1   & 3.0   & 0.13  & 0.01 \\
    Causal Forest & \textbf{1.58} & 0.02  & 0.88  & 0.93  & 0.35  & 0.53  & 9\%   & 0.0   & 3.1   & 0.14  & 0.01 \\
    Causal Forest with local centering & \textbf{1.54} & 0.02  & 0.87  & 0.92  & 0.34  & 0.52  & 3\%   & 0.0   & 3.0   & 0.14  & 0.01 \\
    \textit{Lasso:} &       &       &       &       &       &       &       &       &       &       &  \\
    Conditional mean regression & 1.59  & 0.03  & 1.19  & 0.55  & 0.18  & 1.04  & 5\%   & 0.0   & 3.0   & 0.15  & 0.05 \\
    MOM IPW & 2.33  & 0.02  & 1.16  & 1.18  & 0.44  & 0.51  & 84\%  & -0.3  & 4.8   & -0.07 & 0.00 \\
    MOM DR & 2.06  & 0.03  & 1.09  & 1.11  & 0.41  & 0.47  & 66\%  & 0.0   & 3.4   & 0.00  & 0.00 \\
    MCM   & 2.38  & 0.02  & 1.22  & 1.23  & 0.29  & 0.48  & 67\%  & -0.1  & 4.7   & -0.05 & 0.00 \\
    MCM with efficiency augmentation & 2.09  & 0.02  & 1.09  & 1.12  & 0.41  & 0.46  & 72\%  & 0.0   & 3.6   & -0.01 & 0.00 \\
    R-learning & 2.09  & 0.02  & 1.08  & 1.12  & 0.41  & 0.46  & 63\%  & 0.0   & 3.6   & -0.03 & 0.00 \\
    \bottomrule
    \end{tabular}%
         \begin{tablenotes} \item \textit{Notes:} Table shows the performance measures defined in Sections \ref{sec:performance} and \ref{sec:app-performance} over 2000 replications for the sample size of 1000 observations and 500 replications for the sample size of 4000 observations. Bold numbers indicate the best performing estimators in terms of $\overline{MSE}$ and estimators within two standard (simulation) errors of the lowest $\overline{MSE}$. \end{tablenotes}  
  \label{tab:app-gate2-rarn}%
\end{threeparttable}

\doublespacing

\end{landscape}
\restoregeometry 
\doublespacing

\subsection{Results for ATE estimation}  \label{sec:app-ate}

The Appendices \ref{sec:app-ate-base} to \ref{sec:app-ate-rarn} show the full results for ATE estimation in the 24 DGP-sample size combinations. Compared to IATEs and GATEs, the ATE performance measures require no averaging over several validation observations and provides the standard MSE, bias and SD for a point estimate. Also the summary of the fraction of observations with rejected JB test is not applicable in this case. Instead, we provide the p-value of the JB test to investigate whether the ATE estimators constructed as the average of the different IATEs are normally distributed.

The findings are similar to the findings for the GATE estimation. The differences in SD over all ATE estimators are minor making the bias the decisive component. This means that the estimators that account best for the selection bias perform best in the settings with selectivity. Like for the GATEs, we observe that averaging noisy IATE estimates can provide competitive ATE estimators. In particular, MCM shows consistently small bias for the ATE. As the bias drops close to zero in the randomized settings, there remains hardly any difference between the estimators.

\subsubsection{ATEs from ITE with selection and without random noise} \label{sec:app-ate-base}
    \singlespacing

\begin{threeparttable}[t]
  \centering \footnotesize      
  \caption{Performance measures for ATE of ITE0 with selection and without random noise (baseline)}
    \begin{tabular}{lcccccc}
    \toprule
          & MSE   & Bias & SD  & Skew. & Kurt. & p-value JB \\
              \midrule
          & (1)   & (2)   & (3)   & (4)   & (5)   & (6)   \\
          \midrule
          & \multicolumn{6}{c}{\textbf{1000 observations}} \\
    \midrule
    \textit{Random Forest:} &       &       &       &       &       &  \\
    Conditional mean regression & 0.97  & 0.60  & 0.78  & -0.1  & 3.0   & 0.31 \\
    MOM IPW & 1.18  & 0.71  & 0.82  & 0.0   & 3.1   & 0.34 \\
    MOM DR & \textbf{0.68} & 0.40  & 0.72  & -0.1  & 3.1   & 0.21 \\
    Causal Forest & 1.20  & 0.75  & 0.80  & -0.1  & 3.1   & 0.25 \\
    Causal Forest with local centering & 0.77  & 0.34  & 0.81  & 0.0   & 3.0   & 0.39 \\
    \textit{Lasso:} &       &       &       &       &       &  \\
    Conditional mean regression & 1.00  & 0.60  & 0.80  & 0.0   & 3.0   & 0.42 \\
    MOM IPW & 1.10  & 0.61  & 0.86  & 0.0   & 3.2   & 0.07 \\
    MOM DR & 38.75 & 0.60  & 6.20  & 43.4  & 1921.8 & 0.00 \\
    MCM   & 0.94  & 0.45  & 0.86  & -0.1  & 3.1   & 0.08 \\
    MCM with efficiency augmentation & 0.87  & 0.42  & 0.83  & 0.0   & 3.1   & 0.24 \\
    R-learning & 0.88  & 0.45  & 0.82  & 0.0   & 3.1   & 0.21 \\
    \midrule
          & \multicolumn{6}{c}{\textbf{4000 observations}} \\
    \midrule
    \textit{Random Forest:} &       &       &       &       &       &  \\
    Conditional mean regression & 0.43  & 0.53  & 0.38  & 0.1   & 2.7   & 0.14 \\
    MOM IPW & 0.46  & 0.53  & 0.41  & 0.2   & 2.8   & 0.07 \\
    MOM DR & \textbf{0.21} & 0.28  & 0.37  & 0.1   & 2.8   & 0.17 \\
    Causal Forest & 0.66  & 0.70  & 0.40  & 0.1   & 2.6   & 0.06 \\
    Causal Forest with local centering & 0.22  & 0.24  & 0.40  & 0.1   & 2.8   & 0.24 \\
    \textit{Lasso:} &       &       &       &       &       &  \\
    Conditional mean regression & 0.39  & 0.49  & 0.39  & 0.0   & 2.9   & 0.39 \\
    MOM IPW & 0.37  & 0.45  & 0.41  & 0.2   & 3.0   & 0.13 \\
    MOM DR & 0.26  & 0.31  & 0.41  & 0.1   & 2.9   & 0.36 \\
    MCM   & 0.29  & 0.34  & 0.42  & 0.2   & 2.9   & 0.17 \\
    MCM with efficiency augmentation & 0.23  & 0.26  & 0.40  & 0.1   & 3.0   & 0.38 \\
    R-learning & 0.24  & 0.28  & 0.41  & 0.1   & 3.0   & 0.29 \\
    \bottomrule

    \end{tabular}%
         \begin{tablenotes} \item \textit{Notes:} Table shows the performance measures defined in Sections \ref{sec:performance} and \ref{sec:app-performance} over 2000 replications for the sample size of 1000 observations and 500 replications for the sample size of 4000 observations. \end{tablenotes}  
  \label{tab:app-ate0-base}%
\end{threeparttable}

\doublespacing
    \singlespacing

\begin{threeparttable}[t]
  \centering \footnotesize
  \caption{Performance measures for ATE of ITE1 with selection and without random noise}
    \begin{tabular}{lcccccc}
    \toprule
          & MSE   & Bias & SD  & Skew. & Kurt. & p-value JB \\
              \midrule
          & (1)   & (2)   & (3)   & (4)   & (5)   & (6)   \\
          \midrule
          & \multicolumn{6}{c}{\textbf{1000 observations}} \\
    \midrule
    \textit{Random Forest:} &       &       &       &       &       &  \\
    Conditional mean regression & 1.94  & 1.16  & 0.77  & -0.1  & 3.0   & 0.32 \\
    MOM IPW & 2.01  & 1.16  & 0.81  & -0.1  & 3.1   & 0.20 \\
    MOM DR & \textbf{1.06} & 0.75  & 0.70  & -0.1  & 3.1   & 0.14 \\
    Causal Forest & 2.48  & 1.36  & 0.79  & 0.0   & 3.1   & 0.30 \\
    Causal Forest with local centering & 1.16  & 0.72  & 0.80  & 0.0   & 3.0   & 0.38 \\
    \textit{Lasso:} &       &       &       &       &       &  \\
    Conditional mean regression & 1.67  & 1.01  & 0.80  & 0.0   & 3.0   & 0.40 \\
    MOM IPW & 1.61  & 0.94  & 0.85  & -0.1  & 3.2   & 0.02 \\
    MOM DR & 35.55 & 0.87  & 5.90  & 43.3  & 1913.6 & 0.00 \\
    MCM   & \textbf{1.06} & 0.57  & 0.85  & -0.1  & 3.2   & 0.04 \\
    MCM with efficiency augmentation & 1.17  & 0.70  & 0.83  & 0.0   & 3.2   & 0.13 \\
    R-learning & 1.30  & 0.79  & 0.82  & 0.0   & 3.2   & 0.12 \\
    \midrule
          & \multicolumn{6}{c}{\textbf{4000 observations}} \\
    \midrule
    \textit{Random Forest:} &       &       &       &       &       &  \\
    Conditional mean regression & 1.23  & 1.04  & 0.38  & 0.1   & 2.7   & 0.10 \\
    MOM IPW & 0.83  & 0.82  & 0.41  & 0.1   & 2.8   & 0.15 \\
    MOM DR & 0.38  & 0.50  & 0.36  & 0.1   & 2.9   & 0.23 \\
    Causal Forest & 1.83  & 1.29  & 0.39  & 0.1   & 2.6   & 0.06 \\
    Causal Forest with local centering & 0.46  & 0.55  & 0.40  & 0.0   & 2.8   & 0.27 \\
    \textit{Lasso:} &       &       &       &       &       &  \\
    Conditional mean regression & 0.68  & 0.73  & 0.40  & 0.0   & 2.8   & 0.35 \\
    MOM IPW & 0.61  & 0.67  & 0.41  & 0.2   & 3.0   & 0.10 \\
    MOM DR & 0.38  & 0.47  & 0.40  & 0.1   & 2.9   & 0.30 \\
    MCM   & \textbf{0.33} & 0.39  & 0.42  & 0.2   & 3.1   & 0.13 \\
    MCM with efficiency augmentation & 0.34  & 0.43  & 0.40  & 0.1   & 3.0   & 0.29 \\
    R-learning & 0.44  & 0.53  & 0.40  & 0.1   & 3.1   & 0.25 \\
    \bottomrule
    \end{tabular}%
         \begin{tablenotes} \item \textit{Notes:} Table shows the performance measures defined in Sections \ref{sec:performance} and \ref{sec:app-performance} over 2000 replications for the sample size of 1000 observations and 500 replications for the sample size of 4000 observations. \end{tablenotes}  
  \label{tab:app-ate1-base}%
\end{threeparttable}

\doublespacing

    \singlespacing

\begin{threeparttable}[t]
  \centering \footnotesize
\caption{Performance measures for ATE of ITE2 with selection and without random noise}
    \begin{tabular}{lcccccc}
    \toprule
          & MSE   & Bias & SD  & Skew. & Kurt. & p-value JB \\
              \midrule
          & (1)   & (2)   & (3)   & (4)   & (5)   & (6)   \\
          \midrule
    \toprule
          & \multicolumn{6}{c}{\textbf{1000 observations}} \\
    \midrule
    \textit{Random Forest:} &       &       &       &       &       &  \\
    Conditional mean regression & 5.62  & 2.24  & 0.76  & 0.0   & 3.1   & 0.31 \\
    MOM IPW & 5.70  & 2.25  & 0.79  & -0.1  & 3.0   & 0.33 \\
    MOM DR & 3.17  & 1.64  & 0.68  & -0.1  & 3.0   & 0.18 \\
    Causal Forest & 8.72  & 2.85  & 0.77  & 0.0   & 3.1   & 0.35 \\
    Causal Forest with local centering & 3.13  & 1.59  & 0.77  & 0.0   & 3.0   & 0.42 \\
    \textit{Lasso:} &       &       &       &       &       &  \\
    Conditional mean regression & 2.75  & 1.46  & 0.79  & 0.0   & 3.0   & 0.41 \\
    MOM IPW & 3.85  & 1.77  & 0.85  & -0.2  & 3.3   & 0.00 \\
    MOM DR & 36.56 & 1.32  & 5.90  & 43.3  & 1912.9 & 0.00 \\
    MCM   & \textbf{1.68} & 0.97  & 0.86  & -0.1  & 3.1   & 0.17 \\
    MCM with efficiency augmentation & 2.35  & 1.30  & 0.80  & -0.1  & 3.3   & 0.02 \\
    R-learning & 3.00  & 1.53  & 0.80  & 0.0   & 3.1   & 0.29 \\
    \midrule
          & \multicolumn{6}{c}{\textbf{4000 observations}} \\
    \midrule
    \textit{Random Forest:} &       &       &       &       &       &  \\
    Conditional mean regression & 2.79  & 1.62  & 0.39  & 0.1   & 2.8   & 0.13 \\
    MOM IPW & 2.31  & 1.47  & 0.40  & 0.1   & 2.8   & 0.13 \\
    MOM DR & 1.07  & 0.97  & 0.35  & 0.1   & 2.8   & 0.28 \\
    Causal Forest & 6.24  & 2.47  & 0.40  & 0.1   & 2.6   & 0.05 \\
    Causal Forest with local centering & 1.08  & 0.97  & 0.38  & 0.1   & 2.8   & 0.14 \\
    \textit{Lasso:} &       &       &       &       &       &  \\
    Conditional mean regression & 0.84  & 0.84  & 0.37  & 0.1   & 2.8   & 0.22 \\
    MOM IPW & 1.56  & 1.18  & 0.41  & 0.2   & 3.0   & 0.04 \\
    MOM DR & \textbf{0.52} & 0.62  & 0.38  & 0.1   & 2.9   & 0.19 \\
    MCM   & 0.56  & 0.62  & 0.42  & 0.2   & 3.0   & 0.16 \\
    MCM with efficiency augmentation & 0.65  & 0.71  & 0.38  & 0.2   & 3.0   & 0.10 \\
    R-learning & 0.90  & 0.87  & 0.39  & 0.2   & 3.1   & 0.08 \\
    \bottomrule
    \end{tabular}%
         \begin{tablenotes} \item \textit{Notes:} Table shows the performance measures defined in Sections \ref{sec:performance} and \ref{sec:app-performance} over 2000 replications for the sample size of 1000 observations and 500 replications for the sample size of 4000 observations. \end{tablenotes}  
  \label{tab:app-ate2-base}%
\end{threeparttable}

\doublespacing

\subsubsection{ATEs from ITE with selection and random noise}   \label{sec:app-ate-rn}
    \singlespacing

\begin{threeparttable}[t]
  \centering \footnotesize      
  \caption{Performance measures for ATE of ITE0 with selection and random noise}
    \begin{tabular}{lcccccc}
    \toprule
          & MSE   & Bias & SD  & Skew. & Kurt. & p-value JB \\
              \midrule
          & (1)   & (2)   & (3)   & (4)   & (5)   & (6)   \\
          \midrule
          & \multicolumn{6}{c}{\textbf{1000 observations}} \\
    \midrule
    \textit{Random Forest:} &       &       &       &       &       &  \\
    Conditional mean regression & 0.96  & 0.60  & 0.78  & -0.1  & 3.1   & 0.27 \\
    MOM IPW & 1.17  & 0.71  & 0.82  & -0.1  & 3.1   & 0.22 \\
    MOM DR & \textbf{0.66} & 0.39  & 0.71  & -0.1  & 3.1   & 0.22 \\
    Causal Forest & 1.18  & 0.74  & 0.79  & 0.0   & 3.1   & 0.35 \\
    Causal Forest with local centering & 0.76  & 0.33  & 0.80  & 0.0   & 3.1   & 0.29 \\
    \textit{Lasso:} &       &       &       &       &       &  \\
    Conditional mean regression & 0.98  & 0.60  & 0.79  & 0.0   & 3.0   & 0.39 \\
    MOM IPW & 1.08  & 0.60  & 0.85  & -0.1  & 3.2   & 0.05 \\
    MOM DR & 60.69 & 0.63  & 7.77  & 43.9  & 1949.8 & 0.00 \\
    MCM   & 0.91  & 0.42  & 0.86  & -0.1  & 3.2   & 0.05 \\
    MCM with efficiency augmentation & 0.87  & 0.42  & 0.83  & 0.0   & 3.1   & 0.23 \\
    R-learning & 0.88  & 0.45  & 0.82  & 0.0   & 3.2   & 0.20 \\
    \midrule
          & \multicolumn{6}{c}{\textbf{4000 observations}} \\
    \midrule
    \textit{Random Forest:} &       &       &       &       &       &  \\
    Conditional mean regression & 0.43  & 0.53  & 0.38  & 0.1   & 2.7   & 0.12 \\
    MOM IPW & 0.45  & 0.53  & 0.42  & 0.2   & 2.7   & 0.07 \\
    MOM DR & \textbf{0.21} & 0.27  & 0.37  & 0.1   & 2.8   & 0.16 \\
    Causal Forest & 0.64  & 0.70  & 0.39  & 0.1   & 2.7   & 0.09 \\
    Causal Forest with local centering & \textbf{0.21} & 0.23  & 0.40  & 0.1   & 2.8   & 0.19 \\
    \textit{Lasso:} &       &       &       &       &       &  \\
    Conditional mean regression & 0.39  & 0.49  & 0.39  & 0.1   & 2.8   & 0.35 \\
    MOM IPW & 0.37  & 0.45  & 0.41  & 0.2   & 2.9   & 0.15 \\
    MOM DR & 0.26  & 0.31  & 0.40  & 0.1   & 2.9   & 0.27 \\
    MCM   & 0.27  & 0.31  & 0.42  & 0.2   & 3.0   & 0.16 \\
    MCM with efficiency augmentation & 0.23  & 0.26  & 0.40  & 0.1   & 3.0   & 0.30 \\
    R-learning & 0.24  & 0.28  & 0.41  & 0.1   & 3.0   & 0.22 \\
    \bottomrule
    \end{tabular}%
         \begin{tablenotes} \item \textit{Notes:} Table shows the performance measures defined in Sections \ref{sec:performance} and \ref{sec:app-performance} over 2000 replications for the sample size of 1000 observations and 500 replications for the sample size of 4000 observations. \end{tablenotes}  
  \label{tab:app-ate0-rn}%
\end{threeparttable}

\doublespacing
    \singlespacing

\begin{threeparttable}[t]
  \centering \footnotesize
  \caption{Performance measures for ATE of ITE1 with selection and random noise (baseline)}
    \begin{tabular}{lcccccc}
    \toprule
          & MSE   & Bias & SD  & Skew. & Kurt. & p-value JB \\
              \midrule
          & (1)   & (2)   & (3)   & (4)   & (5)   & (6)   \\
          \midrule
          & \multicolumn{6}{c}{\textbf{1000 observations}} \\
    \midrule
    \textit{Random Forest:} &       &       &       &       &       &  \\
    Conditional mean regression & 1.24  & 0.80  & 0.77  & -0.1  & 3.1   & 0.24 \\
    MOM IPW & 1.42  & 0.87  & 0.81  & 0.0   & 3.1   & 0.31 \\
    MOM DR & \textbf{0.77} & 0.52  & 0.71  & 0.0   & 3.1   & 0.22 \\
    Causal Forest & 1.55  & 0.96  & 0.79  & -0.1  & 3.1   & 0.24 \\
    Causal Forest with local centering & 0.86  & 0.48  & 0.80  & 0.0   & 3.1   & 0.31 \\
    \textit{Lasso:} &       &       &       &       &       &  \\
    Conditional mean regression & 1.20  & 0.76  & 0.79  & 0.0   & 3.0   & 0.44 \\
    MOM IPW & 1.25  & 0.73  & 0.85  & -0.1  & 3.3   & 0.00 \\
    MOM DR & 38.89 & 0.70  & 6.20  & 43.4  & 1922.0 & 0.00 \\
    MCM   & 0.94  & 0.46  & 0.85  & -0.1  & 3.2   & 0.05 \\
    MCM with efficiency augmentation & 0.95  & 0.52  & 0.83  & 0.0   & 3.2   & 0.15 \\
    R-learning & 1.00  & 0.58  & 0.82  & 0.0   & 3.2   & 0.17 \\
    \midrule
          & \multicolumn{6}{c}{\textbf{4000 observations}} \\
    \midrule
    \textit{Random Forest:} &       &       &       &       &       &  \\
    Conditional mean regression & 0.67  & 0.73  & 0.38  & 0.1   & 2.7   & 0.15 \\
    MOM IPW & 0.57  & 0.63  & 0.42  & 0.1   & 2.8   & 0.14 \\
    MOM DR & \textbf{0.26} & 0.35  & 0.37  & 0.1   & 2.9   & 0.27 \\
    Causal Forest & 0.99  & 0.91  & 0.40  & 0.1   & 2.7   & 0.11 \\
    Causal Forest with local centering & 0.29  & 0.35  & 0.40  & 0.1   & 2.8   & 0.26 \\
    \textit{Lasso:} &       &       &       &       &       &  \\
    Conditional mean regression & 0.52  & 0.60  & 0.39  & 0.0   & 2.9   & 0.42 \\
    MOM IPW & 0.44  & 0.53  & 0.41  & 0.2   & 3.0   & 0.12 \\
    MOM DR & 0.30  & 0.37  & 0.40  & 0.1   & 3.0   & 0.34 \\
    MCM   & 0.28  & 0.32  & 0.42  & 0.2   & 3.1   & 0.16 \\
    MCM with efficiency augmentation & 0.27  & 0.33  & 0.40  & 0.1   & 3.0   & 0.36 \\
    R-learning & 0.31  & 0.38  & 0.40  & 0.1   & 3.0   & 0.32 \\
    \bottomrule
    \end{tabular}%
         \begin{tablenotes} \item \textit{Notes:} Table shows the performance measures defined in Sections \ref{sec:performance} and \ref{sec:app-performance} over 2000 replications for the sample size of 1000 observations and 500 replications for the sample size of 4000 observations. \end{tablenotes}  
  \label{tab:app-ate1-rn}%
\end{threeparttable}

\doublespacing

    \singlespacing

\begin{threeparttable}[t]
  \centering \footnotesize
\caption{Performance measures for ATE of ITE2 with selection and random noise (baseline)}
    \begin{tabular}{lcccccc}
    \toprule
          & MSE   & Bias & SD  & Skew. & Kurt. & p-value JB \\
              \midrule
          & (1)   & (2)   & (3)   & (4)   & (5)   & (6)   \\
          \midrule
          & \multicolumn{6}{c}{\textbf{1000 observations}} \\
    \midrule
    \textit{Random Forest:} &       &       &       &       &       &  \\
    Conditional mean regression & 2.13  & 1.24  & 0.77  & -0.1  & 3.1   & 0.24 \\
    MOM IPW & 2.15  & 1.23  & 0.81  & -0.1  & 3.0   & 0.31 \\
    MOM DR & 1.17  & 0.82  & 0.71  & -0.1  & 3.1   & 0.05 \\
    Causal Forest & 2.64  & 1.43  & 0.78  & -0.1  & 3.1   & 0.23 \\
    Causal Forest with local centering & 1.29  & 0.81  & 0.80  & -0.1  & 3.0   & 0.23 \\
    \textit{Lasso:} &       &       &       &       &       &  \\
    Conditional mean regression & 1.88  & 1.12  & 0.80  & -0.1  & 3.0   & 0.31 \\
    MOM IPW & 1.75  & 1.02  & 0.84  & -0.1  & 3.2   & 0.01 \\
    MOM DR & 23.44 & 0.92  & 4.75  & 42.5  & 1867.4 & 0.00 \\
    MCM   & \textbf{0.99} & 0.52  & 0.85  & -0.1  & 3.2   & 0.02 \\
    MCM with efficiency augmentation & 1.29  & 0.79  & 0.82  & -0.1  & 3.2   & 0.07 \\
    R-learning & 1.46  & 0.89  & 0.81  & -0.1  & 3.1   & 0.12 \\
    \midrule
          & \multicolumn{6}{c}{\textbf{4000 observations}} \\
    \midrule
    \textit{Random Forest:} &       &       &       &       &       &  \\
    Conditional mean regression & 1.41  & 1.12  & 0.38  & 0.0   & 2.7   & 0.16 \\
    MOM IPW & 0.94  & 0.88  & 0.41  & 0.0   & 2.8   & 0.31 \\
    MOM DR & 0.44  & 0.56  & 0.36  & 0.0   & 2.9   & 0.39 \\
    Causal Forest & 1.99  & 1.36  & 0.39  & 0.1   & 2.7   & 0.18 \\
    Causal Forest with local centering & 0.55  & 0.63  & 0.39  & 0.0   & 2.9   & 0.42 \\
    \textit{Lasso:} &       &       &       &       &       &  \\
    Conditional mean regression & 0.83  & 0.82  & 0.39  & 0.0   & 2.8   & 0.28 \\
    MOM IPW & 0.71  & 0.74  & 0.41  & 0.1   & 3.0   & 0.26 \\
    MOM DR & 0.46  & 0.54  & 0.41  & 0.1   & 3.1   & 0.37 \\
    MCM   & \textbf{0.29} & 0.34  & 0.42  & 0.3   & 3.0   & 0.03 \\
    MCM with efficiency augmentation & 0.42  & 0.51  & 0.40  & 0.1   & 2.9   & 0.29 \\
    R-learning & 0.56  & 0.63  & 0.40  & 0.1   & 2.9   & 0.27 \\
    \bottomrule
    \end{tabular}%
         \begin{tablenotes} \item \textit{Notes:} Table shows the performance measures defined in Sections \ref{sec:performance} and \ref{sec:app-performance} over 2000 replications for the sample size of 1000 observations and 500 replications for the sample size of 4000 observations. \end{tablenotes}  
  \label{tab:app-ate2-rn}%
\end{threeparttable}

\doublespacing

\subsubsection{ATEs from ITE with random assignment and without random noise}   \label{sec:app-ate-ra}
    \singlespacing

\begin{threeparttable}[t]
  \centering \footnotesize      
  \caption{Performance measures for ATE of ITE0 with random assignment and without random noise}
    \begin{tabular}{lcccccc}
    \toprule
          & MSE   & Bias & SD  & Skew. & Kurt. & p-value JB \\
              \midrule
          & (1)   & (2)   & (3)   & (4)   & (5)   & (6)   \\
          \midrule
          & \multicolumn{6}{c}{\textbf{1000 observations}} \\
    \midrule
    \textit{Random Forest:} &       &       &       &       &       &  \\
    Conditional mean regression & 0.62  & 0.01  & 0.79  & 0.0   & 3.1   & 0.38 \\
    MOM IPW & 0.66  & 0.01  & 0.81  & 0.0   & 3.1   & 0.23 \\
    MOM DR & \textbf{0.55} & 0.01  & 0.74  & 0.0   & 3.0   & 0.36 \\
    Causal Forest & 0.66  & 0.01  & 0.81  & 0.0   & 3.1   & 0.23 \\
    Causal Forest with local centering & 0.60  & 0.02  & 0.78  & 0.0   & 3.0   & 0.47 \\
    \textit{Lasso:} &       &       &       &       &       &  \\
    Conditional mean regression & 0.62  & 0.01  & 0.79  & 0.0   & 3.1   & 0.40 \\
    MOM IPW & 0.67  & 0.01  & 0.82  & 0.0   & 3.1   & 0.29 \\
    MOM DR & 0.63  & 0.01  & 0.79  & 0.0   & 3.0   & 0.48 \\
    MCM   & 0.67  & 0.01  & 0.82  & 0.0   & 3.1   & 0.26 \\
    MCM with efficiency augmentation & 0.61  & 0.02  & 0.78  & 0.0   & 3.0   & 0.44 \\
    R-learning & 0.61  & 0.02  & 0.78  & 0.0   & 3.0   & 0.43 \\
    \midrule
          & \multicolumn{6}{c}{\textbf{4000 observations}} \\
    \midrule
    \textit{Random Forest:} &       &       &       &       &       &  \\
    Conditional mean regression & 0.14  & 0.02  & 0.37  & 0.1   & 2.9   & 0.30 \\
    MOM IPW & 0.14  & 0.03  & 0.38  & 0.1   & 2.9   & 0.18 \\
    MOM DR & \textbf{0.12} & 0.02  & 0.35  & 0.1   & 2.9   & 0.29 \\
    Causal Forest & 0.15  & 0.03  & 0.38  & 0.1   & 2.9   & 0.33 \\
    Causal Forest with local centering & 0.13  & 0.02  & 0.37  & 0.1   & 2.9   & 0.26 \\
    \textit{Lasso:} &       &       &       &       &       &  \\
    Conditional mean regression & 0.14  & 0.02  & 0.37  & 0.1   & 2.9   & 0.39 \\
    MOM IPW & 0.15  & 0.03  & 0.38  & 0.2   & 3.0   & 0.18 \\
    MOM DR & 0.14  & 0.02  & 0.37  & 0.1   & 2.8   & 0.29 \\
    MCM   & 0.15  & 0.03  & 0.38  & 0.2   & 2.9   & 0.17 \\
    MCM with efficiency augmentation & 0.14  & 0.02  & 0.37  & 0.1   & 2.8   & 0.27 \\
    R-learning & 0.14  & 0.02  & 0.37  & 0.1   & 2.8   & 0.28 \\
    \bottomrule
    \end{tabular}%
         \begin{tablenotes} \item \textit{Notes:} Table shows the performance measures defined in Sections \ref{sec:performance} and \ref{sec:app-performance} over 2000 replications for the sample size of 1000 observations and 500 replications for the sample size of 4000 observations. \end{tablenotes}   
  \label{tab:app-ate0-ra}%
\end{threeparttable}

\doublespacing
    \singlespacing

\begin{threeparttable}[t]
  \centering \footnotesize
  \caption{Performance measures for ATE of ITE1 with random assignment and without random noise}
    \begin{tabular}{lcccccc}
    \toprule
          & MSE   & Bias & SD  & Skew. & Kurt. & p-value JB \\
              \midrule
          & (1)   & (2)   & (3)   & (4)   & (5)   & (6)   \\
          \midrule
          & \multicolumn{6}{c}{\textbf{1000 observations}} \\
    \midrule
    \textit{Random Forest:} &       &       &       &       &       &  \\
    Conditional mean regression & 0.62  & 0.05  & 0.78  & 0.0   & 3.1   & 0.30 \\
    MOM IPW & 0.65  & 0.05  & 0.81  & 0.0   & 3.1   & 0.17 \\
    MOM DR & \textbf{0.54} & 0.05  & 0.73  & 0.0   & 3.1   & 0.35 \\
    Causal Forest & 0.65  & 0.05  & 0.81  & 0.0   & 3.1   & 0.19 \\
    Causal Forest with local centering & 0.59  & 0.06  & 0.77  & 0.0   & 3.0   & 0.46 \\
    \textit{Lasso:} &       &       &       &       &       &  \\
    Conditional mean regression & 0.61  & 0.02  & 0.78  & 0.0   & 3.0   & 0.42 \\
    MOM IPW & 0.66  & 0.03  & 0.81  & 0.0   & 3.1   & 0.26 \\
    MOM DR & 0.62  & 0.03  & 0.79  & 0.0   & 3.0   & 0.47 \\
    MCM   & 0.66  & -0.02 & 0.81  & 0.1   & 3.1   & 0.23 \\
    MCM with efficiency augmentation & 0.61  & 0.04  & 0.78  & 0.0   & 3.1   & 0.40 \\
    R-learning & 0.60  & 0.04  & 0.77  & 0.0   & 3.1   & 0.38 \\
    \midrule
          & \multicolumn{6}{c}{\textbf{4000 observations}} \\
    \midrule
    \textit{Random Forest:} &       &       &       &       &       &  \\
    Conditional mean regression & 0.13  & 0.06  & 0.36  & 0.1   & 3.0   & 0.40 \\
    MOM IPW & 0.15  & 0.07  & 0.38  & 0.1   & 3.0   & 0.27 \\
    MOM DR & \textbf{0.12} & 0.05  & 0.34  & 0.1   & 2.9   & 0.29 \\
    Causal Forest & 0.14  & 0.06  & 0.37  & 0.1   & 2.9   & 0.34 \\
    Causal Forest with local centering & 0.13  & 0.06  & 0.36  & 0.1   & 3.0   & 0.36 \\
    \textit{Lasso:} &       &       &       &       &       &  \\
    Conditional mean regression & 0.13  & 0.03  & 0.36  & 0.1   & 3.0   & 0.41 \\
    MOM IPW & 0.15  & 0.05  & 0.38  & 0.1   & 3.0   & 0.25 \\
    MOM DR & 0.13  & 0.04  & 0.36  & 0.1   & 2.8   & 0.33 \\
    MCM   & 0.14  & -0.01 & 0.38  & 0.1   & 3.0   & 0.21 \\
    MCM with efficiency augmentation & 0.14  & 0.04  & 0.37  & 0.1   & 2.9   & 0.31 \\
    R-learning & 0.14  & 0.04  & 0.37  & 0.1   & 2.9   & 0.30 \\
    \bottomrule
    \end{tabular}%
         \begin{tablenotes} \item \textit{Notes:} Table shows the performance measures defined in Sections \ref{sec:performance} and \ref{sec:app-performance} over 2000 replications for the sample size of 1000 observations and 500 replications for the sample size of 4000 observations. \end{tablenotes}  
  \label{tab:app-ate1-ra}%
\end{threeparttable}

\doublespacing

    \singlespacing

\begin{threeparttable}[t]
  \centering \footnotesize
\caption{Performance measures for ATE of ITE2 with random assignment and without random noise}
    \begin{tabular}{lcccccc}
    \toprule
          & MSE   & Bias & SD  & Skew. & Kurt. & p-value JB \\
              \midrule
          & (1)   & (2)   & (3)   & (4)   & (5)   & (6)   \\
          \midrule
          & \multicolumn{6}{c}{\textbf{1000 observations}} \\
    \midrule
    \textit{Random Forest:} &       &       &       &       &       &  \\
    Conditional mean regression & 0.57  & 0.04  & 0.76  & 0.0   & 3.1   & 0.28 \\
    MOM IPW & 0.66  & 0.09  & 0.81  & 0.0   & 3.2   & 0.12 \\
    MOM DR & \textbf{0.52} & 0.08  & 0.72  & 0.0   & 3.1   & 0.25 \\
    Causal Forest & 0.63  & 0.06  & 0.79  & 0.1   & 3.1   & 0.20 \\
    Causal Forest with local centering & 0.56  & 0.06  & 0.74  & 0.0   & 3.1   & 0.27 \\
    \textit{Lasso:} &       &       &       &       &       &  \\
    Conditional mean regression & 0.55  & -0.02 & 0.74  & 0.0   & 3.0   & 0.46 \\
    MOM IPW & 0.66  & 0.03  & 0.81  & 0.1   & 3.1   & 0.07 \\
    MOM DR & 0.59  & 0.02  & 0.77  & 0.0   & 3.1   & 0.34 \\
    MCM   & 0.67  & -0.03 & 0.82  & 0.1   & 3.0   & 0.17 \\
    MCM with efficiency augmentation & 0.58  & 0.02  & 0.76  & 0.1   & 3.1   & 0.24 \\
    R-learning & 0.58  & 0.03  & 0.76  & 0.1   & 3.1   & 0.18 \\
    \midrule
          & \multicolumn{6}{c}{\textbf{4000 observations}} \\
    \midrule
    \textit{Random Forest:} &       &       &       &       &       &  \\
    Conditional mean regression & 0.12  & 0.01  & 0.35  & 0.1   & 3.0   & 0.37 \\
    MOM IPW & 0.14  & 0.08  & 0.37  & 0.1   & 3.0   & 0.28 \\
    MOM DR & \textbf{0.11} & 0.03  & 0.33  & 0.1   & 3.0   & 0.39 \\
    Causal Forest & 0.13  & 0.01  & 0.36  & 0.1   & 2.9   & 0.39 \\
    Causal Forest with local centering & 0.12  & 0.00  & 0.35  & 0.1   & 3.1   & 0.37 \\
    \textit{Lasso:} &       &       &       &       &       &  \\
    Conditional mean regression & 0.12  & -0.02 & 0.35  & 0.1   & 3.1   & 0.38 \\
    MOM IPW & 0.14  & 0.02  & 0.38  & 0.1   & 3.0   & 0.23 \\
    MOM DR & 0.12  & -0.01 & 0.35  & 0.0   & 3.0   & 0.49 \\
    MCM   & 0.14  & -0.03 & 0.38  & 0.1   & 3.0   & 0.24 \\
    MCM with efficiency augmentation & 0.13  & 0.00  & 0.36  & 0.1   & 3.0   & 0.43 \\
    R-learning & 0.13  & 0.00  & 0.35  & 0.1   & 3.0   & 0.43 \\
    \bottomrule
    \end{tabular}%
         \begin{tablenotes} \item \textit{Notes:} Table shows the performance measures defined in Sections \ref{sec:performance} and \ref{sec:app-performance} over 2000 replications for the sample size of 1000 observations and 500 replications for the sample size of 4000 observations. \end{tablenotes}  
  \label{tab:app-ate2-ra}%
\end{threeparttable}

\doublespacing

\subsubsection{ATEs from ITE with random assignment and random noise} \label{sec:app-ate-rarn}
    \singlespacing

\begin{threeparttable}[t]
  \centering \footnotesize      
  \caption{Performance measures for ATE of ITE0 with random assignment and random noise}
    \begin{tabular}{lcccccc}
    \toprule
          & MSE   & Bias & SD  & Skew. & Kurt. & p-value JB \\
              \midrule
          & (1)   & (2)   & (3)   & (4)   & (5)   & (6)   \\
          \midrule
          & \multicolumn{6}{c}{\textbf{1000 observations}} \\
    \midrule
    \textit{Random Forest:} &       &       &       &       &       &  \\
    Conditional mean regression & 0.62  & 0.02  & 0.79  & 0.0   & 3.1   & 0.40 \\
    MOM IPW & 0.66  & 0.01  & 0.81  & 0.0   & 3.2   & 0.18 \\
    MOM DR & \textbf{0.54} & 0.01  & 0.74  & 0.0   & 3.0   & 0.41 \\
    Causal Forest & 0.65  & 0.01  & 0.81  & 0.0   & 3.2   & 0.18 \\
    Causal Forest with local centering & 0.59  & 0.01  & 0.77  & 0.0   & 3.0   & 0.47 \\
    \textit{Lasso:} &       &       &       &       &       &  \\
    Conditional mean regression & 0.61  & 0.01  & 0.78  & 0.0   & 3.1   & 0.40 \\
    MOM IPW & 0.66  & 0.01  & 0.81  & 0.0   & 3.1   & 0.25 \\
    MOM DR & 0.62  & 0.02  & 0.79  & 0.0   & 3.1   & 0.44 \\
    MCM   & 0.67  & -0.02 & 0.82  & 0.0   & 3.1   & 0.23 \\
    MCM with efficiency augmentation & 0.61  & 0.02  & 0.78  & 0.0   & 3.0   & 0.45 \\
    R-learning & 0.60  & 0.02  & 0.78  & 0.0   & 3.1   & 0.42 \\
    \midrule
          & \multicolumn{6}{c}{\textbf{4000 observations}} \\
    \midrule
    \textit{Random Forest:} &       &       &       &       &       &  \\
    Conditional mean regression & 0.13  & 0.03  & 0.36  & 0.1   & 2.9   & 0.32 \\
    MOM IPW & 0.14  & 0.03  & 0.38  & 0.1   & 2.9   & 0.23 \\
    MOM DR & \textbf{0.12} & 0.02  & 0.34  & 0.1   & 2.9   & 0.25 \\
    Causal Forest & 0.14  & 0.02  & 0.38  & 0.1   & 2.9   & 0.27 \\
    Causal Forest with local centering & 0.13  & 0.02  & 0.36  & 0.1   & 2.9   & 0.29 \\
    \textit{Lasso:} &       &       &       &       &       &  \\
    Conditional mean regression & 0.13  & 0.02  & 0.37  & 0.1   & 2.9   & 0.39 \\
    MOM IPW & 0.14  & 0.03  & 0.38  & 0.1   & 2.9   & 0.20 \\
    MOM DR & 0.13  & 0.02  & 0.36  & 0.1   & 2.8   & 0.26 \\
    MCM   & 0.15  & 0.00  & 0.38  & 0.2   & 2.9   & 0.17 \\
    MCM with efficiency augmentation & 0.13  & 0.02  & 0.37  & 0.1   & 2.8   & 0.29 \\
    R-learning & 0.13  & 0.02  & 0.37  & 0.1   & 2.8   & 0.28 \\
    \bottomrule
    \end{tabular}%
         \begin{tablenotes} \item \textit{Notes:} Table shows the performance measures defined in Sections \ref{sec:performance} and \ref{sec:app-performance} over 2000 replications for the sample size of 1000 observations and 500 replications for the sample size of 4000 observations. \end{tablenotes}   
  \label{tab:app-ate0-rarn}%
\end{threeparttable}

\doublespacing
    \singlespacing

\begin{threeparttable}[t]
  \centering \footnotesize
  \caption{Performance measures for ATE of ITE1 with random assignment and random noise}
    \begin{tabular}{lcccccc}
    \toprule
          & MSE   & Bias & SD  & Skew. & Kurt. & p-value JB \\
              \midrule
          & (1)   & (2)   & (3)   & (4)   & (5)   & (6)   \\
          \midrule
          & \multicolumn{6}{c}{\textbf{1000 observations}} \\
    \midrule
    \textit{Random Forest:} &       &       &       &       &       &  \\
    Conditional mean regression & 0.61  & 0.04  & 0.78  & 0.0   & 3.1   & 0.25 \\
    MOM IPW & 0.65  & 0.03  & 0.81  & 0.0   & 3.1   & 0.26 \\
    MOM DR & \textbf{0.54} & 0.03  & 0.73  & 0.0   & 3.1   & 0.32 \\
    Causal Forest & 0.64  & 0.03  & 0.80  & 0.0   & 3.1   & 0.24 \\
    Causal Forest with local centering & 0.59  & 0.03  & 0.77  & 0.0   & 3.0   & 0.46 \\
    \textit{Lasso:} &       &       &       &       &       &  \\
    Conditional mean regression & 0.60  & 0.03  & 0.78  & 0.0   & 3.1   & 0.39 \\
    MOM IPW & 0.65  & 0.02  & 0.81  & 0.0   & 3.1   & 0.27 \\
    MOM DR & 0.62  & 0.03  & 0.79  & 0.0   & 3.0   & 0.49 \\
    MCM   & 0.66  & -0.04 & 0.81  & 0.0   & 3.1   & 0.26 \\
    MCM with efficiency augmentation & 0.60  & 0.03  & 0.78  & 0.0   & 3.0   & 0.43 \\
    R-learning & 0.60  & 0.03  & 0.77  & 0.0   & 3.0   & 0.42 \\
    \midrule
          & \multicolumn{6}{c}{\textbf{4000 observations}} \\
    \midrule
    \textit{Random Forest:} &       &       &       &       &       &  \\
    Conditional mean regression & 0.14  & 0.05  & 0.37  & 0.1   & 2.9   & 0.20 \\
    MOM IPW & 0.15  & 0.04  & 0.38  & 0.2   & 2.8   & 0.12 \\
    MOM DR & \textbf{0.12} & 0.04  & 0.34  & 0.1   & 2.9   & 0.26 \\
    Causal Forest & 0.14  & 0.04  & 0.38  & 0.1   & 2.9   & 0.21 \\
    Causal Forest with local centering & 0.13  & 0.04  & 0.36  & 0.1   & 3.0   & 0.36 \\
    \textit{Lasso:} &       &       &       &       &       &  \\
    Conditional mean regression & 0.13  & 0.03  & 0.37  & 0.1   & 3.0   & 0.30 \\
    MOM IPW & 0.15  & 0.04  & 0.38  & 0.2   & 2.9   & 0.15 \\
    MOM DR & 0.13  & 0.04  & 0.36  & 0.1   & 2.8   & 0.23 \\
    MCM   & 0.15  & -0.02 & 0.38  & 0.2   & 2.9   & 0.17 \\
    MCM with efficiency augmentation & 0.14  & 0.03  & 0.37  & 0.1   & 2.8   & 0.25 \\
    R-learning & 0.14  & 0.03  & 0.37  & 0.1   & 2.8   & 0.24 \\
    \bottomrule
    \end{tabular}%
         \begin{tablenotes} \item \textit{Notes:} Table shows the performance measures defined in Sections \ref{sec:performance} and \ref{sec:app-performance} over 2000 replications for the sample size of 1000 observations and 500 replications for the sample size of 4000 observations. \end{tablenotes}  
  \label{tab:app-ate1-rnra}%
\end{threeparttable}

\doublespacing

    \singlespacing

\begin{threeparttable}[t]
  \centering \footnotesize
\caption{Performance measures for ATE of ITE2 with random assignment and random noise}
    \begin{tabular}{lcccccc}
    \toprule
          & MSE   & Bias & SD  & Skew. & Kurt. & p-value JB \\
              \midrule
          & (1)   & (2)   & (3)   & (4)   & (5)   & (6)   \\
          \midrule
          & \multicolumn{6}{c}{\textbf{1000 observations}} \\
    \midrule
    \textit{Random Forest:} &       &       &       &       &       &  \\
    Conditional mean regression & 0.61  & 0.12  & 0.77  & 0.0   & 3.1   & 0.28 \\
    MOM IPW & 0.65  & 0.12  & 0.80  & 0.0   & 3.2   & 0.12 \\
    MOM DR & \textbf{0.54} & 0.10  & 0.73  & 0.0   & 3.0   & 0.34 \\
    Causal Forest & 0.64  & 0.11  & 0.79  & 0.0   & 3.1   & 0.17 \\
    Causal Forest with local centering & 0.60  & 0.12  & 0.76  & 0.0   & 3.1   & 0.28 \\
    \textit{Lasso:} &       &       &       &       &       &  \\
    Conditional mean regression & 0.60  & 0.11  & 0.77  & 0.0   & 3.1   & 0.39 \\
    MOM IPW & 0.65  & 0.10  & 0.80  & 0.1   & 3.1   & 0.18 \\
    MOM DR & 0.62  & 0.11  & 0.78  & 0.0   & 3.1   & 0.40 \\
    MCM   & 0.66  & -0.08 & 0.81  & 0.0   & 3.1   & 0.26 \\
    MCM with efficiency augmentation & 0.60  & 0.11  & 0.77  & 0.0   & 3.0   & 0.36 \\
    R-learning & 0.60  & 0.11  & 0.76  & 0.0   & 3.1   & 0.29 \\
    \midrule
          & \multicolumn{6}{c}{\textbf{4000 observations}} \\
    \midrule
    \textit{Random Forest:} &       &       &       &       &       &  \\
    Conditional mean regression & 0.15  & 0.13  & 0.36  & 0.1   & 2.7   & 0.11 \\
    MOM IPW & 0.15  & 0.12  & 0.37  & 0.2   & 2.8   & 0.10 \\
    MOM DR & \textbf{0.12} & 0.10  & 0.34  & 0.1   & 2.8   & 0.22 \\
    Causal Forest & 0.15  & 0.12  & 0.37  & 0.1   & 2.9   & 0.27 \\
    Causal Forest with local centering & 0.14  & 0.11  & 0.36  & 0.1   & 2.7   & 0.15 \\
    \textit{Lasso:} &       &       &       &       &       &  \\
    Conditional mean regression & 0.14  & 0.09  & 0.36  & 0.1   & 2.8   & 0.18 \\
    MOM IPW & 0.15  & 0.12  & 0.37  & 0.1   & 2.9   & 0.17 \\
    MOM DR & 0.14  & 0.11  & 0.36  & 0.1   & 2.7   & 0.14 \\
    MCM   & 0.15  & -0.06 & 0.38  & 0.1   & 3.0   & 0.32 \\
    MCM with efficiency augmentation & 0.14  & 0.11  & 0.36  & 0.1   & 2.6   & 0.10 \\
    R-learning & 0.14  & 0.11  & 0.36  & 0.1   & 2.6   & 0.09 \\
    \bottomrule
    \end{tabular}%
         \begin{tablenotes} \item \textit{Notes:} Table shows the performance measures defined in Sections \ref{sec:performance} and \ref{sec:app-performance} over 2000 replications for the sample size of 1000 observations and 500 replications for the sample size of 4000 observations. \end{tablenotes}  
  \label{tab:app-ate2-rnra}%
\end{threeparttable}

\doublespacing

\clearpage

\subsection{Computation time} \label{sec:comp-time}
\doublespacing\textbf{}
This appendix shows the average computation times (in seconds) of the different estimation approaches. We computed all our results on a \href{https://www.switch.ch/de/engines/}{SWITCHengines} cloud with 8 cores and 8GB RAM. It is difficult to compare the computation times between Random Forests and Lasso, because they depend strongly on the selection of the tuning parameters. The Lasso becomes slow when it selects many variables. The Causal Forests and MOM approaches differ in the way how they estimate the nuisance parameters. 

\begin{table}[htbp]
  \centering  \footnotesize
  \caption{Average computation time of one replication in seconds}
    \begin{tabular}{lccc}
    \toprule
          & \multicolumn{1}{c}{ITE0 w/o noise} & \multicolumn{1}{l}{ITE1 w/ noise} & \multicolumn{1}{c}{ITE2 w/ noise} \\
            \midrule
          & \multicolumn{1}{c}{(1)} & \multicolumn{1}{c}{(2)} & \multicolumn{1}{c}{(3)} \\
        \midrule
          & \multicolumn{3}{c}{\textbf{1000 observations}} \\
    \midrule
    \textit{Random Forest:} &       &       &  \\
    Infeasible & 1.1   & 2.6   & 2.8 \\
    Conditional mean regression & 4.0   & 4.1   & 4.0 \\
    MOM IPW & 5.2   & 5.1   & 5.2 \\
    MOM DR & 8.2   & 8.2   & 8.1 \\
    Causal Forest & 3.9   & 3.9   & 3.9 \\
    Causal Forest with local centering & 5.2   & 5.2   & 5.2 \\
    \textit{Lasso:} &       &       &  \\
    Infeasible & - & 26.8  & 29.5 \\
    Conditional mean regression & 7.6   & 7.7   & 7.7 \\
    MOM IPW & 12.4  & 12.3  & 12.3 \\
    MOM DR & 17.9  & 17.9  & 17.9 \\
    MCM   & 11.3  & 11.3  & 11.3 \\
    MCM with efficiency augmentation & 17.4  & 17.4  & 17.4 \\
    R-learning & 17.4  & 17.4  & 17.4 \\
    \midrule
          & \multicolumn{3}{c}{\textbf{4000 observations}} \\
    \midrule
    \textit{Random Forest:} &       &       &  \\
    Infeasible & 3.2   & 8.6   & 9.7 \\
    Conditional mean regression & 11.2  & 11.4  & 11.3 \\
    MOM IPW & 17.0  & 17.0  & 17.0 \\
    MOM DR & 32.4  & 33.1  & 32.8 \\
    Causal Forest & 11.6  & 11.8  & 11.7 \\
    Causal Forest with local centering & 18.3  & 18.3  & 18.3 \\
    \textit{Lasso:} &       &       &  \\
    Infeasible & - & 40.5  & 46.4 \\
    Conditional mean regression & 24.2  & 24.1  & 24.2 \\
    MOM IPW & 49.6  & 49.4  & 49.2 \\
    MOM DR & 68.0  & 67.9  & 67.9 \\
    MCM   & 51.8  & 51.7  & 51.5 \\
    MCM with efficiency augmentation & 67.4  & 67.2  & 67.2 \\
    R-learning & 67.4  & 67.2  & 67.3 \\
    \bottomrule
    \end{tabular}%
  \label{tab:time}%
\end{table}%

\end{appendices}

\end{document}